\documentclass[
aps,
prb,
superscriptaddress,
reprint,
onecolumn,
amsfonts,
amssymb,
amsmath,
]{revtex4-2}

\usepackage[utf8]{inputenc}
\usepackage{hyperref} 
\hypersetup{pdfpagemode={UseOutlines},
	bookmarksopen=true,
	bookmarksopenlevel=0,
	hypertexnames=false,
	colorlinks=true, 
	citecolor=blue, 
	linkcolor=red, 
	urlcolor=black, 
	pdfstartview={FitV},
	unicode,
	breaklinks=true,
}
\usepackage{graphicx}
\usepackage{grffile}
\usepackage{color}
\usepackage{array}
\usepackage{hhline}
\usepackage[]{algorithm2e}
\usepackage{booktabs}
\usepackage{ulem}
\usepackage{siunitx}
\usepackage{float}

\graphicspath{{../tmp/}}

\newlength{\figwidth}
\newcommand{\R}{\mathcal{R}}

\newcommand{\epsK}{{\varepsilon_{\!\scriptscriptstyle K}}}
\newcommand{\epsKK}{{\varepsilon_{\!\scriptscriptstyle K 2}}}
\newcommand{\epsKKKK}{{\varepsilon_{\!\scriptscriptstyle K 4}}}
\newcommand{\epsA}{{\varepsilon_{\!\scriptscriptstyle P}}}

\newcommand{\kk}{\boldsymbol{k}}
\newcommand{\eek}{\boldsymbol{e}_{\boldsymbol{k}}}

\newcommand{\eep}{\boldsymbol{e}_{p\kk}}
\newcommand{\eet}{\boldsymbol{e}_{t\kk}}
\newcommand{\eex}{\boldsymbol{e}_x}
\newcommand{\eey}{\boldsymbol{e}_y}
\newcommand{\eez}{\boldsymbol{e}_z}

\newcommand{\vv}{\boldsymbol{v}}
\newcommand{\hatvv}{\hat{\boldsymbol{v}}}
\newcommand{\hatvp}{\hat{v}_{p}}
\newcommand{\hatvt}{\hat{v}_{t}}
\newcommand{\hatb}{\hat{b}}
\newcommand{\hata}{\hat{a}}
\newcommand{\vvs}{\hat{\boldsymbol{v}}_{s}}
\newcommand{\ff}{\boldsymbol{f}}
\newcommand{\hatf}{\hat{f}}
\newcommand{\hatff}{\hat{\boldsymbol{f}}}

\newcommand{\bnabla}{\boldsymbol{\nabla}}

\newcommand{\p}{\partial}

\newcommand{\diff}{\text{d}}
\newcommand{\dxdydz}{\diff x \diff y \diff z}
\newcommand{\bv}{Brunt-V\"ais\"al\"a }
\newcommand{\kmax}{k_{\max}}
\newcommand{\thk}{\theta_{\kk}}
\newcommand{\phk}{\varphi_{\kk}}

\newcommand{\ok}{\omega_{\kk}}

\newcommand{\bOmega}{\boldsymbol{\Omega}}

\newcommand{\odoppler}{\delta \omega_{\text{doppler}}}

\newcommand{\Ivelo}{I_{\rm kin}}
\newcommand{\Idiss}{I_{\text{diss}}}

\newcommand{\Pikin}{\Pi_{\rm kin}}
\newcommand{\Pipot}{\Pi_{\text{pot}}}

\newcommand{\Etoro}{E_{\rm toro} }
\newcommand{\Epolo}{E_{\rm polo} }
\newcommand{\Epot}{E_{\rm pot} }
\newcommand{\Eequi}{E_{\rm equi} }

\newcommand{\Ewave}{E_{\rm wave}}
\newcommand{\Ewaver}{\tilde{E}_{\rm wave}}

\newcommand{\kb}{k_{\rm b}}
\newcommand{\ko}{k_{\rm O}}
\newcommand{\kd}{k_{\rm d}}

\begin{document}

\title{Internal gravity waves in stratified flows with and without vortical modes}

\author{Vincent Labarre}
\email[]{vincent.labarre@oca.eu}
\affiliation{Universit\'{e} C\^{o}te d'Azur, Observatoire de la C\^{o}te d'Azur, CNRS,
Laboratoire Lagrange, Nice, France}

\author{Pierre Augier}
\email[]{pierre.augier@univ-grenoble-alpes.fr}
\affiliation{Laboratoire des Ecoulements G\'eophysiques et Industriels, Universit\'e
Grenoble Alpes, CNRS, Grenoble-INP, F-38000 Grenoble, France}

\author{Giorgio Krstulovic}
\email[]{giorgio.krstulovic@oca.eu}
\affiliation{Universit\'{e} C\^{o}te d'Azur, Observatoire de la C\^{o}te d'Azur, CNRS,
Laboratoire Lagrange, Nice, France}

\author{Sergey Nazarenko}
\email[]{sergey.nazarenko@unice.fr}
\affiliation{Universit\'{e} C\^{o}te d'Azur, CNRS, Institut de Physique de Nice -
INPHYNI, Nice, France}

\begin{abstract}
The comprehension of stratified flows is important for geophysical and astrophysical
applications. The Weak Wave Turbulence theory aims to provide a statistical description
of internal gravity waves propagating in the bulk of such flows. However, internal
gravity waves are usually perturbed by other structures present in stratified flow,
namely the shear modes and the vortical modes. In order to check whether a weak
internal gravity wave turbulence regime can occur, we perform direct numerical
simulations of stratified turbulence without shear modes, and with or without vortical
modes at various Froude and buoyancy Reynolds numbers. We observe that removing
vortical modes naturally helps to have a better overall balance between poloidal
kinetic energy, involved in internal gravity waves, and potential energy. However,
conversion between kinetic energy and potential energy does not necessarily show
fluctuations around zero in our simulations, as we would expect for a system of weak
waves. A spatiotemporal analysis reveals that removing vortical modes helps to
concentrate the energy around the wave frequency, but it is not enough to observe a
weak wave turbulence regime. Yet, we observe that internal gravity waves whose
frequency are large compared to the eddy turnover time are present, and we also find
evidences for slow internal gravity waves interacting by Triadic Resonance
Instabilities in our strongly stratified flows simulations. Finally, we propose
conditions that should be fulfilled in order to observe a weak internal gravity waves
turbulence regime in real flows.
\end{abstract}


\maketitle


\section{Introduction}
\label{sec:introduction}

As eddies in classical hydrodynamic turbulence, waves in nonlinear systems interact and
transfer conserved quantities along scales in a cascade process. The Weak-Wave
Turbulence (WWT) theory aims to provide a statistical description of the system when
the nonlinearity is small \cite{zakharov_kolmogorov_1992,nazarenko_wave_2011,
nazarenko_wave_2015}. The applications of this theory encompass capillary-gravity waves
\cite{falcon_experiments_2022}, gravito-inertial waves in rotating and stratified
fluids \cite{caillol_kinetic_2000, galtier_weak_2003, medvedev_turbulence_2007}, 2D
acoustic waves \cite{griffin_energy_2022}, elastic plates \cite{during_weak_2006},
Alfvén waves in magnetohydrodynamics (MHD) \cite{galtier_weak_2000}, Kelvin waves in
superfluids \cite{lvov_weak_2010}, density waves in Bose-Einstein condensates
\cite{dyachenko_optical_1992}, and gravitational waves \cite{galtier_turbulence_2017}.

Three key hypotheses are used in weak-wave turbulence (WWT) theory. The first one is
weak nonlinearity of the dynamical equations. This condition is often translated in
terms of spatial scale separation: the considered scale on one side and the saturation
scale on the other. The saturation scale is here defined as the scale at which the weak
nonlinearity no longer holds, inducing wave breaking. This hypothesis is quite similar
to the separation between the integral scale and the dissipative scale in the classical
picture of 3D hydrodynamic turbulence, which is necessary for observing an inertial
range. The second hypothesis corresponds to a separation of time scales. It requires
that the linear time, given by the wave period $\tau_{\rm L}$, is much shorter than the
nonlinear time of interactions between waves  $\tau_{\rm NL}$
\cite{nazarenko_wave_2011}. The third hypothesis is that the nonlinear broadening
$\delta \omega \sim 1/\tau_{\rm NL}$, which measures the frequency of  nonlinear
interactions, must be larger than the frequency gap between wave modes $\Delta \omega$
in the discrete Fourier space. This last condition is necessary to permit enough
nonlinear interactions among waves such that we can consider Fourier space as being
continuous \cite{lvov_discrete_2010}.

For many physical situations, at least one of the three hypotheses of WWT is broken in
some range of scales, which reduces the validity of the theory
\cite{biven_breakdown_2001, lvov_discrete_2010}. Yet, when scale separation is observed
in space and time, a WWT range can emerge in wave energy spectra. Besides the practical
difficulties for obtaining scale separation, testing WWT in isotropic systems is
conceptually simpler. For this reason, main progress have been made in the experimental
and numerical verification of WWT for elastic plates \cite{miquel_nonstationary_2011,
yokoyama_identification_2014}, capillary-gravity waves \cite{pan_direct_2014,
falcon_experiments_2022}, density waves in Bose-Einstein condensates
\cite{zhu_testing_2022}, and 2D acoustic waves \cite{griffin_energy_2022}.

Anisotropic turbulence is, generally speaking, more difficult to investigate than
isotropic turbulence because of the reduced number of symmetries of the considered
system. For example, it has been shown that studying 2D spatial spectra instead of 1D
integrated spectra is essential to investigate stratified turbulence
\cite{yokoyama_energy-based_2019}. Anisotropy makes the problem multidimensional in
Fourier space, which makes the notion of spectral energy fluxes, and other relevant
quantities, more difficult to define and calculate than in isotropic turbulence
\cite{yokoyama_energy-flux_2021}. An additional difficulty is that scale separations
required by WWT can also be anisotropic. For linearly stratified flows, the weak
nonlinearity of the Navier-Stokes equations requires that the spatial scale separation
\begin{equation}
k/\kb \ll 1
\end{equation}
is satisfied,  where $\kb = N/U_h$ is the buoyancy wave-vector, $k$ is the wave-vector
modulus, $U_h$ is the rms of the horizontal velocity, and $N$ is the \bv frequency.
This condition should be fulfilled in order to avoid wave-breaking
\cite{waite_stratified_2011}. The temporal scale separation between linear waves and
eddies leads to the different condition
\begin{equation}
\tau_{\rm L}/\tau_{\rm NL} = \frac{(\epsK k^2)^{1/3}}{N k_h/k} \ll 1,
\end{equation}
where $\epsK$ is the kinetic energy dissipation rate, $k_h$ is the horizontal
wave-vector modulus, $\tau_{\rm L} = 2\pi k /(N k_h)$ is the period of internal gravity
waves, and $\tau_{\rm NL} = 2\pi/(k^2 \epsK)^{1/3}$ is the eddy turnover time
\cite{yokoyama_energy-based_2019}. Physically, when $\tau_{\rm L}/\tau_{\rm NL} \ll 1$,
the waves are faster than the typical time of their nonlinear interactions. Due to the
anisotropic dispersion relation of internal gravity waves, time scale separation is
less valid for small $k_h$, and even impossible for modes with $k_h=0$. Consequently,
separation of times scales can be violated even if a large separation of spatial scales
is observed (i.e. $k/\kb \ll 1$). A similar situation takes place in plasmas under the
effect of a strong magnetic field, or strongly rotating and stratified flows. It is
therefore more difficult to observe signatures of WWT in anisotropic systems.

Stratified turbulence is not only an interesting conceptual problem, but it is also
essential for understanding geophysical flows. \cite{staquet_internal_2002,
vallis_atmospheric_2017}. In particular, sub-grid parameterizations in climate models
require an understanding of the role of waves in mixing \cite{mackinnon_climate_2017,
gregg_mixing_2018}. It is therefore not surprising that stratified turbulence received
a particular attention from both the ``strong'' turbulence community
\cite{billant_self-similarity_2001, waite_stratified_2004, waite_stratified_2006,
lindborg_energy_2006, brethouwer_scaling_2007, waite_stratified_2011,
kimura_energy_2012, bartello_sensitivity_2013, brunner-suzuki_upscale_2014,
augier_stratified_2015, maffioli_vertical_2017}, and the WWT community
\cite{caillol_kinetic_2000, lvov_hamiltonian_2001, lvov_weak_2010,
dematteis_downscale_2021, dematteis_origins_2022}. It turns out that internal waves are
effectively important in the dynamics and mixing in stratified flows
\cite{maffioli_signature_2020, lam_partitioning_2020, lam_energy_2021}, and that
three-wave resonant interactions are responsible for slow energy transfers between
different scales \cite{brouzet_internal_2016, davis_succession_2020,
rodda_experimental_2022}. Internal waves can be excited, for example, by tides
\cite{reun_parametric_2018} or by linearly unstable wave attractors
\cite{brouzet_internal_2016}. However, many questions and issues remain about the
applicability of WWT to stratified flows. In particular, it is not yet clear under
which conditions a weak wave turbulence regime could occur.

It has been showed that a wave dominated region should lie in the spectral region
defined by $\tau_{\rm L} / \tau_{\rm NL} < 1/3$ \cite{yokoyama_energy-based_2019}, in
agreement with observations made for MHD \cite{meyrand_direct_2016}. However, the
observation of a system of weakly interacting internal gravity waves is still delicate
due to non-wave structures such as shear modes (purely vertical shear) and vortical
modes (vertical vorticity). The same problem arises in rotating flows, in which the
geostrophic modes play the role of the non-propagative structures. It has been
ingeniously shown that tidal forcing is very efficient at triggering weakly nonlinear
internal gravity waves \cite{reun_parametric_2018}, in a way reminiscent of the
inertial waves' excitation by libration in rotating fluids \cite{le_reun_inertial_2017,
reun_experimental_2019}. Experimentalists overcame this difficulty by using honeycomb
grids at the top and bottom boundaries of a rotating tank to dissipate geostrophic
modes efficiently, allowing them to observe weak inertial wave turbulence
\cite{brunet_shortcut_2020, monsalve_quantitative_2020}, in a similar fashion to the
numerical study of \cite{le_reun_inertial_2017}. More recently, it has been observed
that resonance with the container modes also prevent to observe weak internal gravity
wave turbulence, and that the introduction of slightly tilted panels at the top and at
the bottom of the fluid domain allows to inhibit the emergence of these modes
\cite{lanchon_internal_2023}. All these works point out that some wave systems can
generate non-wave motions that can severely affect the wave dynamics and should be
suppressed in experiments aiming to observe wave turbulence. In stratified flows these
modes correspond to vertical shear, vertical vorticity, an eventually the container
modes.

The most commonly used prediction of WWT is the scale-invariant stationary solution to
the kinetic equation. This solution gives the expected spatial energy spectra of waves
in the statistically steady state. There are two types of solutions: the thermodynamic
equilibrium solution (Rayleigh-Jeans spectra), and the non-equilibrium solutions that
are linked to the cascade of the dynamical invariants of the system along scales
(Kolmogorov-Zakharov spectra) \cite{nazarenko_wave_2011}. It is important to note that
the Kolmogorov-Zakharov spectra are obtained after the Zakharov transformation, which
can add spurious solutions. Therefore, these spectra can be considered as valid only if
the collision integral in the original wave kinetic equation converges. The importance
of the nonlinear interactions among internal gravity waves was recognized early,
leading to several derivations of waves kinetic equations (see
\cite{muller_nonlinear_1986, lvov_resonant_2012}). Derivation of the
Kolmogorov-Zakharov spectra in the limit of large vertical wave-numbers $k_z \simeq k
\gg k_h$ can be found in \cite{caillol_kinetic_2000, lvov_hamiltonian_2001}. This
candidate solution corresponds to an energy cascade, and is given by the energy spectra
\begin{equation}
E(k_h, k_z) \sim k_h^{-3/2} k_z^{-3/2} ~~~~ \text{and} ~~~~ E(\omega, k_z) \sim \omega^{-3/2} k_z^{-2},
\end{equation}
where the change of coordinates is defined by the dispersion relation $E(\omega, k_z) =
E(k_h,k_z) \left( \partial \ok / \partial k_h \right)^{-1}$. Yet, it was noted that
this candidate does not satisfy the requirement of locality, i.e. the collision
integral diverges on it. In the other words, it is not a valid mathematical solution of
the kinetic equation. Later, it was shown that power law solutions $E(k_h,k_z) \sim
k_h^{-\alpha_h} ~ k_z^{-\alpha_z}$ have convergent collision integral's contributions
only on the segment $\alpha_h \in ]2, 3[, \alpha_z=1$ \cite{lvov_oceanic_2010,
dematteis_downscale_2021}. The collision integral was then computed numerically on this
segment, and it was deduced that the only scale invariant stationary solution to the
kinetic equation was close to
\begin{equation}
\label{eq:WWTpredictions}
E(k_h, k_z) \sim k_h^{-1.69} k_z^{-1} ~~~~ \text{and} ~~~~ E(\omega, k_z) \sim \omega^{-1.69} k_z^{-1.69}.
\end{equation}
It was also shown that the dominant contributions to the collision integral corresponds
to non-local transfers first identified by McComas \cite{mccomas_resonant_1977},
notably the Parametric Subharmonic Instability (PSI) \cite{mccomas_resonant_1977,
muller_nonlinear_1986} observed in oceans \cite{mackinnon_parametric_2013} and
experiments \cite{rodda_experimental_2022}, consisting in the resonant interaction of a
primary wave and two smaller-scale waves of nearly half the frequency.

When $k/\kb \gg 1$, nonlinearity is not small and WWT is not valid. Based on the idea
that there exists a range of spatial scales where buoyancy force has the same order of
magnitude than inertia, the following 1D integrated energy spectra
\begin{equation}
\label{eq:CriticalBalance}
E_{\rm 1D}(k_h) \sim \epsK^{2/3} k_h^{-5/3} ~~~~ \text{and} ~~~~ E_{\rm 1D}(k_z) \sim N^2 k_z^{-3}
\end{equation}
where predicted \cite{lindborg_energy_2006}. In the strong wave turbulence context,
these predictions can also be obtained using critical balance arguments
\cite{nazarenko_critical_2011, nazarenko_wave_2011}. At even smaller scales, where
stratification is negligible, an isotropic range with the energy spectra
\begin{equation}
\label{eq:Isotropic}
E_{\rm 1D}(k) \sim \epsK^{2/3} k^{-5/3}
\end{equation}
is expected. It happens when $\kb < \ko \ll k \ll k_{\eta}$ where $\ko$ is the Ozmidov
wave-vector and $k_{\eta}$ is the Kolmogorov wave-vector.

The present study first deals with the existence and properties of a weak internal
gravity wave turbulence regime. To this end, we perform numerical simulations of
stratified turbulence at various \bv frequency and viscosity. In the same spirit as
simulations and experiments of rotating flows \cite{le_reun_inertial_2017,
brunet_shortcut_2020, monsalve_quantitative_2020}, we remove shear modes in all of our
simulations. For each values of the control parameters, we perform two ``twin''
simulations: one where the vortical modes remain, and one where vortical modes are
removed from the dynamics by a projection in spectral space
\cite{craya_contribution_1957}. It is worth noting that numerical simulations of
reduced dynamical equations (i.e., without non-wave structures) of stratified rotating
flows in the hydrostatic balance have already been conducted to remove non-wave
structures \cite{lvov_nonlinear_2009}. Despite the simplifications, these simulations
reproduced some key features of oceanic internal-wave spectra, such as the accumulation
of energy at near-inertial waves and realistic frequency and horizontal wave-number
dependencies of spatiotemporal spectra. In the present work, we do not account for
rotation. However, our simulations allows to investigate the role of vortical modes on
the dynamics of stratified flows outside the hydrostatic balance approximation.

The manuscript is organized as follows. In section \ref{sec:methods}, we present our
methodology including a presentation of the code and the simulations. Our results are
presented in section \ref{sec:results}. Subsection \ref{subsec:global} is devoted to
the study of flow regimes in the control parameter space. It shows, as expected, that
WWT is naturally more likely to occur at high stratification and without vortical
modes. Subsections \ref{subsec:khkz} and \ref{subsec:khkzomega} deal with the
spatiotemporal analysis of a couple of strongly stratified simulations to investigate
further the presence of linear waves in spatial scales. In the last subsection
\ref{subsec:waves_energy}, we propose a simple diagnostic to evaluate the predominance
of waves in the parameter space. We give concluding remarks in section
\ref{sec:conclusions}.

\section{Methods}
\label{sec:methods}

We use the 3D Navier-Stokes equations under the Boussinesq approximation:
\begin{align}
\label{eq:Continuity}
\bnabla \cdot \vv &= 0 \\
\label{eq:Impulsion}
\p_t\vv + \vv \cdot \bnabla \vv &= b ~ \boldsymbol{e}_z - \bnabla p +
\nu \nabla^2\vv + \ff, \\
\label{eq:Buoyancy}
\p_t{b} + \vv \cdot \bnabla b &= -N^2 v_z + \kappa\nabla^2{b} ,
\end{align}
where $(x,y,z)$ will represent the three spatial coordinates in the cartesian frame
$(O, \eex, \eey, \eez)$, $\eez$ is the stratification axis, $\vv=(v_x, v_y, v_z)$ is
the velocity, $b$ the buoyancy, $p$ the total kinematic pressure, $N$ the \bv
frequency, $\nu$ the viscosity, $\kappa$ the diffusivity and $\ff$ is the velocity
forcing. The buoyancy is defined as $b = -g \rho' / \rho_0$, where $g$ is the
acceleration due to gravity, $\rho_0$ is the average density of the fluid at $z=0$, and
$\rho'$ is the density perturbation with respect to the average linear density profile
$\bar{\rho}(z) = \rho_0 + (\mathrm{d}\bar{\rho}/\mathrm{d}z)z$. The Schmidt number $Sc
= \nu/\kappa$ is fixed to one.

We simulate forced-dissipated flows with the pseudo-spectral solver \texttt{ns3d.strat}
from the FluidSim software \cite{mohanan_fluidsim_2019} (an open-source Python package
of the FluidDyn project \cite{fluiddyn} using Fluidfft \cite{fluidfft} to compute the
Fast Fourier Transforms). The forcing, which will be described in details at the end of
this section and in Appendix~\ref{appendix:forcing}, is computed in spectral space such
that the kinetic energy injection rate $P_K$ is constant and equal to unity. The
physical input parameters are the \bv frequency $N$ and the diffusive coefficients
$\nu=\kappa$, but in practice, we identify our simulations with the couple
$(N,\,\R_i)$, where $\R_i \equiv P_K / (\nu N^2)$ is the input buoyancy Reynolds
number. The turbulent non-dimensional numbers characterizing the statistically
stationarity flow are the horizontal turbulent Froude number and the buoyancy Reynolds
number \cite[]{brethouwer_scaling_2007} that are respectively
\begin{equation}
\label{eq:FhR} F_h = \frac{\epsK}{{U_h}^2 N} ~~~~ \text{and} ~~~~ \R = \frac{\epsK}{\nu N^2},
\end{equation}
where $\epsK$ is the kinetic energy dissipation rate and $U_h$ the rms of the
horizontal velocity. Note that the turbulent Reynolds number is given by $Re = \R /
F_h^2$. We also compute the buoyancy and the Ozmidov wave-vectors $\kb \equiv N/U_h$
and $\ko \equiv \sqrt{N^3 / P_K}$. All the quantities presented in this manuscript are
computed from averaging when stationarity is reached. A list of the simulations, with
relevant parameters and physical quantities, is given in
Appendix~\ref{table-better-simuls}.

In this study, we consider a periodic domain of horizontal size $L_x = L_y = L_h = 3$
and vertical size $L_z$. We note $(n_x, n_y, n_z)$ the numbers of collocations points
in the three spatial directions, with $n_x = n_y \equiv n_h$. We chose $n_z$ in order
to have an isotropic mesh in physical space, i.e. $L_z/n_z = L_h/n_h$. We decrease
$L_z$ with $N$. Typically, $L_z \propto 1/N$, while $L_h$ is kept constant for all
simulations. More precisely, the aspect ratio $L_z/L_h$ is $1/2$ for $N \leq 20$, $1/4$
for $N \leq 60$, and $1/8$ for $N \geq 80$. This choice is motivated by the fact that
the unforced and undissipated Boussinesq equations are self-similar in the limit $F_h
\rightarrow 0$, with similarity variable $zN/U$, where $U$ is the typical velocity
\cite{billant_self-similarity_2001}. In that way, we simulate few layers (of height
$L_b = U/N$) for all our simulations.

To reduce the computational costs, we first simulate the transient state with a coarse
resolution with $n_h = 320$. For these simulations, two hyperdiffusive terms $-\nu_4
\nabla^4 \vv$ and $-\kappa_4\nabla^4{b}$ are added to (\ref{eq:Impulsion}) and
(\ref{eq:Buoyancy}), respectively, in order to keep the dissipative range in the
simulated scales and avoid thermalisation at small scales. Once a statistically
steady-state is reached, we increase the resolution of the simulation while decreasing
hyper-viscosity. Then the simulation is run until reaching a new statistically steady
state. The previous step is repeated until reaching the highest resolution. We measure
the turbulent kinetic dissipation rates $\epsKK$ and $\epsKKKK$ based on both
viscosities, and the total kinetic energy dissipation rate $\epsK = \epsKK + \epsKKKK$.
The product of the maximal wave-vector $\kmax$ with the Kolmogorov scale $\eta \equiv
(\nu^3 / \epsK)^{1/4}$ is computed to quantify how close our simulations are from true
Direct Numerical Simulations (DNS). In practice, it is common to consider that
simulations are proper DNS with well-resolved small scales when $\kmax\eta > 1$
\cite[]{deBruynKops1998,brethouwer_scaling_2007}. For the statistically stationarity
state, the time average of the total energy dissipation rate is equal to the injection
rate $P_K$ so that the product $\kmax\eta$ depends mostly on $\nu$ and $n_h$. For most
couples $(N,\,\R_i)$, the resolution of the larger simulation is fine enough
($\kmax\eta \gtrsim 1$) so that the hyperdiffusion is zero or negligible. For example,
the two simulations analyzed in details in the next section
(Figures~\ref{fig:buoyancy_fields} to \ref{fig:ratioEwaves_khkz}) are proper DNS with
$\kmax\eta$ equal to 0.99 and 1.05, respectively (see table~\ref{table-better-simuls}).
There are also few simulations with $0.45 < \kmax\eta < 1$ (19 out of 78 simulations),
which remain slightly under-resolved and affected by hyper-viscosity. In that case,
small-scales and flow statistics should be analyzed carefuly. We checked that these
simulations do not change the results presented here.

\begin{figure}
\centering
\includegraphics[width=0.4\textwidth]{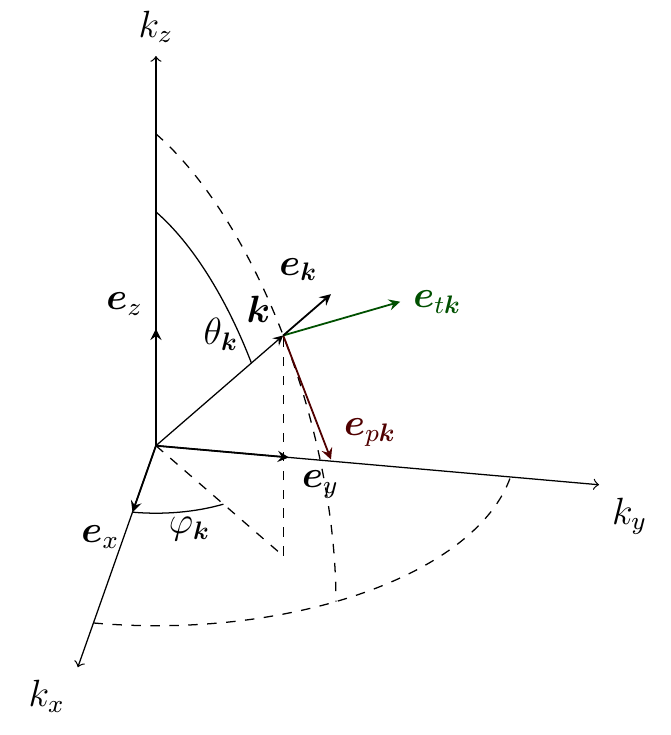}
\caption{Illustration of the poloidal-toroidal basis $(\eek, \eep, \eet)$ defined by
equations (\ref{eq:poloidal-toroidal}). $\thk$ is the angle between $\eez$ and $\eek$.
$\phk$ is the angle between the horizontal projection of $\kk$ and $\eex$.
\label{fig:poloidal-toroidal}}
\end{figure}

The Fourier transform of the velocity field $\hatvv = (\hat{v}_x, \hat{v}_y,
\hat{v}_z)$ can be written using the poloidal-toroidal-shear decomposition (see e.g.
\cite{craya_contribution_1957, smith_generation_2002, laval_forced_2003,
godeferd_toroidalpoloidal_2010, kimura_energy_2012, maffioli_vertical_2017})
\begin{equation}
\hatvv = \begin{cases} \hatvp ~ \eep + \hatvt ~ \eet ~~~~ \text{if} ~ k_h \neq 0, \\
	\vvs = \hat{v}_x ~ \eex + \hat{v}_y ~ \eey ~~~~ \text{if} ~ k_h = 0,
\end{cases}
\end{equation}
where
\begin{equation}
\label{eq:poloidal-toroidal}
\eek = \frac{\kk}{k}, ~~~~
\eep = \frac{\kk \times (\kk \times \eez)}{|\kk \times (\kk \times \eez)|}, ~~~~
\eet = \frac{\eez \times \kk}{|\eez \times \kk|}.
\end{equation}
$\hatvp$ is the poloidal component, $\hatvt$ the toroidal component, $\vvs$ the shear
modes component, $\kk=(k_x, k_y, k_z)$ denotes the wave-vector, $k =|\kk| = \sqrt{k_x^2
+ k_y^2 + k_z^2}$ is its modulus, and $k_h = \sqrt{k_x^2 + k_y^2}$ is the modulus of
the horizontal component of the wave-vector (Figure~\ref{fig:poloidal-toroidal}). Since
the toroidal component $\hatvt$ corresponds only to the vertical vorticity
($\hat{\Omega}_z = i k \hatvt \sin \thk$, with $\bOmega = \bnabla \times \vv$ being the
vorticity), we also denote it as the ``vortical" velocity.

In spectral space, once projected, the equations of motion
(\ref{eq:Continuity}-\ref{eq:Buoyancy}) read
\begin{align}
\label{eq:StratifiedSpectralPoloidalToroidal}
\begin{cases}
	\dot{\hat{v}}_p &= - (\widehat{\vv \cdot \bnabla \vv}) \cdot \eep - \hatb \sin \thk - \nu k^2 \hatvp + \hatff \cdot \eep \\
	\dot{\hat{v}}_t &= - (\widehat{\vv \cdot \bnabla \vv}) \cdot \eet - \nu k^2 \hatvt  + \hatff \cdot \eet \\
	\dot{\hatb} &= - \widehat{\vv \cdot \bnabla b} + N^2 \hatvp \sin \thk - \kappa k^2 \hatb
\end{cases}
~~~~~~~~ \text{for }~ k_h \neq 0
\end{align}
and
\begin{align}
\label{eq:StratifiedSpectralShearModes}
\begin{cases}
	\dot{\hat{\vv}}_{s} &= - \widehat{\vv \cdot \bnabla \vv_h} - \nu k^2 \vvs, \\
	\dot{\hatb} &= - \widehat{\vv \cdot \bnabla b} - \kappa k^2 \hatb
\end{cases}
~~~~~~~~ \text{for }~ k_h = 0,
\end{align}
where $\vv_h = (v_x,v_y,0)$ is the horizontal velocity and $\widehat{\cdot}$ denotes
the Fourier transform. We will also note $k_i \in \Delta k_i ~ \mathbb{Z}$ with $\Delta
k_i = 2\pi /L_i$ for $i =x,y,z$, and hereafter assume  $\Delta k_x = \Delta k_y \equiv
\Delta k_h$. Without forcing and dissipation, the linearized equations
(\ref{eq:StratifiedSpectralPoloidalToroidal}-\ref{eq:StratifiedSpectralShearModes}) can
be written as

\begin{align}
\label{eq:StratifiedSpectralPoloidalToroidalLinear}
\dot{\hata} &= -i \ok \hata, ~~~~ \dot{\hata}^{(0)} = 0, ~~~~ \text{if} ~~ k_h \neq 0 \\
\label{eq:StratifiedSpectralShearModesLinear}
\dot{\hatvv}_{s} &= \mathbf{0}, ~~~~
\dot{\hatb} = 0 ~~~~ \text{if} ~~ k_h = 0,
\end{align}
where
\begin{equation}
\hata = \frac{\hatvp - i  \frac{\hatb}{N}}{\sqrt{2\ok}}, ~~~~ \text{and} ~~~~ \ok = N \frac{k_h}{k} = N \sin \thk
\end{equation}
are the waves modes and the pulsation of the waves, $\thk$ is the angle between $\kk$
and the stratification axis $\eez$, and $\hata^{(0)} \propto \hatvt$ are the vortical
modes. Equations
(\ref{eq:StratifiedSpectralPoloidalToroidalLinear}-\ref{eq:StratifiedSpectralShearModesLinear})
show that both shear modes and vortical modes have zero frequency. This means that
linear waves can only exist in the poloidal velocity and the buoyancy, but not in the
toroidal velocity nor the shear velocity.

We are motivated by forcing internal gravity waves, which only involve the poloidal
part of the velocity field and have an anisotropic dispersion relation. Therefore, we
use an anisotropic, poloidal velocity forcing $\hatff = \hatf \, \eep$. The flow is
forced at large spatial scales $ \left\{\kk ~ | ~ 5 \leq k/\Delta k_h \leq 20 \right\}$
and small angle $\left\{\kk ~ | ~ |\ok /N - \sin \theta_f| \leq 0.05 \right\}$ where
$\sin \theta_f = 0.3$, meaning that relatively slow internal waves are forced. The
forcing scheme is described in Appendix~\ref{appendix:forcing}. It is neither harmonic
nor given by a stochastic differential equation. Instead, a time correlated forcing is
computed via generations of pseudo random numbers and time interpolations. Its
correlation time is equal to the period of the forced waves $T_c = 2\pi /(N \sin
\theta_f)$. The forcing is normalized such that the kinetic injection rate $P_K$ is
always equal to 1. Forcing slow waves is motivated by oceanic applications, where waves
are generated, among other processes, by slow tides \cite{mackinnon_climate_2017,
nikurashin_legg_mechanism_2011}. Low frequency forcing is also used in order to have a
scale separation between forced frequencies and the \bv frequency so that one can
potentially reproduce features of the oceanic temporal spectra close to $N$.

The time advancement is performed using the $4^{th}$ order Runge-Kutta scheme. All
modes with wave-number modulus larger than $\kmax = 0.8 (n_h/2) \Delta k_h$ are
truncated to limit aliasing. We checked that this 0.8 spherical truncation is a good
compromize consistent with our other numerical choices. Shear modes and vertically
invariant vertical velocity (internal waves at $\omega = N$), which are absent in flows
bounded by walls, are also removed in our simulations by fixing nonlinear transfers to
these modes to zero. For the simulations without vortical modes, the toroidal
projection of the velocity and of the nonlinear transfers are set to zero.
\citet{smith_generation_2002} used a similar procedure in order to distangle the roles
of potential vorticity modes and waves modes in rotating stratified turbulence.

Without dissipation and forcing, equations (\ref{eq:Continuity}-\ref{eq:Buoyancy})
conserve the total energy $E = \int ~ \left[ \vv^2 / 2 + b^2 / (2N^2) \right] ~ \diff x
\diff y \diff z$ and the potential vorticity $\Pi = \bOmega \cdot \left( N^2 \eez +
\bnabla b \right)$ is a Lagrangian invariant \cite{bartello_geostrophic_1995}. It
follows that the spatial average of any function of $\Pi$ is conserved. As a special
case, the potential enstrophy
\begin{align}
V &\equiv \frac{1}{2} \int ~ \Pi^2 ~ \dxdydz \\ &=
\frac{1}{2} \int ~ N^4 \Omega_z^2 ~ \dxdydz + \int ~ N^2 \Omega_z
\bOmega \cdot \bnabla b ~ \dxdydz + \frac{1}{2} \int ~ \left( \bOmega
\cdot \bnabla b \right)^2 ~ \dxdydz \\ &\equiv V_2 + V_3 + V_4
\end{align}
is an invariant of equations (\ref{eq:Continuity}-\ref{eq:Buoyancy}) in absence of
dissipation and forcing. For a flow without vertical vorticity $V_2 = V_3 =0$.

\section{Results}
\label{sec:results}

\subsection{Regimes and global energy distribution}
\label{subsec:global}

\begin{figure}
\includegraphics[width=1.0\textwidth]{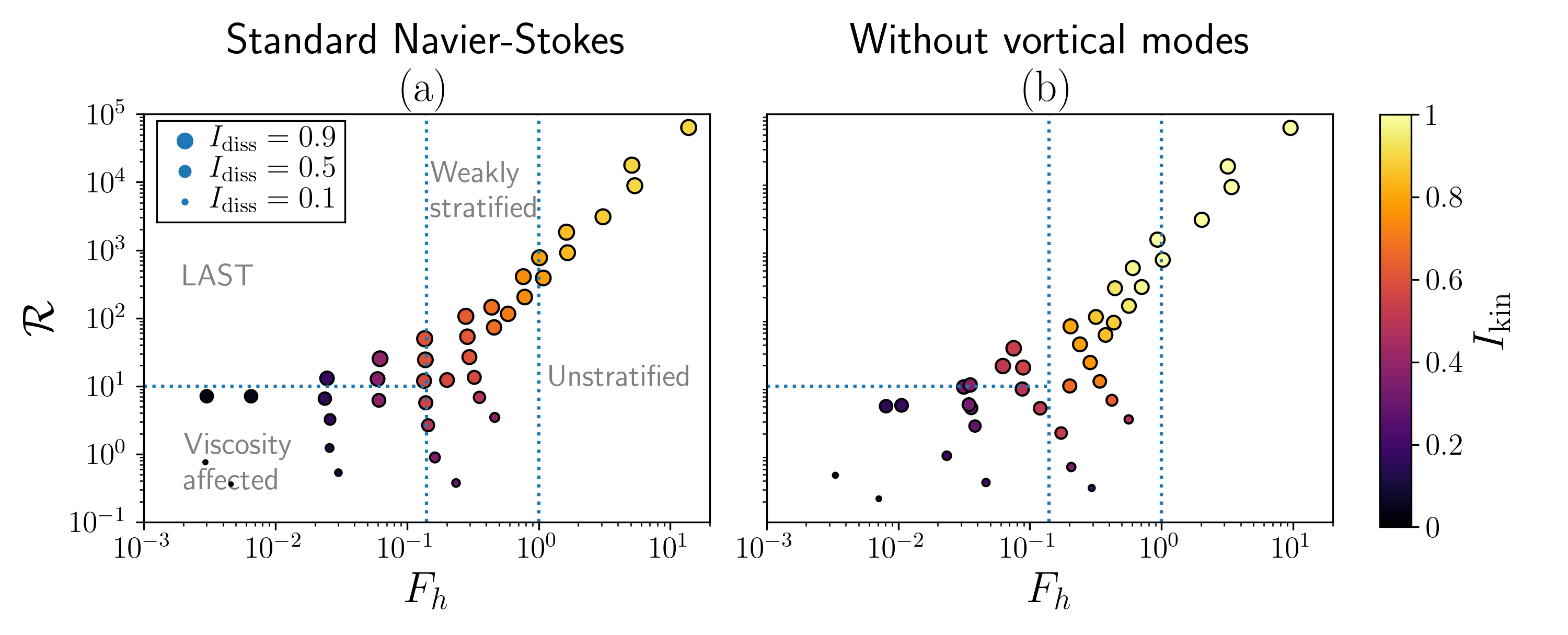}
\caption{Classification of regimes in our simulations using the isotropy coefficients.
The large scale isotropy coefficient $\Ivelo$ is given by the color-scale, and the
small scale isotropy coefficient $\Idiss$ by the size of the symbols. For simulations
with vortical modes $\rm (a)$, definitions (\ref{eq:Ivelo}-\ref{eq:Idiss}) are used.
For simulations without vortical modes $\rm (b)$, definitions (\ref{eq:IcoeffsProj})
are used. The dotted blue lines correspond to $\R = 10$, $F_h = 0.14$, and $F_h=1$.
\label{fig:regimes}}
\end{figure}

In this subsection, we study the anisotropy of the flow and global energy budget in the
$(F_h, \R)$ plane. To discuss stratified turbulence regimes, it is useful to introduce
a large and a small scale isotropy coefficients \cite{linares_numerical_2020}. The
former is based on the kinetic energy components
\begin{equation}
\label{eq:Ivelo}
\Ivelo = \frac{3 E_{\rm kin,z}}{E_{\rm kin}}
\end{equation}
where $E_{\rm kin,z}$ is the vertical velocity energy and $E_{\rm kin}$ is the total
kinetic energy. The small scale isotropy coefficient is computed using the kinetic
energy dissipation rates, namely
\begin{equation}
\label{eq:Idiss}
\Idiss = \frac{1 - \varepsilon_{\rm kin,z}/\varepsilon_{\rm kin}}{(1 - 1/3)}
\end{equation}
where $\varepsilon_{\rm kin,z}$ is the energy dissipation rate due to vertical
gradients. Both $\Ivelo$ and $\Idiss$ are equal to unity for an isotropic flow.
Conversely, $\Ivelo$ and $\Idiss$ should be close to zero if the flow is strongly
anisotropic. When vortical modes are removed, we rather use
\begin{equation}
\label{eq:IcoeffsProj}
	\Ivelo  = \frac{2 E_{\rm kin,z}}{E_{\rm kin}}
\end{equation}
because it is expected that only half of the kinetic energy is contained in the
vertical velocity field. Figure~\ref{fig:regimes} shows the variations of $\Ivelo$ and
$\Idiss$ with $(F_h,\R)$. Small points correspond to strong anisotropic energy
dissipation, while large points correspond to isotropic energy dissipation. Dark points
correspond to strong large-scale isotropy, while light points correspond to isotropic
flows. The combination of $\Ivelo$ and $\Idiss$ allows to distinguish between four
regimes (Figure~\ref{fig:regimes}$\rm (a)$): the passive scalar regime where $\Ivelo
\simeq 1$, the weakly stratified regime where $0.5 \lesssim \Ivelo \lesssim 1$, and the
strongly stratified regimes where $\Ivelo \leq 0.5$. As explained in
\cite{brethouwer_scaling_2007}, the strongly stratified flows fall in two regimes: The
Layered Anisotropic Stratified Turbulence (LAST) regime where the dissipation is
isotropic because a 3D turbulence range can develop, typically when $\R \geq 10$; The
viscosity affected regime where the small dissipative scales remain affected by the
anisotropy. WWT is foreseen at small $F_h$ at some unknown values of $\R$. We observe
that removing vortical modes does not modify this picture (Figure~\ref{fig:regimes}$\rm
(b)$). However, for a given $(F_h,\R)$, the isotropy coefficients tend to have bigger
values when the vortical modes are removed.

\begin{figure}
\includegraphics[width=1.0\textwidth]{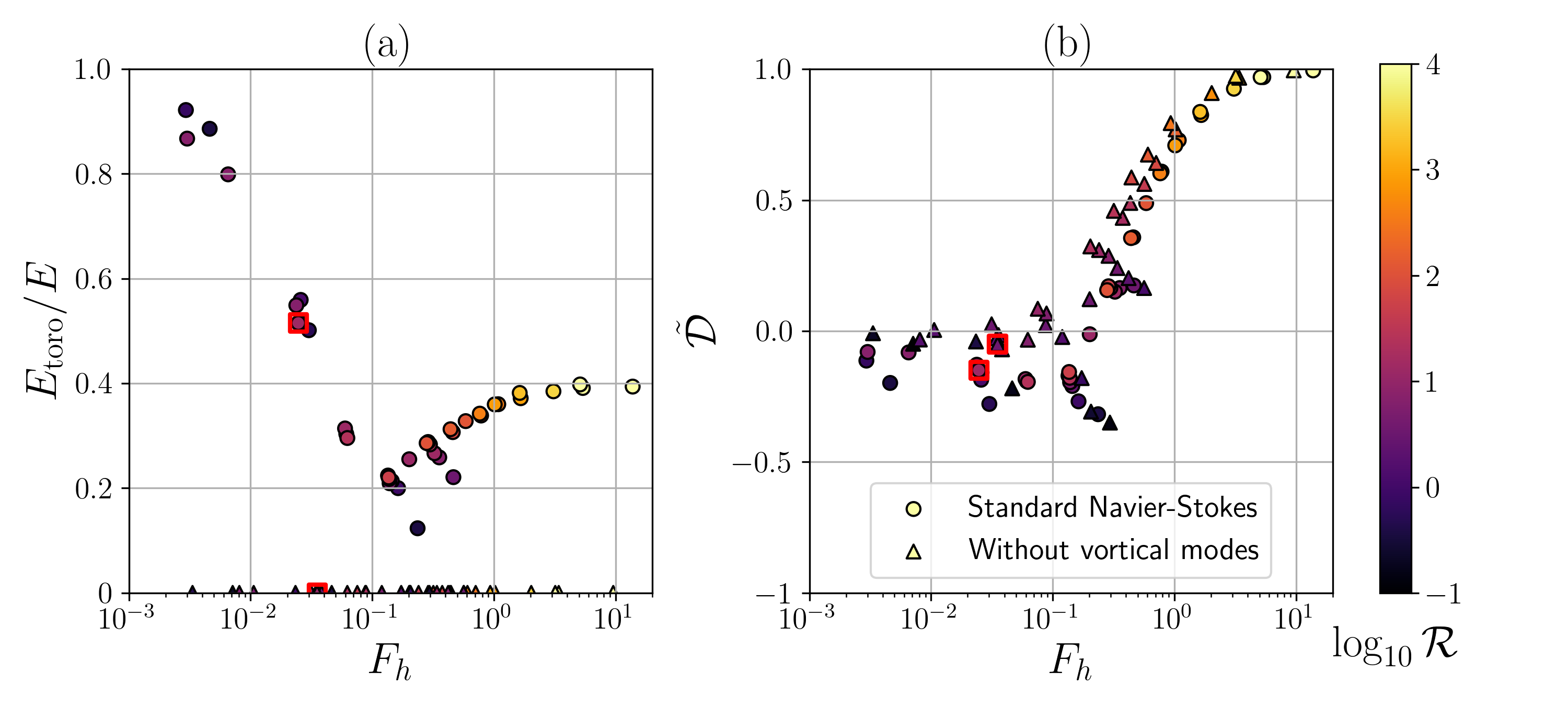}
\caption{Vortical modes to total energy ratio (a) and relative difference between
poloidal and potential energies $\tilde{\mathcal{D}} = (E_{\rm polo}  - E_{\rm
pot})/(E_{\rm polo}  + E_{\rm pot})$ (b) vs $F_h$ for simulations with or without
vortical modes. The red boxes indicate the simulations with $(N,\R_i)=(40,20)$
investigated in the next subsections. \label{fig:global-energy}}
\end{figure}

\begin{figure}
\includegraphics[width=1.0\textwidth]{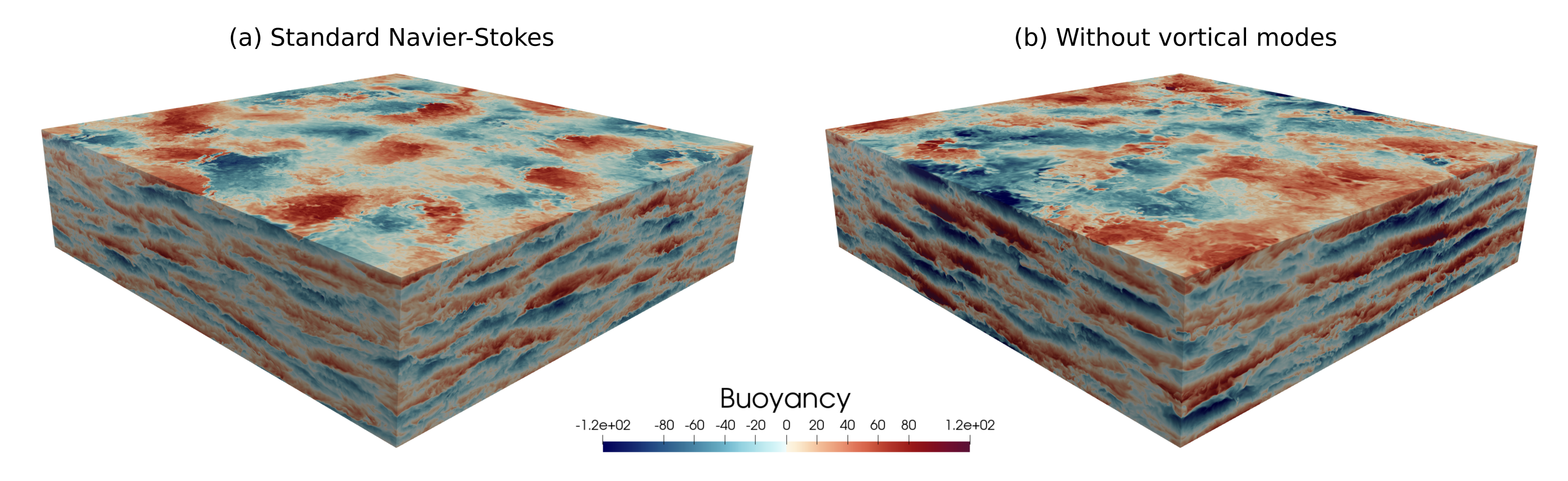}
\caption{Snapshots of the buoyancy fields for simulations $(N,\R_i)=(40,20)$ with $\rm
(a)$ and without $\rm (b)$ vortical modes. \label{fig:buoyancy_fields}}
\end{figure}

We denote
\begin{equation}
\label{eq:energies} E_{\rm pot}  = \frac{1}{2N^2} \sum\limits_{\kk}
|\hatb|^2, ~~~~ E_{\rm polo}  = \frac{1}{2} \sum\limits_{\kk} |\hatvp|^2, ~~~~ E_{\rm
toro}  = \frac{1}{2} \sum\limits_{\kk} |\hatvt|^2, ~~~~ \tilde{\mathcal{D}} =
\frac{E_{\rm polo}  - E_{\rm pot} }{E_{\rm polo}  + E_{\rm pot} }
\end{equation}
which are, respectively, the potential energy, the poloidal kinetic energy, the
vortical modes (toroidal) kinetic energy, and the relative difference between poloidal
and potential energy. We also note $E = E_{\rm pot}  + E_{\rm polo}  + E_{\rm toro} $
the total energy. Figure~\ref{fig:global-energy} shows the vortical modes energy ratio
$E_{\rm toro} /E$ and $\tilde{\mathcal{D}}$  as a function of $F_h$ and $\R$. For
waves, we expect to observe an equipartition between the poloidal kinetic energy and
potential energy. Consequently, both $E_{\rm toro} /E$ and $\tilde{\mathcal{D}}$ should
be close to zero for a system mainly composed by internal gravity waves. We observe
that vortical modes energy becomes dominant at high stratification (low $F_h$) if they
are not removed from the dynamics in these simulations with $\R \geq 0.1$
(Figure~\ref{fig:global-energy} (a)). The ratio $E_{\rm toro} /E$ is the lowest at
intermediate stratification $F_h \simeq 0.1-0.3$ and low values of $\R$. However, the
same simulations are marked by a predominance of potential energy over poloidal energy,
$\tilde{\mathcal{D}} < 0$ (Figure~\ref{fig:global-energy} (b)), meaning that these
weakly stratified flows do contain other structures than waves. When vortical modes are
removed, we can obtain flows with a global balance between poloidal and potential
energies ($\tilde{\mathcal{D}} \simeq 0$) at high stratification. Consequently, a flow
governed by weak nonlinear interactions between waves may be obtained at high
stratification by removing vortical modes. In the two next subsections, we perform a
spatiotemporal analysis of a couple of stongly stratified turbulent simulations with
$(N,\,\R_i\equiv P_K/\nu N^2)=(40,20)$ and aspect ratio $L_z/L_h=1/4$.
Figure~\ref{fig:buoyancy_fields} shows the buoyancy fields for these two simulations
with the same color scale. We observe that the flow is layered in the vertical
direction and that overturning (horizontal vorticity) is present with or without
vortical modes. It is a standard feature of strongly stratified turbulence
\cite{laval_forced_2003, lindborg_energy_2006, brethouwer_scaling_2007,
waite_stratified_2011}. With vortical modes, the vertical vorticity is not zero
(Figure~\ref{fig:buoyancy_fields}(a)) so the buoyancy has a different structure than
when vortical modes are absent (Figure~\ref{fig:buoyancy_fields}(b)). Without vortical
modes the dynamics in the horizontal direction is irrotational and the buoyancy field
has a larger amplitude.

\subsection{Energy budget in the strongly stratified regime}
\label{subsec:khkz}

It is expected that the spatial energy budget will depend on the ratio of temporal
timescales. In particular the nonlinearity parameter used for several physical systems,
including stratified and rotating turbulence \cite{nazarenko_critical_2011,
yokoyama_energy-based_2019} and MHD \cite{meyrand_direct_2016, cerri_turbulent_2022} is
\begin{equation}
\label{eq:NonLinearityParameter} \chi_{\kk} \equiv \frac{\tau_{\rm L}}{\tau_{\rm NL}} =
\frac{(\epsK k^2)^{1/3}}{N k_h / k} =   \frac{k}{k_h} \left(\frac{k}{\ko} \right)^{2/3}
= \frac{1}{\sin \thk} \left(\frac{k}{\ko} \right)^{2/3}.
\end{equation}
It represents the ratio between the period of the linear wave, $\tau_{\rm L} = 2 \pi k
/ (N k_h)$, and the eddy turnover time $\tau_{\rm NL} = 2 \pi / (\varepsilon_{\rm kin}
k^2)^{1/3}$ for a given wave-vector. In a similar way, we introduce the wave
dissipation parameter
\begin{equation}
\label{eq:DissipationParameter} \gamma_{\kk} \equiv \frac{\tau_{\rm
L}}{\tau_{\nu}} = \frac{\nu k^2}{N k_h / k} =   \frac{k}{k_h} \left(\frac{k}{\kd}
\right)^{2} = \frac{1}{\sin \thk} \left(\frac{k}{\kd} \right)^{2}
\end{equation}
which represents the ratio between the period of the linear wave and the dissipation
time $\tau_{\nu} = 2 \pi / (\nu k^2)$. In the last equation, we have introduced the
wave dissipation wave-vector $\kd \equiv \sqrt{N/\nu}$.  The kinetic energy and
potential energy budgets for one Fourier mode read
\begin{equation}
\label{eq:seb}
	\left\langle\frac{1}{2}\frac{\mathrm{d} |\hatvv|^2}{\mathrm{d}t} \right \rangle =
\mathcal{I}_{\kk} + \mathcal{T}_{\rm kin,\kk} - \mathcal{B}_{\kk} - \varepsilon_{\rm
kin,\kk},  ~~~~ \text{and} ~~~~
\left\langle\frac{1}{2N^2}\frac{\mathrm{d}|\hatb|^2}{\mathrm{d}t} \right\rangle=
\mathcal{T}_{\text{pot},\kk} + \mathcal{B}_{\kk} - \varepsilon_{\text{pot},\kk},
\end{equation}
where
\begin{align}
\nonumber &\mathcal{I}_{\kk} = \left\langle \Re
\left(\hatff  \cdot \hatvv^* \right)\right \rangle, ~~~~ \mathcal{T}_{\rm kin, \kk} =
- \left\langle \Re \left(\hatvv^* \cdot \left[\bar{\bar{P}}_{\kk}  \cdot (\widehat{\vv \cdot \bnabla \vv}) \right]\right)\right\rangle, \\
\nonumber &\mathcal{T}_{\text{pot},\kk} = - \left\langle \Re \left(\hatb^* (\widehat{\vv \cdot \bnabla b}) \right)/ N^2 \right\rangle,
~~~~ \mathcal{B}_{\kk} = - \left\langle \Re \left(\hat{v}_{z}^* \hatb \right) \right\rangle, \\
&\varepsilon_{\rm kin, \kk} =  (\nu k^2 + \nu_4 k^4) \frac{|\hatvv|^2}{2}, ~~~~ \varepsilon_{\text{pot},\kk} = (\kappa k^2 + \kappa_4 k^4) \frac{|\hatb|^2}{2N^2},
\end{align}
are respectively the kinetic energy injection rate, the kinetic energy transfer, the
potential energy transfer, the conversion of kinetic energy to potential energy, the
kinetic energy dissipation, and the potential energy dissipation. In the last
equations, $(\cdot)^*$ denotes the complex conjugate, $\Re(\cdot)$ the real part,
$\left\langle\cdot \right\rangle$ stands for the averaging operator, and
$\bar{\bar{P}}_{\kk} = \mathbb{I} - \eek \otimes \eek$ is the projector onto the plane
orthogonal to $\kk$.  In this section, we study azimuthal average of the energy budget.
Namely, we computed quantities like:

\begin{equation}
F(k_h, k_z) = \frac{1}{\Delta k_h ~ \Delta k_z} \mathop{\sum \sum}_{\substack{k_h \leq k_h'< k_h + \Delta k_h \\ k_z \leq |k_z'| < k_z + \Delta k_z}} ~ F_{\kk'},
\end{equation}
where $F$ can be $E_{\rm pot} $, $E_{\rm polo} $, $E_{\rm toro} $, $E_{\rm kin} =
E_{\rm polo}  + E_{\rm toro} $, $E = E_{\rm kin} + \Epot$, $\mathcal{T}_{\text{pot}}$,
$\mathcal{T}_{\rm kin}$, $\mathcal{B}$, $\mathcal{I}$, $\epsK$, or $\epsA$.

Following \cite{yokoyama_energy-based_2019}, we use the energy ratios like
\begin{equation}
\frac{E_{\rm toro} (k_h, k_z)}{E(k_h, k_z)} ~~~~ \text{and}  ~~~~ \tilde{\mathcal{D}}(k_h,k_z) = \frac{\Epolo(k_h,k_z)-\Epot(k_h,k_z)}{\Epolo(k_h,k_z)+\Epot(k_h,k_z)},
\end{equation}
in order to quantify the energy content scale by scale. In the same spirit, we also
introduce
\begin{equation}
\label{eq:transfers}
\tilde{\mathcal{T}}_{\rm kin}(k_h,k_z) = \frac{\mathcal{T}_{\rm kin}(k_h,k_z)}{\mathcal{T}_{\text{tot}}(k_h,k_z)}, ~~~~
\tilde{\mathcal{T}}_{\text{pot}}(k_h,k_z) = \frac{\mathcal{T}_{\text{pot}}(k_h,k_z)}{ \mathcal{T}_{\text{tot}}(k_h,k_z)} ~~~~
\text{and} ~~~~
\tilde{\mathcal{B}}(k_h,k_z) = \frac{\mathcal{B}(k_h,k_z)}{\mathcal{T}_{\text{tot}}(k_h,k_z)},
\end{equation}
where
\begin{equation}
\mathcal{T}_{\text{tot}}(k_h,k_z) \equiv |\mathcal{T}_{\rm kin}(k_h,k_z)| + |\mathcal{T}_{\text{pot}}(k_h,k_z)| + |\mathcal{B}(k_h,k_z)| + \epsK(k_h,k_z) + \epsA(k_h,k_z).
\end{equation}
These quantities are useful for tracking the energy pathways scale by scale. By
construction, $\tilde{\mathcal{T}}_{\text{pot}}$, $\tilde{\mathcal{T}}_{\rm kin}$, and
$\tilde{\mathcal{B}}$ vary between $-1$ and $1$ depending on the amplitude and
direction of the energy transfer or conversion ($\tilde{\mathcal{B}}>0$ corresponds to
conversion of kinetic energy to potential energy). In a statistically stationary state
($\langle\cdot  \rangle = 0$) and in the inertial range ($\mathcal{I}(k_h,k_z) = 0$,
$\epsK(k_h,k_z) \simeq 0$, and $\epsA(k_h,k_z) \simeq 0$), we should have
$\mathcal{T}_{\rm kin}(k_h,k_z) \simeq - \mathcal{T}_{\text{pot}}(k_h,k_z) \simeq
\mathcal{B}(k_h,k_z)$. For this reason, we will present only
$\tilde{\mathcal{B}}(k_h,k_z)$.

\subsubsection{Energy spectra}

\begin{figure}
\includegraphics[width=1.0\textwidth]{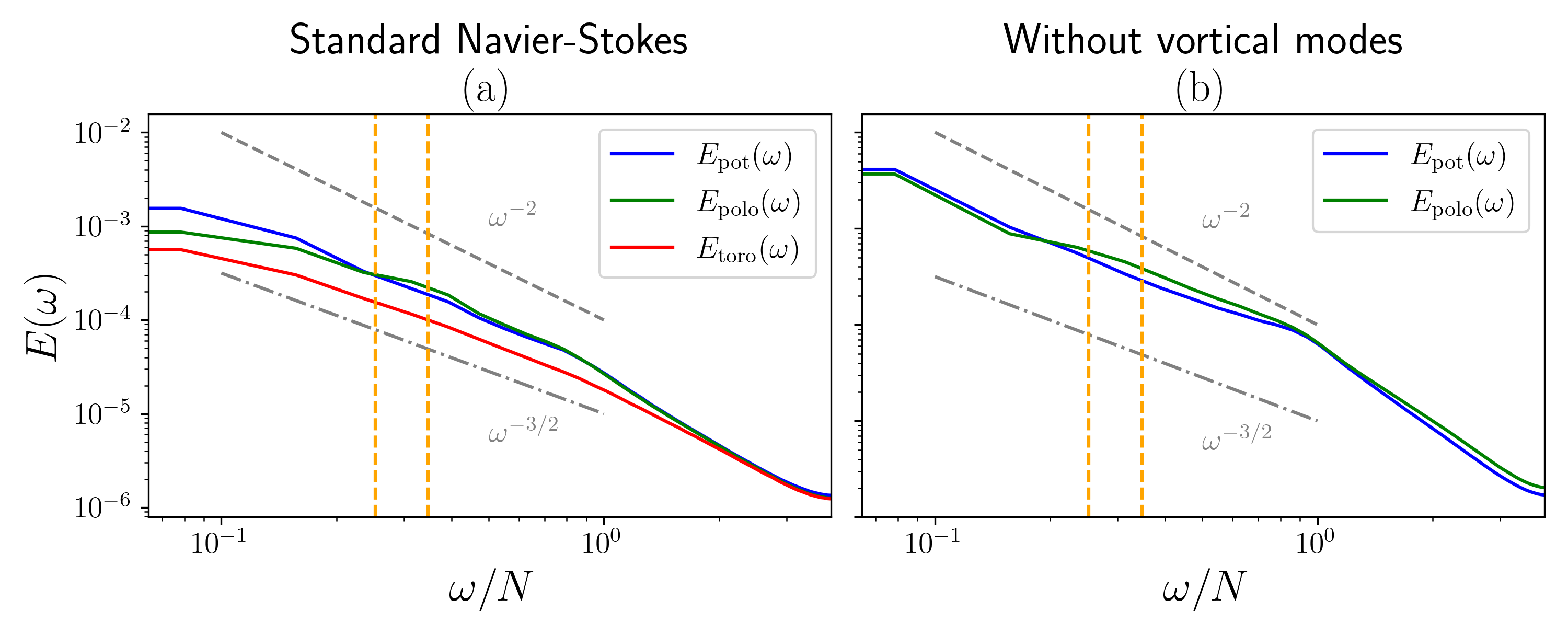}
\caption{Temporal energy spectra for simulations $(N,\R_i)=(40,20)$ with $\rm (a)$ and
without $\rm (b)$ vortical modes. The orange dotted lines correspond to the minimal and
maximal frequencies of the linear internal gravity waves in the forcing region.
\label{fig:spectra_omega}}
\end{figure}

Figure~\ref{fig:spectra_omega} displays the temporal energy spectra for simulations
with $(N,\R_i)=(40,20)$. In the case where vortical modes are present
(Figure~\ref{fig:spectra_omega}$\rm (a)$), we observe that potential and poloidal
energies dominate over toroidal energy for frequencies smaller than the \bv frequency,
i.e. $\omega/N <1$. For $\omega/N \geq 1$, the three components of the energy have very
similar spectra. When vortical modes are removed (Figure~\ref{fig:spectra_omega}$\rm
(b)$), potential energy is almost in perfect equipartition with poloidal energy at a
given frequency. The potential and poloidal energy spectra behave very similarly wether
vortical modes are present or not. This might be explained by the fact that vortical
modes energy is smaller than potential an poloidal energy at large temporal scales
$\omega \leq N$. Interestingly, in Figure~\ref{fig:spectra_omega}$\rm (b)$ temporal
spectra get closer to the Kolmogorov-Zakharov spectra $E(\omega) \sim \omega^{-3/2}$
\cite{caillol_kinetic_2000, lvov_hamiltonian_2001} than the the high frequency limit of
the Garrett-Munk spectra $E(\omega) \sim \omega^{-2}$ \cite{garrett_internal_1979}. The
inertial range is larger when vortical modes are removed.

\begin{figure}
\includegraphics[width=1.0\textwidth]{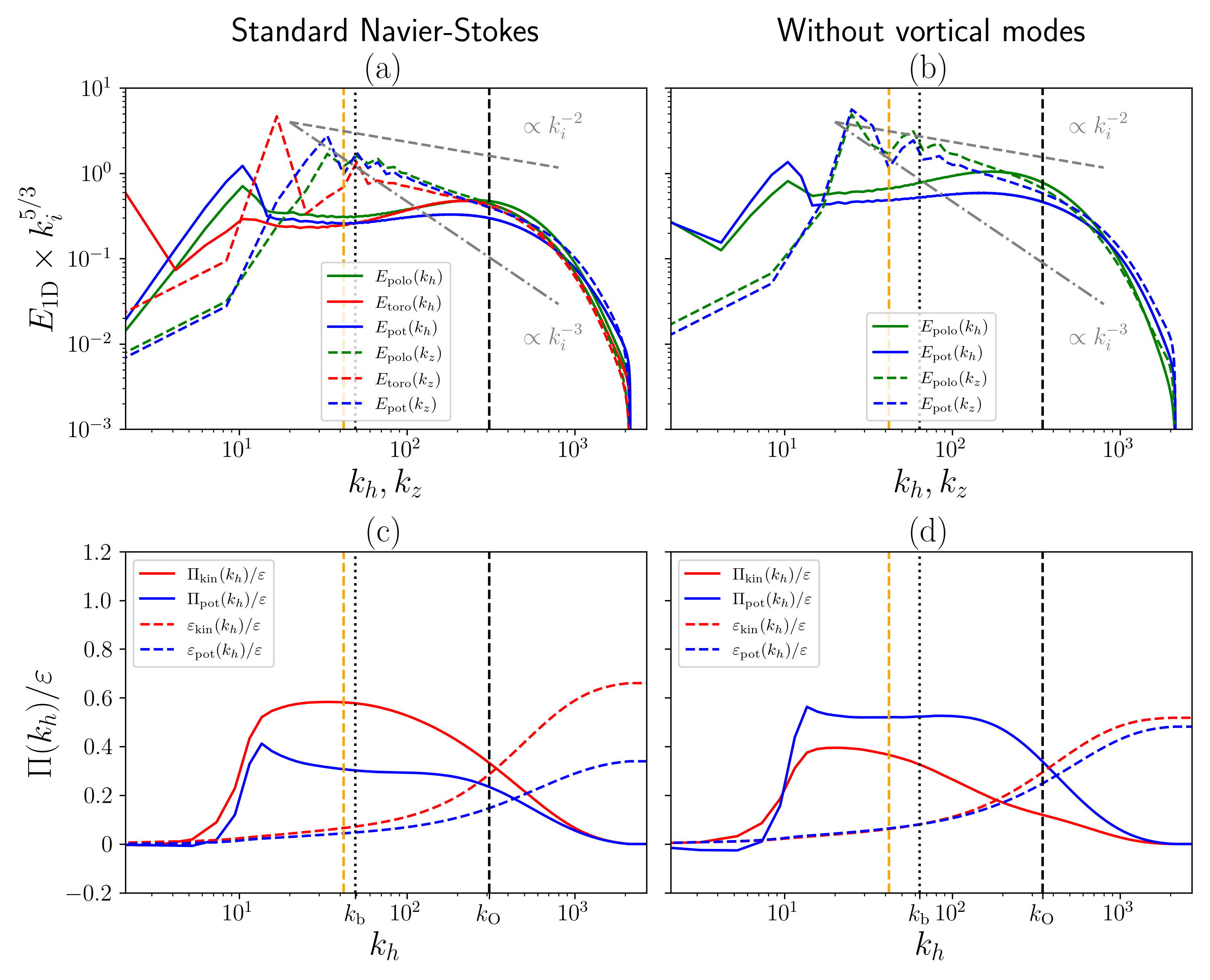}
\caption{Compensated 1D spatial energy spectra for simulations $(N,\R_i)=(40,20)$ with
$\rm (a)$ and without $\rm (b)$ vortical modes. Normalized $k_h$ energy fluxes for the
same simulations with $\rm (c)$ and without $\rm (d)$ vortical modes. The orange dotted
line corresponds to the maximal wave-vector modulus of the forcing region, the black
dotted line to $\kb$, the black dashed line to $\ko$. \label{fig:spectra_1D}}
\end{figure}

Figure~\ref{fig:spectra_1D} shows the 1D integrated spatial spectra, and the normalized
(i.e. divided by the total energy dissipation $\varepsilon \equiv \epsK + \epsA$)
energy fluxes along $k_h$ for the same simulations. We first observe that removing
vortical modes does not change the behavior of the poloidal an potential energy spectra
(Figure~\ref{fig:spectra_1D}$\rm (a)$-$\rm (b)$). Up to the the buoyancy wave-vector,
the poloidal and potential energies dominate with spectra that are slightly steeper
than $k_h^{-5/3}$. After the buoyancy scale (i.e. $k=\kb$), the toroidal energy
spectrum starts to be of the same order than the poloidal energy spectrum, which are
shallower than $k_h^{-5/3}$. Such changes in the spectral slope around the buoyancy
scale were reported in earlier studies \cite{kimura_energy_2012,
bartello_sensitivity_2013, augier_stratified_2015}. They could be due to shear
instabilities occurring at those scales, leading to non-local energy transfers toward
small scales \cite{brethouwer_scaling_2007, waite_stratified_2011,
augier_stratified_2015}. The vertical spectra has a spectral slope between $-3$ and
$-2$, which is similar to earlier simulations \cite{maffioli_vertical_2017}. Therefore,
these 1D spectra do not show strongly stratified turbulence scaling $~k_h^{-5/3}$ and
$k_z^{-3}$ for $\kb \leq k_h,k_z \leq \ko$ \cite{billant_self-similarity_2001,
lindborg_energy_2006}. This might be due to the fact that these simulations do not have
a sufficient scale separation between $\kb$ and $\ko$, which can be attained only at
very small $F_h$ \cite{lindborg_energy_2006, bartello_sensitivity_2013}, or when
considering only the largest horizontal scales \cite{maffioli_vertical_2017}. Indeed,
we observe that vertical spectra tends to be steeper as $F_h$ is decreased, and are
steeper than the scaling $k_z^{-2}$ observed is 2D numerical simulations at comparable
$F_h$ \cite{linares_numerical_2020}. The kinetic energy flux $\Pikin(k_h)$ and the
potential energy flux $\Pipot(k_h)$ start to show a plateau over almost a decade in
these simulations (Figure~\ref{fig:spectra_1D}$\rm (c)$-$\rm (d)$). We observe that the
dissipation starts to be important at the Ozmidov scale, meaning that these simulations
lie between the LAST regime and the viscosity affected regime
\cite{brethouwer_scaling_2007}. This explains why we do not observe an isotropic
turbulence range (with energy spectra $\sim k^{-5/3}$) in these simulations. The main
effect of removing vortical modes is to make $\Pipot(k_h)$ larger than $\Pikin(k_h)$.
Also, the kinetic energy dissipation $\epsK(k_h)$ is almost equal to the potential
energy dissipation $\epsA(k_h)$ when vortical modes are removed, as we expect for a
system of internal gravity waves with unit Schmidt number \cite{reun_parametric_2018}.

\begin{figure}
\includegraphics[width=1.0\textwidth]{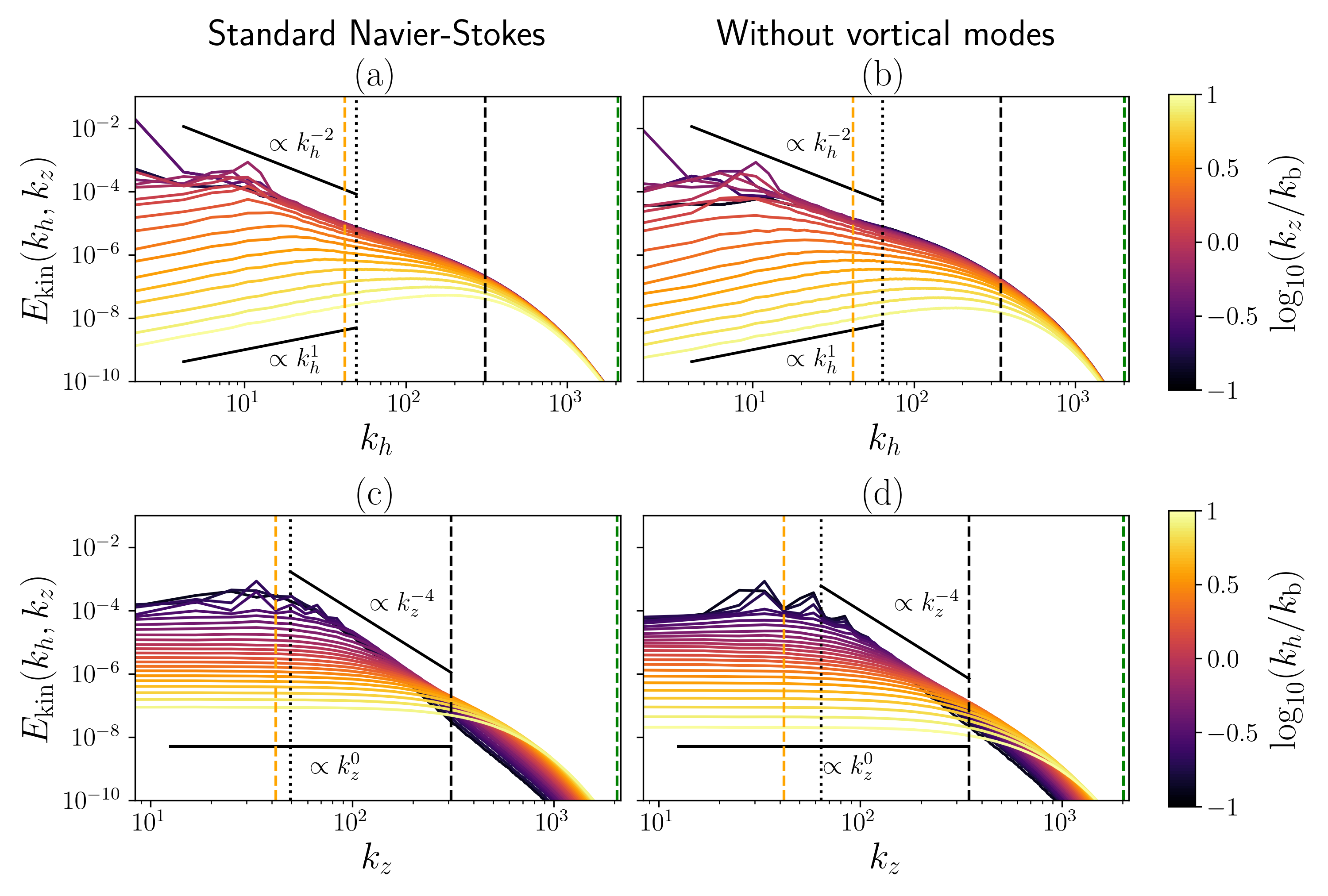}
\caption{Slices of the kinetic energy spectrum $E_{\rm kin}(k_h,k_z)$ for simulations
$(N,\R_i)=(40,20)$ with and without vortical modes. The orange dotted line corresponds
to the maximal wave-vector modulus of the forcing region, the black dotted line to
$\kb$, the black dashed line to $\ko$, and the green dashed line to the dissipative
wave-vector. (a) $E_{\rm kin}$ vs $k_h$ with vortical modes. (b) $E_{\rm kin}$ vs $k_h$
without vortical modes. (c) $E_{\rm kin}$ vs $k_z$ with vortical modes. (d) $E_{\rm
kin}$ vs $k_z$ without vortical modes. \label{fig:spectra_khkz_kin}}
\end{figure}

Figure~\ref{fig:spectra_khkz_kin} shows slices of the $(k_h, k_z)$ kinetic energy
spectrum. The potential and kinetic energy spectra have the same trends with respect to
$k_h$ and $k_z$, so the potential energy spectrum is not presented. As in
\cite{yokoyama_energy-based_2019}, we observe different scaling laws depending on the
region in the $(k_h,k_z)$ plane. Namely, important differences between small $k_z$ and
large $k_z$ for a given $k_h$ (Figure~\ref{fig:spectra_khkz_kin}$\rm (a)$-$\rm (b)$):
\begin{itemize}
\item For $k_h \ll \kb$, energy is accumulated at the lowest available $k_h$ and small
$k_z$, as usual in stratified turbulence \cite{smith_generation_2002,
laval_forced_2003, herbert_waves_2016}. At $k_z \ll \kb$ the spectra are close to a
$k_h^{-2}$ dependency, which is consistent with integrated energy spectra reported in
earlier studies \cite{waite_stratified_2011, kimura_energy_2012}, but not with the WWT
predictions (\ref{eq:WWTpredictions}). For a fixed $k_z \gg \kb$, we rather observe
$E_{\rm kin} \sim k_h^1$ which would correspond to an equipartition of energy in
horizontal scales.

\item For $k_h \gg \kb$, the spectrum starts to depend less and less on $k_z$, and the
$E_{\rm kin}(k_h,k_z)$ slices eventually merge around $k_h \sim \ko$, when the eddy
turnover time becomes less than the period of any linear waves, i.e. $\chi_{\kk} > 1$.
\end{itemize}

Important differences can also be noted when looking at $k_z$ slices
(Figure~\ref{fig:spectra_khkz_kin}$\rm (c)$-$\rm (d)$):
\begin{itemize}
\item For $k_z \ll \kb$, the spectrum is close to $\sim k_z^{0}$, indicating that
horizontal layers of height larger than the buoyancy scale are in equipartition of
energy in these strongly stratified simulations.

\item For $\kb \ll k_z \ll \ko$, the spectrum is very steep at small horizontal
wave-vectors, while it remains flat at large horizontal wave-vectors.

\item For $k_z \gg \ko$, the spectral slope starts to depend less and less on $k_h$.
Yet, the $E_{\rm kin}(k_h,k_z)$ does not merge around $k_h \sim \ko$ since small
horizontal scales are more energetic compared to the large horizontal scales.
\end{itemize}

Due to the reduced aspect ratio of our simulations, we do not observe a net separation
between the forcing and the buoyancy scales. Still, the kinetic energy spectrum remains
very similar to the one presented in \cite{yokoyama_energy-based_2019}. Interestingly,
the trends of the kinetic energy spectra appear to be very similar, regardless of
whether vortical modes are present or not. This is a first indication that the presence
of vortical modes is not the only obstacle to observing internal gravity wave
turbulence in realistic flows.

\begin{figure}
\includegraphics[width=1.0\textwidth]{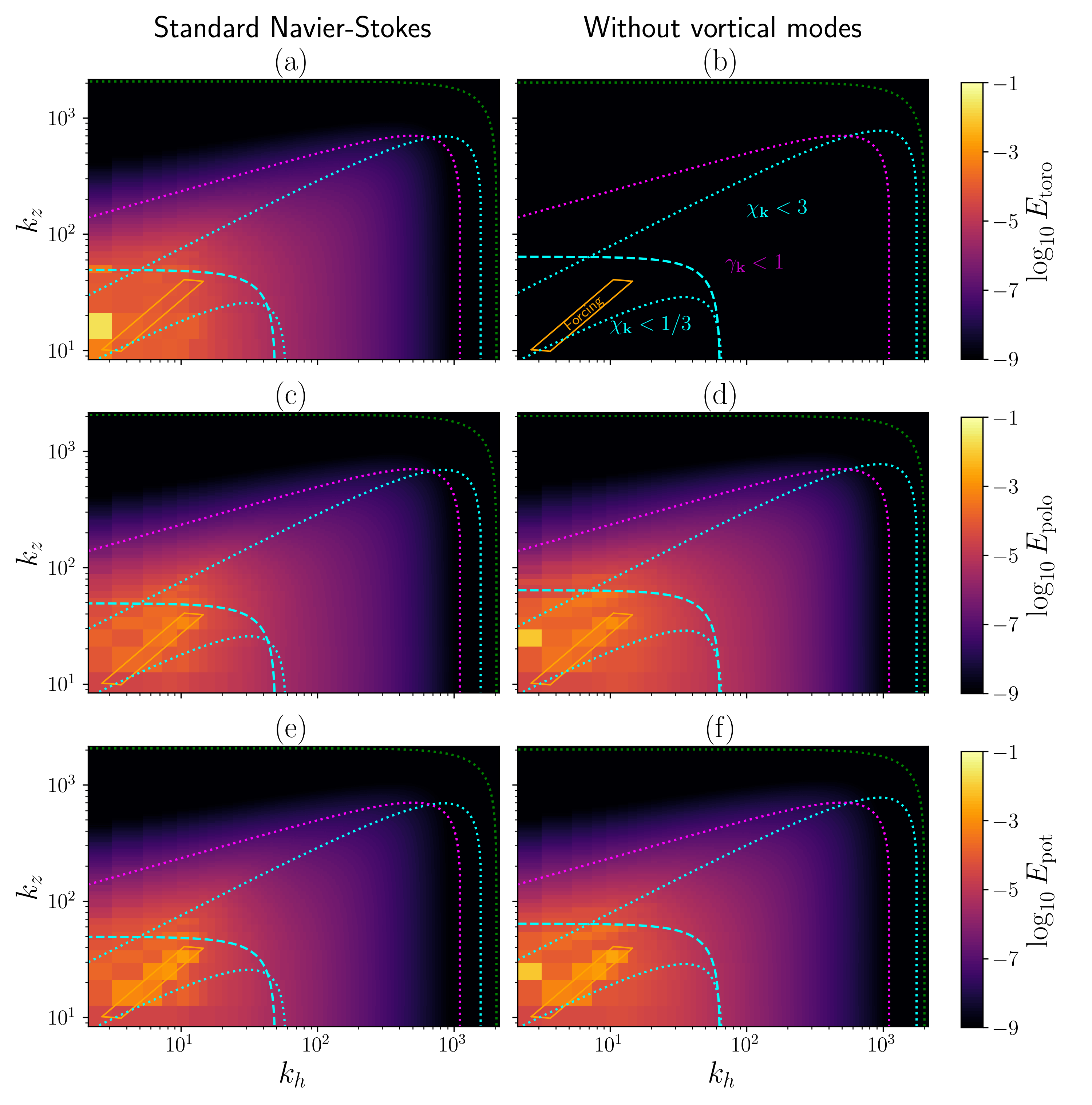}
\caption{$(k_h,k_z)$ spectra for the simulations at $(N,\R_i)=(40,20)$, with and
without vortical modes. The cyan dotted lines correspond to $\chi_{\kk}=1/3$ and
$\chi_{\kk}=3$, the cyan dashed line to $k=\kb$, the magenta dotted line to
$\gamma_{\kk}=1$, and the green dotted line to the dissipative scale. The orange box
corresponds to the forcing region. $\rm (a)$ $\Etoro$ with vortical modes. $\rm (b)$
$\Etoro$ without vortical modes. $\rm (c)$ $\Epolo$ with vortical modes. $\rm (d)$
$\Epolo$ without vortical modes. $\rm (e)$ $\Epot$ with vortical modes. $\rm (f)$
$\Epot$ without vortical modes. \label{fig:spectra_khkz}}
\end{figure}

Figure~\ref{fig:spectra_khkz} shows the $(k_h,k_z)$ spectra. Obviously, the simulation
without vortical modes has no energy in the toroidal velocity. When vortical modes are
present, an important part of the energy is contained in one vortical mode with
$(k_h,k_z) = (\Delta k_h, 2 \Delta k_z)$, corresponding to large, nearly vertically
stacked shear layers (Figure~\ref{fig:spectra_khkz}$\rm (a)$). Energy then tends to be
accumulated at the smallest horizontal wave vectors, close to shear modes. When
vortical modes are removed, energy is sill concentrated in the same wave-vectors, but
in the form of poloidal an potential energy (Figure~\ref{fig:spectra_khkz}$\rm
(d)$-$\rm (f)$). Except for this qualitative difference, the toroidal, poloidal, and
potential energy spectra show the same trends.

\begin{figure}
\includegraphics[width=1.0\textwidth]{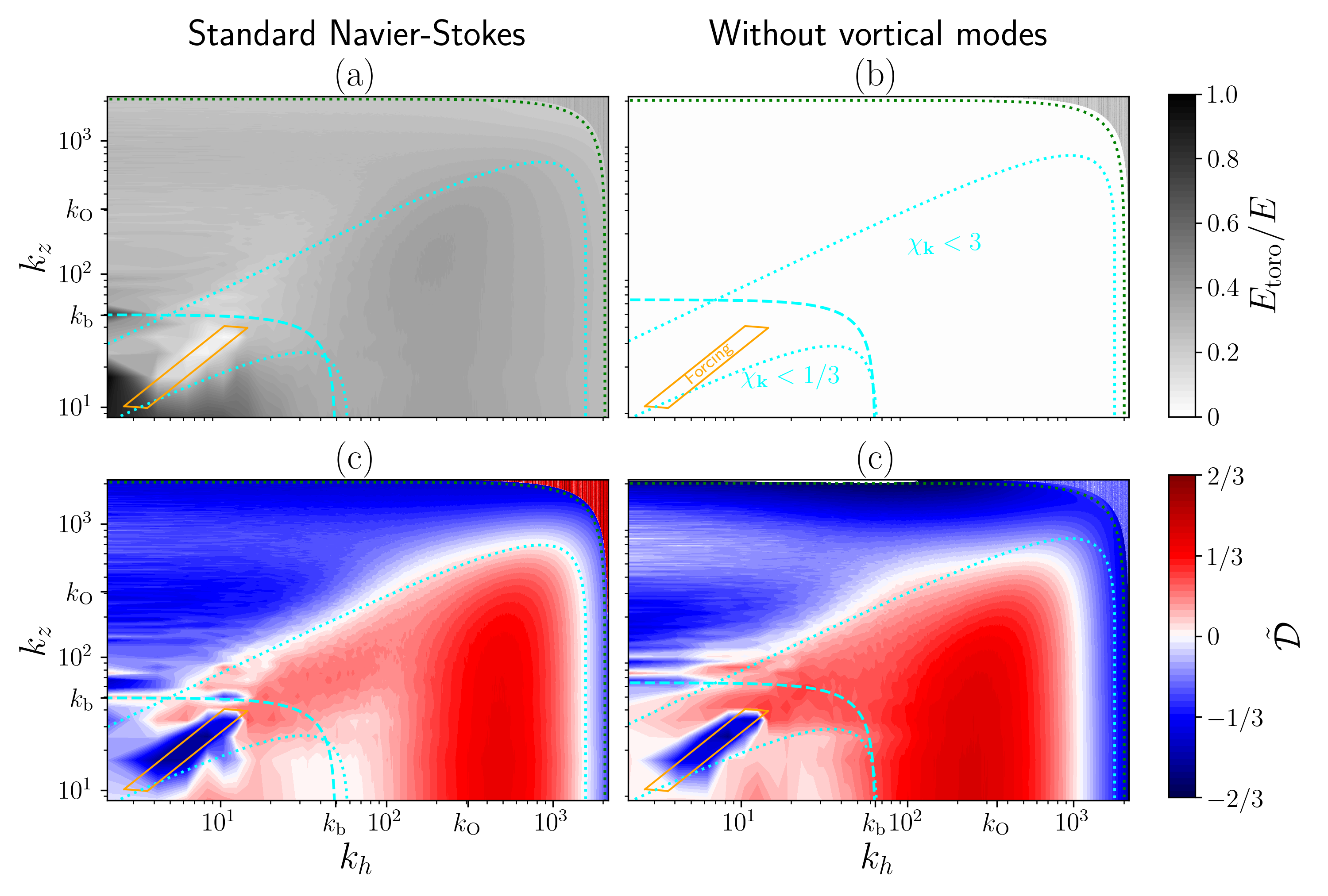}
\caption{Ratios of energy for the simulations at $(N,\R_i)=(40,20)$, with and without
vortical modes. The cyan dotted lines correspond to $\chi_{\kk}=1/3$ and
$\chi_{\kk}=3$, the cyan dashed line to $k=\kb$, and the green dotted line to the
dissipative scale. The orange box corresponds to the forcing region. $\rm (a)$
$\Etoro/E$ with vortical modes. $\rm (b)$ $\Etoro/E$ without vortical modes. $\rm (c)$
$\tilde{\mathcal{D}} = (\Epolo -\Epot)/(\Epolo+\Epot)$ with vortical modes. $\rm (d)$
$\tilde{\mathcal{D}}$ without vortical modes. A wave dominated region is expected for
the simulation without vortical where $\chi_{\kk} < 1/3$, for which $\Etoro = 0$ and
$\tilde{\mathcal{D}} \ll 1$. \label{fig:spectra_ratio_khkz}}
\end{figure}

In order to have a finer analysis of the distribution of energy in the $(k_h, k_z)$
plane, it is useful to look at energy ratios $\Etoro/E$ and $\tilde{\mathcal{D}}$
(\ref{eq:energies}) shown on Figure~\ref{fig:spectra_ratio_khkz}. One clearly observes
that $\Etoro$ is never negligible away from the forcing region, and it is very
important for small wave-vectors when the vortical modes are not removed
(Figure~\ref{fig:spectra_ratio_khkz}$\rm (a)$-$\rm (b)$). Apart from this important
difference, the spectral energy budgets of the two simulations appear to share striking
similarities. Outside the forcing region, the poloidal energy is dominant
($\tilde{\mathcal{D}} > 0$) between the cyan dotted lines $\chi_{\kk} = 1/3$ and
$\chi_{\kk} = 3$ (\ref{eq:NonLinearityParameter}), corresponding to a region where the
critical balance condition $\chi_{\kk} \sim 1$ is fulfilled
(Figure~\ref{fig:spectra_ratio_khkz}$\rm (c)$-$\rm (d)$). The potential energy is
dominant ($\tilde{\mathcal{D}} < 0$) when $\chi_{\kk} > 3$, i.e. in a region where
eddies are faster than waves. We observe a good equipartition between potential and
poloidal energy ($\tilde{\mathcal{D}} \simeq 0$) when $\chi_{\kk} < 1/3$, i.e. in the
region where waves are faster than eddies. This indicates that the ratio of temporal
scales $\chi_{\kk}$ is important when identifying ranges in anisotropic turbulence, as
already explained in \cite{yokoyama_energy-based_2019}. In particular, it shows that a
wave dominated range cannot lie above $\chi_{\kk} > 1/3$.

\subsubsection{Conversion of kinetic to potential energy}

\begin{figure}
\includegraphics[width=1.0\textwidth]{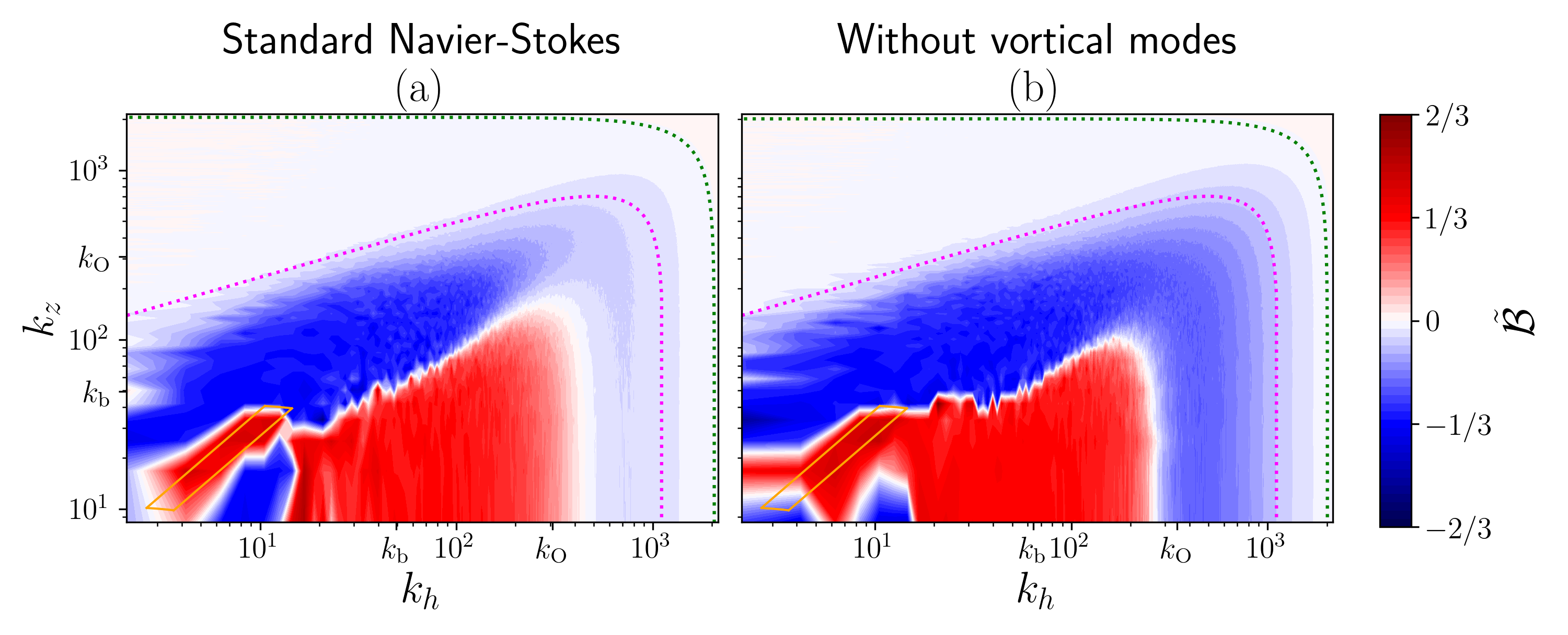}
\caption{Normalized conversion to potential energy $\tilde{\mathcal{B}}$
(\ref{eq:transfers}) for the simulations at $(N,\R_i)=(40,20)$, with $\rm (a)$ and
without $\rm (b)$ vortical modes. The magenta dotted line corresponds to
$\gamma_{\kk}=1$, and the green dotted line to the dissipative scale. The orange box
corresponds to the forcing region. \label{fig:conversion}}
\end{figure}

We can observe, in Figure~\ref{fig:conversion}, that the conversion between potential
energy and kinetic energy $\tilde{\mathcal{B}}$ is very similar with or without
vortical modes. This is not surprising since the vertical velocity is fully contained
in waves modes, and not in vortical modes. Naturally, the amplitude of
$\tilde{\mathcal{B}}$ is important only when $\gamma_{\kk} < 1$, meaning that waves
need to not be damped too strongly by viscosity or diffusivity in order to convert
potential energy to kinetic energy (or conversely). Yet, we observe that
$\tilde{\mathcal{B}}$ is non-zero and has a definite sign when $\gamma_{\kk} > 1$.
Consequently, if waves certainly exist in this range, they cannot persist because their
kinetic energy is converted into potential energy or vice-versa. Therefore, these
simulations are unlikely to correspond to a WWT regime, even in the buoyancy range and
without vortical modes. Nevertheless, we observe that $\tilde{\mathcal{B}}$ fluctuates
in time in the buoyancy range ($k\leq \kb$).

\subsection{Spatiotemporal analysis in the buoyancy range}
\label{subsec:khkzomega}

From the previous subsection, it is expected that waves can dominate (but cannot
necessarily sustain) not too far from the buoyancy range ($k \leq \kb$). To further
assess the presence and degree of nonlinearity of waves, we performed a spatiotemporal
analysis for small k. As explained in the previous subsection, waves are marked by an
equipartition between poloidal kinetic energy and potential energy. This motivates us
to introduce the equipartition energy as
\begin{equation}
\Eequi(k_h, k_z) = 2 \min \left\{\Epolo(k_h, k_z), \Epot(k_h, k_z)\right\},
\end{equation}
in order to track the presence of waves in the $(k_h, k_z)$ plane. Indeed, $\Eequi$
corresponds to the energy contained in $\Epot$ and $\Epolo$ that is in
potential-kinetic equipartition. Consequently, $\Eequi$ encompass the energy of the
waves. However, it is important to note that $\Eequi$ can also contain the energy of
structures that are not linear waves. A more careful separation between waves and
non-wave structures requires a 4D spatiotemporal filtering \cite{lam_partitioning_2020,
lam_energy_2021}, which is expensive in term of computational time and data storage.
Yet, the ratio $\Eequi(k_h,k_z) /E(k_h,k_z)$ can be used to track waves in the spectral
space at a lower cost.

\begin{figure}
\includegraphics[width=1.0\textwidth]{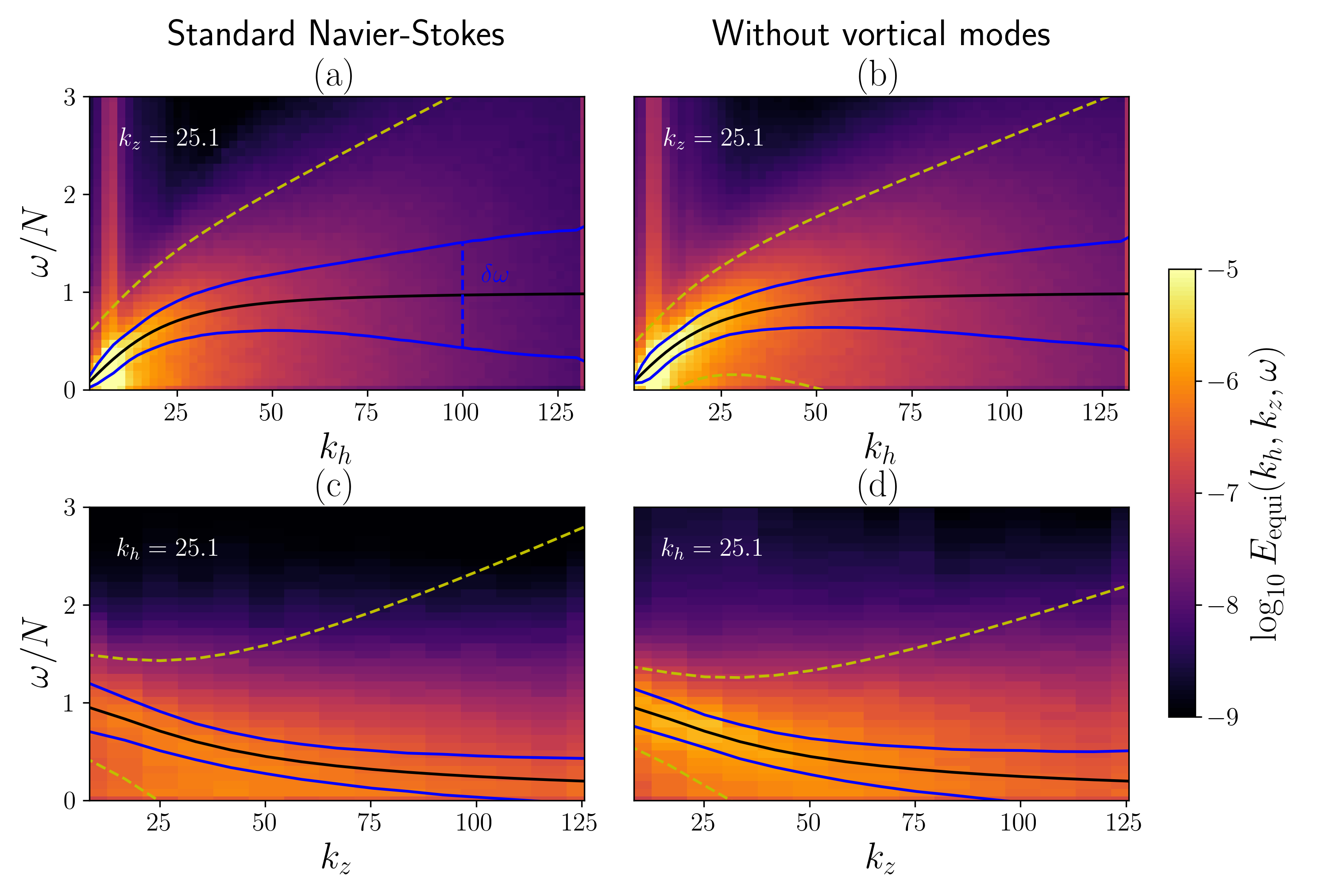}
\caption{Slices of $\Eequi(k_h,k_z,\omega)$ for the simulations at $(N,\R_i)=(40,20)$,
with and without vortical modes. (a) $k_z=25.1$ with vortical modes. (b) $k_z=25.1$
without vortical modes. (c) $k_h=25.1$ with vortical modes. (b) $k_h=25.1$ without
vortical modes. The linear dispersion relation $\omega_{\boldsymbol{k}} = Nk_h/k$ is
plotted in black, the lines $\ok \pm \delta \omega/2$ (\ref{eq:delta-omega}) in blue,
and the yellow dashed lines correspond to the lines $\ok \pm \odoppler$
(\ref{eq:Doppler}). \label{fig:omega-k}}
\end{figure}

Figure~\ref{fig:omega-k} shows slices of the temporal spectrum of the equipartition
energy $\Eequi(k_h,k_z,\omega)$ at large spatial scales, namely $k_h = 25.1$ and
$k_z=25.1$. We observe that the energy tends to be concentrated around the linear wave
frequency only at large spatial scales, but is dispersed at smaller scales even in the
simulations without vortical modes. It has been shown that this dispersion could be
quantified by considering a Doppler effect due to the advection of internal waves by
shear modes \cite{maffioli_signature_2020}. Despite the fact that shear modes are
removed in our simulations, we can try to test this prediction by considering that a
Doppler shift is due to a horizontal mean flow $\boldsymbol{U}$ of amplitude $U_h$
\begin{equation}
\label{eq:Doppler}
\odoppler = \max\limits_{\phk} ~ \boldsymbol{U} \cdot\kk = k U_h \sin \thk = \frac{k}{\kb} \ok
\end{equation}
where $\phk$ is the angle between the horizontal projection of $\kk$ and $\eex$
(Figure~\ref{fig:poloidal-toroidal}). This gives a reasonable explanation of the
important dispersion of the energy in temporal scale at large horizontal wave-vectors
(Figure~\ref{fig:omega-k}).

\begin{figure}
\includegraphics[width=1.0\textwidth]{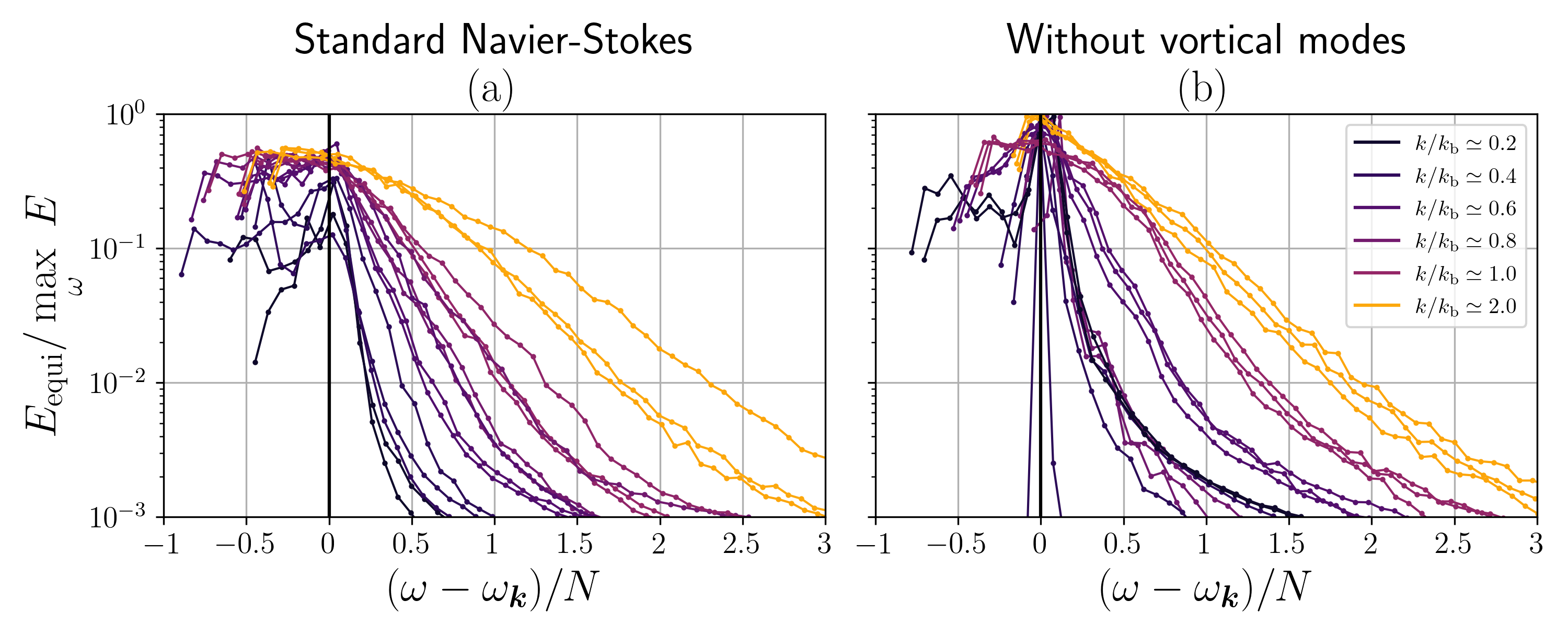}
\caption{Spatiotemporal analysis of the simulations with $(N,\R_i)=(40,20)$.
$\Eequi(k_h, k_z, \omega)/\max\limits_\omega E(k_h, k_z, \omega)$ for the simulation
with vortical modes $\rm (a)$ and for the simulation without vortical modes $\rm (b)$.
Only some couples $(k_h, k_z)$ are shown. Line colors corresponds to different values
of $k/\kb$. The vertical black lines corresponds to $\omega = \ok$.
\label{fig:omega-L}}
\end{figure}

To observe more precisely the dispersion in temporal scales, we represent the ratio
$\Eequi(k_h, k_z, \omega)/ \max\limits_{\omega} E(k_h, k_z, \omega)$ for some
$(k_h,k_z)$ on Figure~\ref{fig:omega-L}, which indicates how the energy is distributed
among temporal scales for a given spatial scale. For a system of weakly nonlinear
internal gravity waves, we should observe a peak of $\Eequi/\max\limits_{\omega} E$
around $\omega = \ok$ (for the energy to be concentrated around the linear frequency),
whose maxima should be close to unity (for the energy to be concentrated in waves
modes). We observe that the concentration of energy around the linear frequency mostly
depends on the ratio $k/\kb$. As expected, the pics of energy around $\omega = \ok$ is
more pronounced when $k \ll \kb$. This is an indication that linear waves are important
if $k \ll \kb$, with or without vortical modes. When $k \simeq \kb$, the energy spreads
over a broader range of temporal scales. We note that removing vortical modes from the
dynamics helps to concentrate $\Eequi$ in temporal scales since temporal spectra tend
to be more sharp. Simulations without vortical modes naturally have bigger
$\Eequi/\max\limits_{\omega} E$ (Figure~\ref{fig:omega-L}$\rm (b)$) when compared to
simulations with vortical modes (Figure~\ref{fig:omega-L}$\rm (a)$).

A way to measure the dispersion of energy in temporal scales is to compute the
deviation from linear waves frequency $\delta \omega$. We estimate it using the
equipartition energy spectrum and defining the following measure,

\begin{equation}
\label{eq:delta-omega}
\delta \omega(k_h, k_z) = \sqrt{ \frac{\sum\limits_{\omega} ~ (\omega - \ok)^2 ~ E_{\rm equi}(k_h, k_z, \omega)}{\sum\limits_{\omega} ~ E_{\rm equi}(k_h, k_z, \omega)}}.
\end{equation}

\begin{figure}
\includegraphics[width=1.0\textwidth]{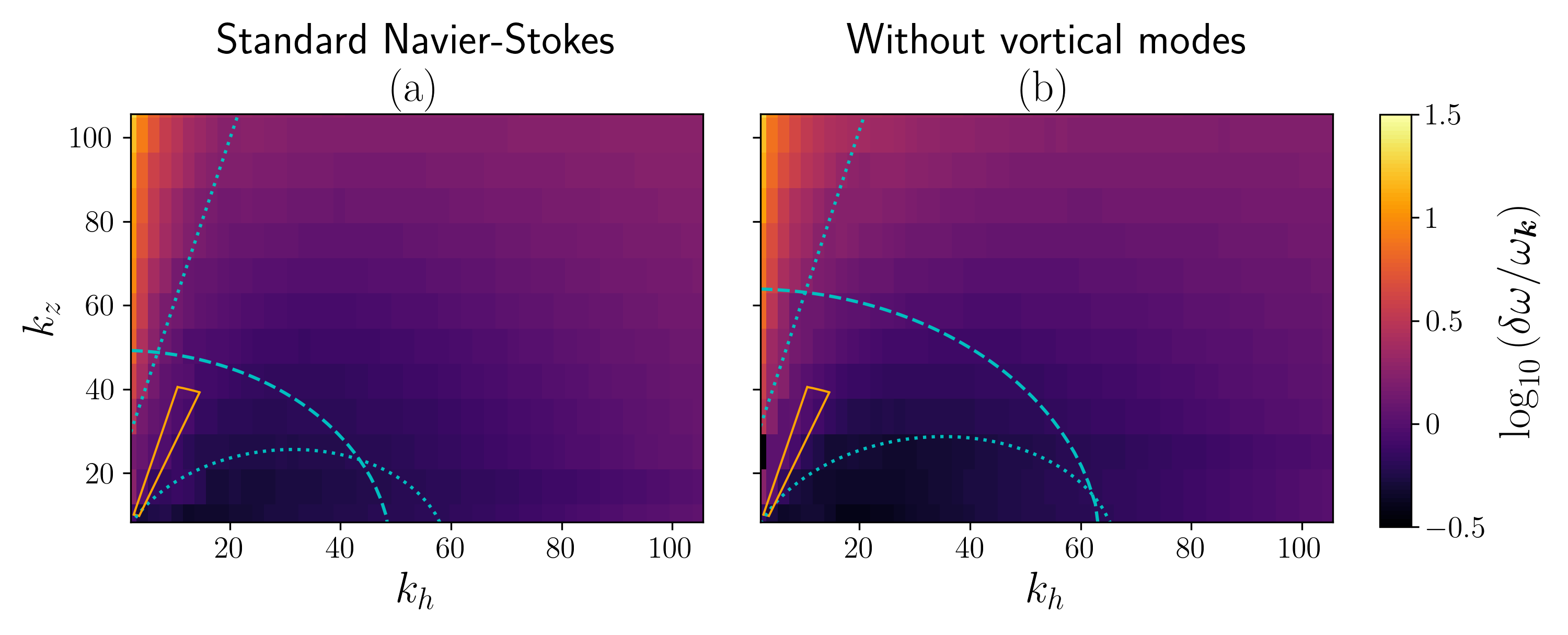}
\caption{$\delta \omega/\ok$ for simulations with $(N,\R_i) = (40,20)$ with vortical
modes (a) and without vortical modes (b). Dotted lines correspond to $\chi_{\kk} = 1/3$
and $\chi_{\kk} = 3$, the dashed line to $k=\kb$, and the orange box to the forcing
region.  \label{fig:nonlinear}}
\end{figure}

The quantity $\delta \omega / \ok$ is another way to estimate the strength of nonlinear
interactions. Unlike the nonlinearity parameter $\chi_{\kk}$, which is defined by
dimensional analysis, $\delta \omega/ \ok$ requires knowledge of the spatio-temporal
spectra. $\delta \omega/ \ok$ is called nonlinear broadening, and is particularly
important in the context of WWT: for $\delta \omega/ \ok \gg 1$, waves' dynamics is
strongly affected by nonlinear interactions, while when $\delta \omega/ \ok \ll 1$,
waves can propagate with only weak nonlinear perturbations and the theory can hold
\cite{nazarenko_wave_2011}. For a system of non-interacting linear waves, $\delta
\omega/\ok$ should be zero. Figure~\ref{fig:nonlinear} shows this quantity in the
$(k_h, k_z)$ plane. It shows that the nonlinear broadening is small if $k \ll \kb$ and
$\chi_{\kk} \leq 1/3$, which is consistent with previous results
\cite{yokoyama_energy-based_2019}. Conversely, $\delta \omega/\ok$ is large when $k \gg
\kb$ or $\chi_{\kk} > 3$. We observed that removing vortical tends to decrease slightly
$\delta \omega/\ok$ at small $k$ and large $\thk$. This might be explained by the fact
that the simulations without vortical modes tend to have a larger buoyancy range when
compared to the simulations with vortical modes.

In order to allow a continuum of interaction between waves, it is also important that
the nonlinear broadening remains much larger than the frequency gap between discrete
modes in the Fourier space \cite{lvov_discrete_2010}. This leads to the condition
$\Delta \omega \equiv \bnabla_{\kk} \ok \cdot \Delta \kk \ll \delta \omega$ where
$\Delta \kk = (\Delta k_x, \Delta k_y, \Delta k_z)$. Taking $\Delta k_x = \Delta k_y =
\Delta k_z = \Delta k$ leads to $|\cos \thk| \Delta k/ k \ll \delta \omega /N$, which
is satisfied in most of our simulations for sufficiently large $k$.

\subsection{Wave energy in the parameters space}
\label{subsec:waves_energy}

The previous subsections presented the dispersion in temporal scales for two
simulations, but we did not quantify the amount of waves in the $(k_h,k_z)$ plane, nor
in the $(F_h,\R)$ plane so far. This is the goal of this subsection.

\begin{figure}
\includegraphics[width=1\textwidth]{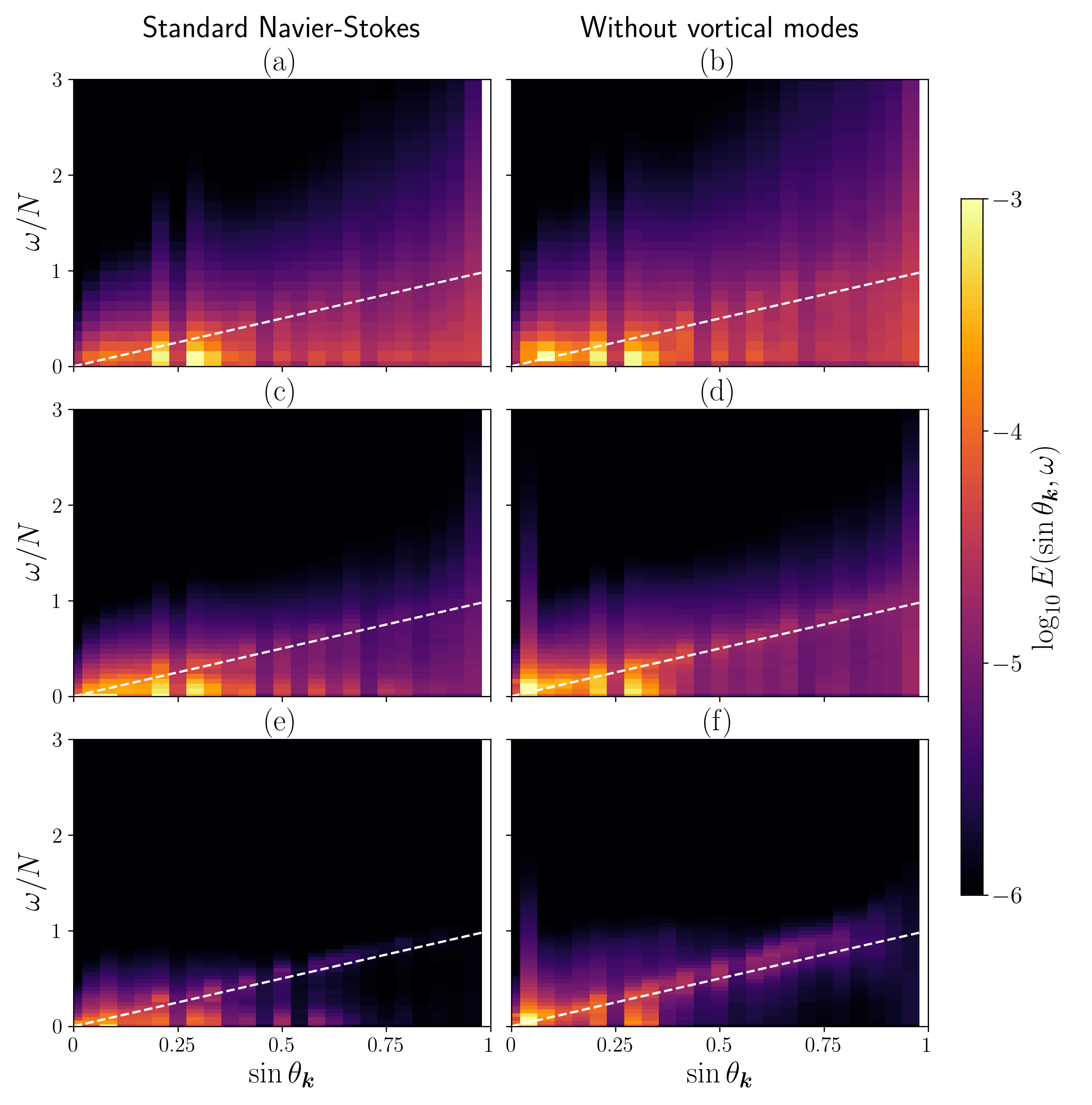}
\caption{Integrated spatiotemporal spectra of the total energy as a function of $\ok/N
= \sin \thk$ and $\omega$ for different simulations at the same viscosity $\nu = 1/(N^2
\R_i)$. $\rm (a)$ $(N,\R_i)=(20,160)$ with vortical modes, $\rm (b)$
$(N,\R_i)=(20,160)$ without vortical modes, $\rm (c)$ $(N,\R_i)=(40,40)$ with vortical
modes, $\rm (d)$ $(N,\R_i)=(40,40)$ without vortical modes, $\rm (e)$
$(N,\R_i)=(80,10)$ with vortical modes, and $\rm (f)$ $(N,\R_i)=(80,10)$ without
vortical modes. The white dotted line represents $\omega  = \ok$, i.e. the linear
dispersion relation. \label{fig:omega_vs_omegak}}
\end{figure}

To discuss the effect of stratification, we show the integrated spatiotemporal total
energy spectra as a function of $\thk/N = \sin \thk$ and $\omega$ for different
simulations in Figure~\ref{fig:omega_vs_omegak}. At low $N$, we observe no big
difference in $E(\ok, \omega)$ between simulations with and without vortical modes
(Figure~\ref{fig:omega_vs_omegak}$\rm (a)$-$\rm (b)$). We observe that, for these
weakly stratified flows, the energy is mostly contained in small frequency modes, but
present over a large range of temporal scales for a given angle of the wave-vector
$\thk$. At larger $N$, energy concentrate around the linear dispersion relation $\omega
= \ok$ and in slow modes $\omega = 0$ for simulation with vortical modes
(Figure~\ref{fig:omega_vs_omegak}$\rm (c)$). When vortical modes are absent, energy
tends to accumulate around $\omega = \ok$ only (Figure~\ref{fig:omega_vs_omegak}$\rm
(d)$). At the highest $N$, the two branches $\omega = \ok$ and $\omega = 0$ are visible
for the simulation with vortical modes (Figure~\ref{fig:omega_vs_omegak}$\rm (e)$)
while only the branches $\omega = \ok$ is clearly observed in the simulation without
vortical modes (Figure~\ref{fig:omega_vs_omegak}$\rm (f)$).

In order to quantify wave energy in a flow, we propose to simply filter the
spatiotemporal spectra $\Eequi$ by a Gaussian weight with mean $\ok$ and
root-mean-square  $\epsilon \ok$, where $\epsilon$ is an arbitrary but small parameter,
such that the wave energy at a given $(k_h,k_z,\omega)$ is
\begin{equation}
\label{eq:Ewave}
\Ewave(k_h,k_z,\omega) =  \Eequi(k_h,k_z,\omega) ~ \exp \left[-\frac{1}{2} \left(\frac{\omega - \ok(k_h,k_z)}{ \epsilon \ok(k_h,k_z)} \right)^2 \right].
\end{equation}
We observed that choosing $\epsilon=0.1$ gives a sufficiently narrow window to keep
only the spatiotemporal region corresponding to internal gravity waves, while keeping
the window sufficiently wide to avoid binning effects. Therefore, we kept this value
for the analysis. This window with variable width allows us to get rid of the energy of
non-wave structures at low frequency. We first look at the distribution of the wave
energy in the $(k_h,k_z)$ plane. In order to allow a fair comparison between
simulations with and without vortical modes, it is necessary to introduce the wave
energy ratio
\begin{equation}
\label{eq:WaveRatio}
\Ewaver(k_h,k_z) =  \frac{\Ewave(k_h,k_z)}{\Epolo(k_h,k_z) + \Epot(k_h,k_z)} ~~~~ \text{where} ~~~~ \Ewave(k_h,k_z) = \sum\limits_{\omega} ~ \Ewave(k_h,k_z,\omega).
\end{equation}

\begin{figure}
\includegraphics[width=1.0\textwidth]{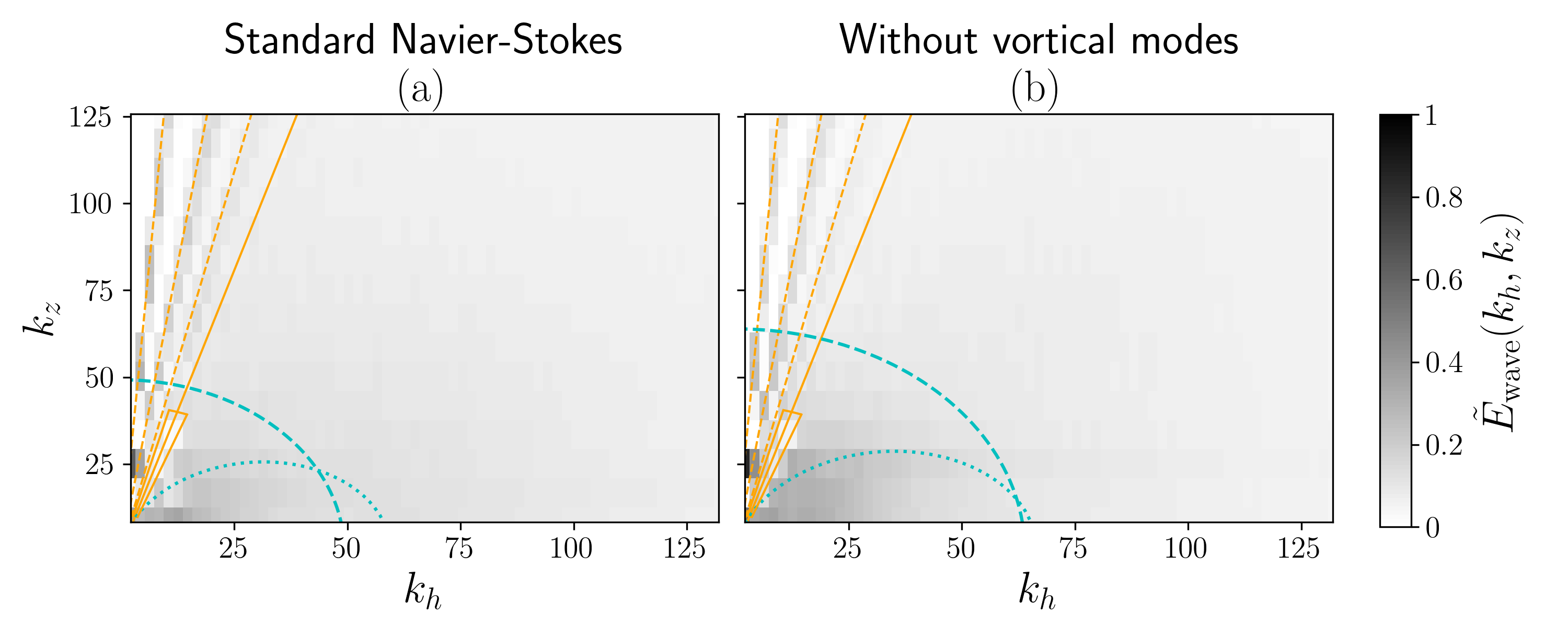}
\caption{Wave energy ratio $\Ewaver(k_h,k_z)$ (\ref{eq:WaveRatio}) for simulations with
$(N,\R_i)=(40,20)$ with vortical modes $\rm (a)$, and without vortical modes $\rm (b)$.
The cyan dotted lines correspond to $\chi_{\kk}=1/3$, and the cyan dashed line to
$k=\kb$. The orange box corresponds to the forcing region. The forcing angle $\theta_f$
is represented by a full orange line, while first harmonics
(\ref{eq:harmonic1}-\ref{eq:harmonic3}) excited by the TRI are indicated by dashed
orange lines. \label{fig:ratioEwaves_khkz}}
\end{figure}

Normalizing the wave energy by the total energy (i.e. including vortical modes energy)
would have mask the presence of waves in the simulation with vortical modes, since they
represent an important part of the energy (Figure~\ref{fig:spectra_ratio_khkz}$\rm
(a)$). The wave energy ratio $\Ewaver(k_h,k_z)$ is depicted on
Figure~\ref{fig:ratioEwaves_khkz}$\rm (a)$-$\rm (b)$. We observe that $\Ewaver$ tends
to be higher for simulations without vortical modes, but removing vortical modes does
not change the variations of $\Ewaver$ in the $(k_h,k_z)$ plane for these simulations.
As we could have expected from previous works \cite{yokoyama_energy-based_2019} and the
previous subsections, $\Ewaver$ is higher in the buoyancy range, in particular where
$\chi_{\kk} \leq 1/3$. Yet, $\Ewaver$ tends to be bigger at very specific angles,
smaller than the forcing angle $\theta_f$. This accumulation of internal gravity wave
energy at specific propagation angles has been reported in earlier studies (see e.g.
\cite{maffioli_signature_2020}). In the present simulations, we can explain this by
invoking Triadic Resonance Instabilities (TRI) between internal gravity waves
\cite{brouzet_internal_2016}, in particular the PSI
\cite{mccomas_resonant_1977,muller_nonlinear_1986}. Considering that a primary wave of
frequency $\omega_f = \omega_0^* = N \sin \theta_f$ is excited by the forcing, and then
decays into two daughter waves of the same frequency $\omega_1^*$ according to the PSI
mechanism, we can deduce $\omega_1^*$ and the associated propagation angle
$\theta_1^*$:
\begin{equation}
\label{eq:harmonic1}
\omega_1^* = \frac{\omega_0^*}{2} = 0.15~N ~~~~ \Rightarrow
~~~~\theta_1^* = \arcsin \left(\frac{\omega_1^*}{N} \right)\simeq 0.15.
\end{equation}
Invoking, for a second time, the PSI mechanism for a wave of frequency $\omega_1^*$, we
can deduce a second harmonic
\begin{equation}
\label{eq:harmonic2}
\omega_2^* = \frac{\omega_1^*}{2} = 0.075~N ~~~~ \Rightarrow
~~~~\theta_2^* = \arcsin \left(\frac{\omega_2^*}{N} \right)\simeq 0.075.
\end{equation}
A third harmonic is obtained by considering that two waves of frequencies $\omega_1^*$
and $\omega_2^*$ interact to give a third wave with frequency
\begin{equation}
\label{eq:harmonic3}
\omega_3^* = \omega_1^* + \omega_2^* = 0.225~N ~~~~ \Rightarrow
~~~~\theta_3^* = \arcsin \left(\frac{\omega_3^*}{N} \right)\simeq 0.225.
\end{equation}
$\theta_f$, $\theta_1^*$, $\theta_2^*$, and $\theta_3^*$ are reported on
Figure~\ref{fig:ratioEwaves_khkz}$\rm (a)$-$\rm (b)$. We see that they reproduce well
with the higher values of $\Ewaver$ observed at small angles. This agreement suggests
that waves at small frequencies are excited by TRI in our strongly stratified
simulations.

The spatiotemporal analysis performed in this subsection allows to give a more precise
measure of the dominance of waves in the spectral space at a given $(F_h,\R)$. The
total wave energy ratio is defined as
\begin{equation}
\Ewaver = \frac{\sum\limits_{k_h,k_z,\omega} ~ \Ewave(k_h,k_z,\omega)}{\sum\limits_{k_h,k_z} ~ \Epolo(k_h,k_z) + \Epot(k_h,k_z)}.
\end{equation}

\begin{figure}
\includegraphics[width=1\textwidth]{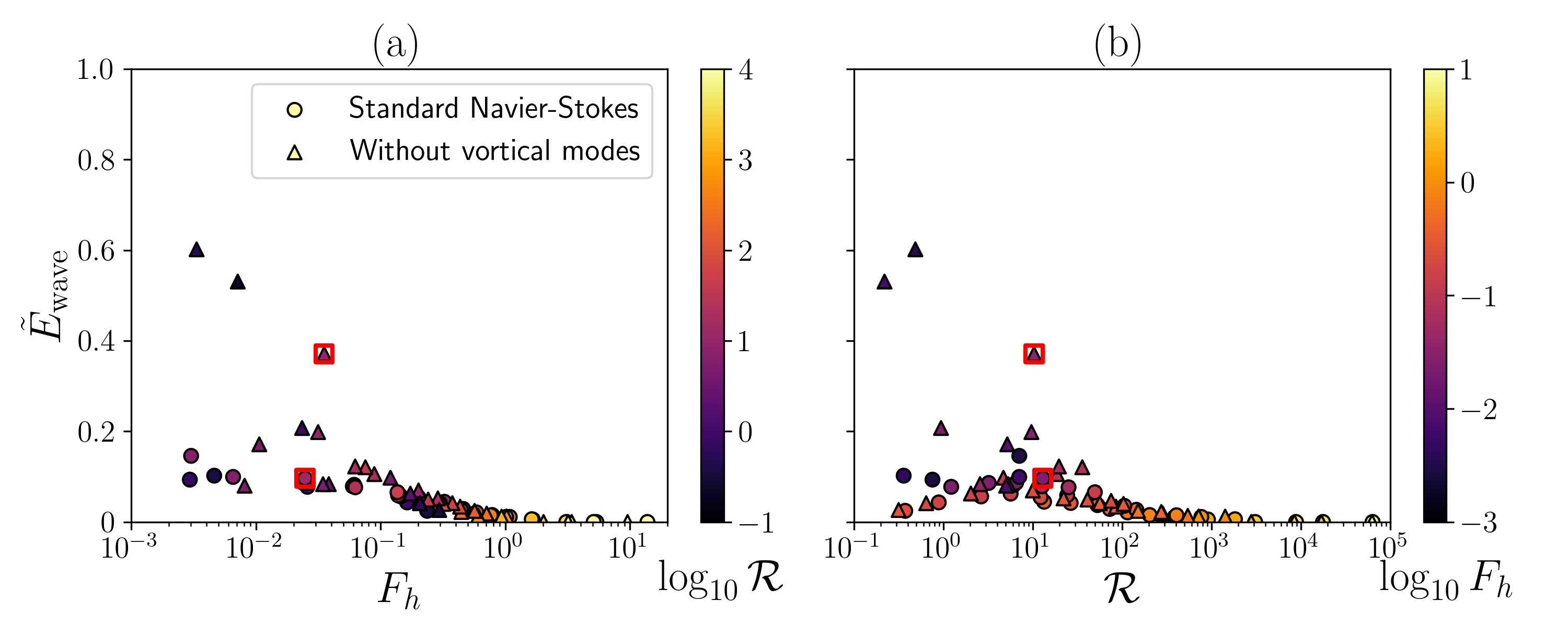}
\caption{$\Ewaver$ as a function of $F_h$ $\rm (a)$ and $\R$ $\rm (b)$ for all our
simulations with or without vortical modes. The red boxes indicate the simulations with
$(N,\R_i)=(40,20)$ investigated in the previous subsections.
\label{fig:ratioEwaves_Fh}}
\end{figure}

Figure~\ref{fig:ratioEwaves_Fh} shows $\Ewaver$ in the plane $(F_h, \R)$. Consistently
with the previous subsections, we observe that removing the vortical modes helps to get
higher values of $\Ewaver$, and thus, in principle, to get closer to a WWT regime. Yet,
unstratified flows ($F_h \geq 1$) have almost no energy in waves modes (i.e. $\Ewave
\ll 1$), and increasing the stratification (decreasing $F_h$) is required to increase
the wave energy ratio (Figures~\ref{fig:ratioEwaves_Fh}$\rm (a)$ and
\ref{fig:omega_vs_omegak}), but it also tends to increases the vortical mode energy
(Figure~\ref{fig:global-energy}$\rm (a)$) when they are not artificially removed from
the dynamics. Figure~\ref{fig:ratioEwaves_Fh}$\rm (b)$ shows that $\Ewaver$ tends to
decrease with $\R$ for sufficiently low $F_h$. This is consistent with the numerical
simulations of stratified flows forced by tides \cite{reun_parametric_2018}, which show
convincing signatures of a WWT regime at relatively small $\R$ when compared to more
usual strongly stratified simulations. Some of our simulations at low $F_h$ and high
$\R$ remain affected by hyperviscosity, so $\R$ may not accurately quantify the effect
of dissipation in this case. However, discarding these simulations from the analysis
still leads to the same conclusions.

\section{Discussions and conclusions}
\label{sec:conclusions}

In order to investigate the conditions under which a weak internal gravity wave
turbulence regime could occur, we performed direct numerical simulations of stratified
turbulence without shear modes, and with or without vortical modes, at various Froude
and buoyancy Reynolds numbers.

We observed that removing vortical modes helps to have a better overall balance between
poloidal kinetic energy and potential energy. However, the spatial spectra appear to
behave very similarly with or without vortical modes in our simulations. The spectral
energy budget reveals that there is no range in $(k_h,k_z)$ for which the conversion
between kinetic energy and potential energy fluctuates around zero, as we would expect
for a system of statistically stationary waves. Additionally, the conversion between
potential energy and kinetic energy becomes small where the wave dissipation parameter
$\gamma_{\kk}$ (\ref{eq:DissipationParameter}) is larger than unity. Due to the
anisotropy of stratified flows, this means that it exists a range of scales where waves
are efficiently dissipated by viscosity, but not necessarily the vortices.

A spatiotemporal analysis in the buoyancy range showed that increasing stratification
is necessary to decrease the nonlinear broadening to the level required for a weak wave
turbulence regime. As in \cite{yokoyama_energy-based_2019}, we observed that waves are
present in a region delimited by the nonlinearity parameter $\chi_{\kk}$
(\ref{eq:NonLinearityParameter}). More precisely, waves are more likely to dominate
where $\chi_{\kk} < 1/3$, if the vortical modes are removed. We also observed evidences
of the presence of slow waves ($k_z \gg k_h$) subject to TRI where $\chi_{\kk} \geq
1/3$ in simulations with or without vortical modes. However, the nonlinear broadening
is shown to be large for $k_z \gg k_h$, so slow waves are less susceptible to be
described by the WWT theory. Although removing vortical modes is not enough to observe
a weak wave turbulence regime, simulations without vortical modes produce $E(\ok,
\omega)$ plots (Figure~\ref{fig:omega_vs_omegak}) characterized by a better
concentration of energy in wave modes. Additionally, temporal spectra of strongly
stratified simulations without vortical modes exhibit a larger inertial range in
temporal scales (Figure~\ref{fig:spectra_omega}). The spectra are also compatible with
the Kolmogorov-Zakharov spectrum offered by WWT. However, this spectrum is known to be
mathematically unrealizable, so the agreement between this prediction and our
simulations remains to be explained.

Using a simple diagnostic to quantify wave energy contained in a stratified flow, we
showed that that the buoyancy Reynolds number should not be too large in order to
observe a WWT regime. To understand this, we propose the following conditions in order
to limit the generation of non wave structures and to ensure the weak nonlinearity for
the evolution of all energetic modes:
\begin{itemize}
\item The waves and eddies should be dissipated in the buoyancy range which leads to
the conditions
\begin{equation}
\label{eq:DissInBuoyancy}
\kd \leq \kb ~~~~ \Rightarrow ~~~~ \R \leq F_h ~~~~ \text{and} ~~~~ k_{\eta} \leq \kb ~~~~ \Rightarrow ~~~~ \R \leq F_h^{2/3}.
\end{equation}
if we assume that dissipation of eddies occurs at the Kolmogorov wave-vector $k_{\eta}
= (\epsK/\nu^3)^{1/4}$. These conditions are necessary to avoid wave breaking and the
development of 3D small-scale eddies.

\item The nonlinearity parameter $\chi_{\kk}$ should remain small for all energetic
modes. If we assume that dissipation occurs at $k \lesssim k_\eta$, this leads to the
condition
\begin{equation}
\chi_{\rm max} \equiv \max\limits_{\substack{\kk \\ k \leq k_{\eta}}} \chi_{\kk}\leq 1 ~~~~ \Rightarrow ~~~~ \max\limits_{\substack{\kk \\ k \leq k_{\eta}}} \frac{1}{\sin \thk} \left(\frac{k}{\ko} \right)^{2/3} = \frac{k_\eta}{k_{\rm h, min}} \left(\frac{k_\eta}{\ko} \right)^{2/3} \leq 1
\end{equation}
where $k_{\rm h, min}$ represents the minimal horizontal wave-vector. If we consider a
flow of finite size with no shear modes, $k_{\rm h, min} = 2\pi / L_h$ such that
\begin{equation}
\label{eq:smallchi}
\chi_{\rm max} = \frac{k_\eta L_h}{2 \pi} \left(\frac{k_\eta}{\ko} \right)^{2/3} \leq 1 ~~~~ \Rightarrow ~~~~  \left(\frac{\epsK L_h^4}{\nu^3} \right)^{1/4} \R^{1/2} \lesssim 1.
\end{equation}
\end{itemize}
Finally, we require the flow to be in a fully turbulent regime such that
\begin{equation}
\label{eq:LargeRe}
Re = \R F_h^{-2} \gg 1 ~~~~ \text{and} ~~~~ \epsK \sim U_h^3 / L_h.
\end{equation}
Then, substituting the definition of $Re$ from equation (\ref{eq:FhR}) and condition
(\ref{eq:LargeRe}) into equation (\ref{eq:smallchi}) yields
\begin{equation}
\label{eq:SmallChi}
\R \lesssim F_h^{6/5}.
\end{equation}

\begin{figure}
\includegraphics[width=0.66\textwidth]{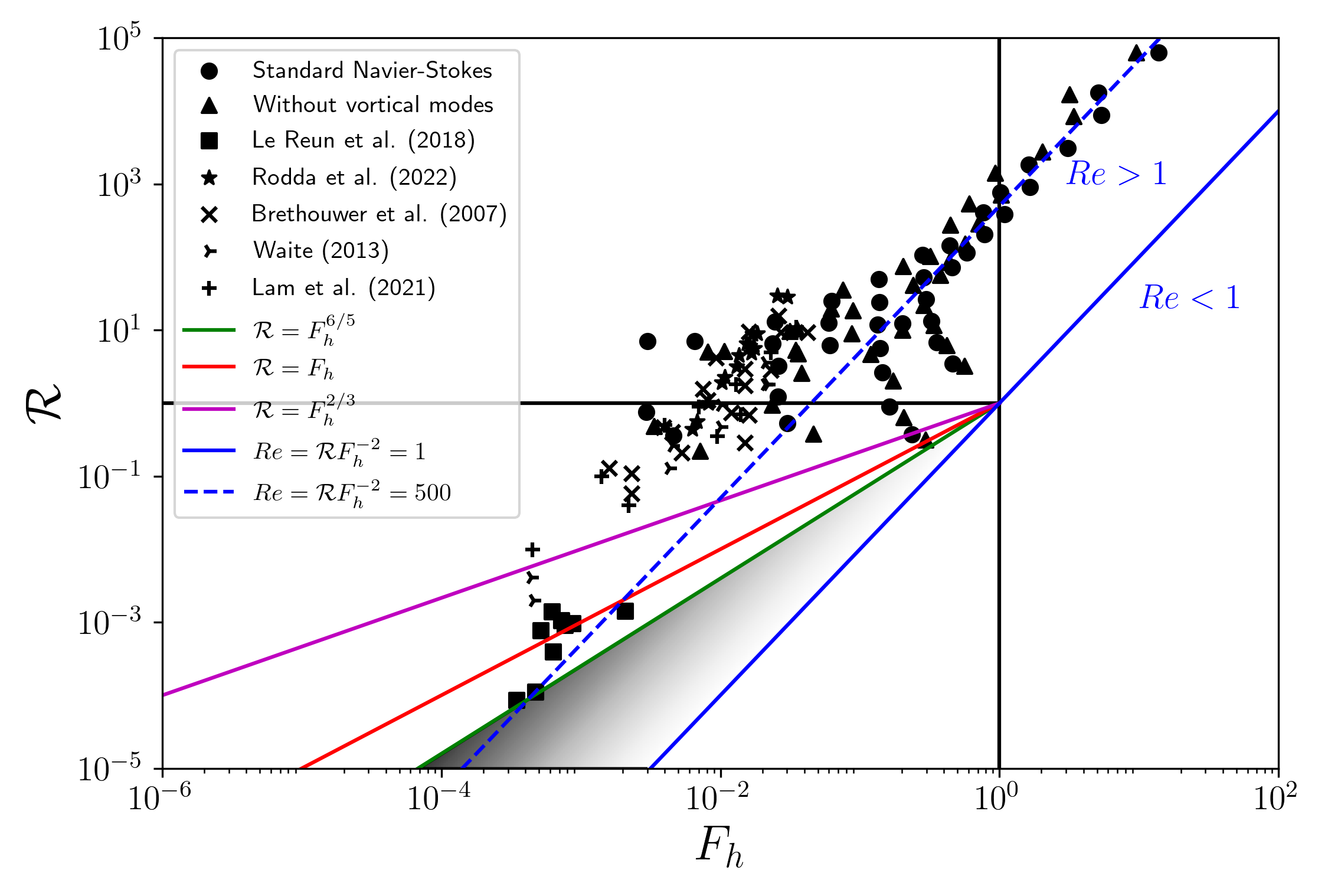}
\caption{Simulations of the present study and
\cite{brethouwer_scaling_2007,waite_potential_2013,reun_parametric_2018,lam_energy_2021},
and the experiments \cite{rodda_experimental_2022} in the $(F_h, \R)$ parameters space.
The colored full lines corresponding to conditions (\ref{eq:DissInBuoyancy}),
(\ref{eq:LargeRe}), and (\ref{eq:SmallChi}) (see legend). The blue dashed line
corresponds to $Re = 500$. The colored region is where a weak wave turbulence regime is
expected. \label{fig:discussion}}
\end{figure}

To summarize, we propose that a weak internal gravity waves turbulence regime could
occur only if $F_h$ and $\R$ are such that conditions (\ref{eq:DissInBuoyancy}),
(\ref{eq:LargeRe}), and (\ref{eq:SmallChi}) are fulfilled. Note that the last
thresholds are obtained by dimensional analysis, so they are defined up to
dimensionless factors, which are set to one for simplicity. On
Figure~\ref{fig:discussion}, we plot our simulations in the $(F_h,\R)$ plane, as well
as the simulations of
\cite{brethouwer_scaling_2007,waite_potential_2013,reun_parametric_2018,lam_energy_2021}
and some recent experiments \cite{rodda_experimental_2022}. For the simulations of
\cite{reun_parametric_2018}, the Froude number is computed using our definition
(\ref{eq:FhR}), and the buoyancy Reynolds number as $\R = Re F_h^2$ in order to compare
to our work. Namely, we have used Table~1 in \cite{reun_parametric_2018} with $U_h =
u_{\rm rms}$, $\epsK = \varepsilon_k$, and $Re = Re_0$. This leads to different values
of $F_h$ and $\R$ than the ones presented by the authors. We observe that our
simulations do not lie in the region corresponding to (\ref{eq:DissInBuoyancy}),
(\ref{eq:LargeRe}), and (\ref{eq:SmallChi}). On the contrary,
\cite{reun_parametric_2018} attained such a region in their simulations, which provides
a tangible explanation for why they obtained better signatures of WWT. It is worth
mentioning that \cite{reun_parametric_2018} did not have to remove shear nor vortical
modes in their simulations to observe signatures of internal wave turbulence. It
indicates that there must be some threshold below which both shear and vortical modes
do not grow if not directly forced. Such a situation would be analogous to rotating
flows where there is a threshold below which geostrophic modes do not grow
\cite{reun_experimental_2019}. We observe that \cite{waite_potential_2013} also
attained very small $\R$. In this study, the authors forced vortical modes so their
simulations are not well suited for WWT. Yet, they showed that potential enstrophy
tends to be quadradic (i.e. $V \simeq V_2$) for $F_h, \R \ll 1$ and that $V_2 \propto
\int ~ \Omega_z^2 ~ \dxdydz$ increases with $\R$ in their simulations. It suggests that
vortical modes energy increases with $\R$ when $\R \lesssim 1$, as explained by
\cite{lam_energy_2021}. Yet, the study of the wave energy ratio is not done while
keeping $F_h$ or $\R$ constant in \cite{lam_energy_2021}. To our knowledge, additional
studies are needed to confirm if the wave energy ratio is for $\R\ll 1$ and $F_h \ll 1$
a decreasing function of $\R$, and if a threshold below which both shear and vortical
modes are stable exists.

If the latter claims are true, they constitute very prohibitive conditions for
observing WWT in numerical simulations and experiments. To illustrate this point, let
us consider a turbulent flow at $Re = 10^3$ that typically requires a $\sim 1024^3$
resolution. Then (\ref{eq:SmallChi}) stipulates that the flow need to attain $F_h \leq
Re^{-5/4} = 1.78 \times 10^{-4}$ for the nonlinearity parameter to be small for all
energetic modes. Yet, the kinetic time for a system of internal gravity waves is
expected to grow as $F_h^{-2}$, meaning that very long simulations are required to
attain the steady state, even for relatively small domains.

\begin{acknowledgments}
We thank two anonymous reviewers for their constructive feedback. This project was
supported by the Simons Foundation through the Simons collaboration on wave turbulence.
Part of the computations have been done on the ``Mesocentre SIGAMM'' machine, hosted by
Observatoire de la Cote d'Azur. The authors are grateful to the OPAL infrastructure
from Université Côte d'Azur and the Université Côte d’Azur's Center for
High-Performance Computing for providing resources and support. This work was granted
access to the HPC/AI resources of IDRIS under the allocation 2022-A0122A13417 made by
GENCI.
\end{acknowledgments}

\appendix

\section{Forcing scheme}
\label{appendix:forcing}

Our forcing is designed to excite waves and computed via generation of pseudo random
numbers with uniform distribution and time interpolation. Namely, we use the following
algorithm:

\begin{algorithm}[H]
$t = 0$; $t_0 = 0$; \\
Generate two complex random fields $\hat f_0(\kk)$ and $\hat f_1(\kk)$; \\
$\hatff = \hat f_0 \, \eep$;\\
Normalize $\hatff$ to ensure
$P_K(t) = \sum\limits_{\kk} ~ \Re \left[ \hatff \cdot \hatvv^* + \frac{\Delta t}{2} |\hatff|^2 \right] =  1$;\\
\While{$t \leq T$}{
	$t = t + \Delta t$; \\
	\If{$t - t_0 \geq T_c$}{
		$t_0 = t$; \\
		$\hat f_0 = \hat f_1$; \\
		Generate $\hat f_1$;
	}{}
	$\hatff = \left\{ \hat f_0 - \dfrac{(\hat f_1 - \hat f_0)}{2} \left[\cos\left(\dfrac{\pi(t-t_0)}{T_c} \right)+ 1 \right]\right\}\, \eep$; \\
	Normalize $\hatff$;
}
\end{algorithm}
where $\Delta t$ is the time increment at each time step and $T$ is the final time of
the simulation. The random complex fields are built such that their inverse Fourier
transform is real and they are null for unforced wavenumbers.

\section{List of simulations}

\begin{table}
\caption{Overview of the numerical and physical parameters used in the simulations.
$\R_i = P_K/ \nu N^2$. $F_h$ and $\R$ are the turbulent horizontal Froude number and
buoyancy Reynolds number, respectively, defined in (\ref{eq:FhR}).}
\begin{tabular}{m{4cm}m{5.5cm}m{5.5cm}}
\textbf{Control parameters} & \textbf{With vortical modes} & \textbf{Without vortical modes} \\
\end{tabular}
\begin{tabular}{|lll|}
	\toprule
	 $N$ & $\R_i$ & $L_h/L_z$ \\
	\midrule
	0.25 &  64000 &         2 \\
	0.66 &   9000 &         2 \\
	0.66 &  18000 &         2 \\
	1.12 &   3200 &         2 \\
	   2 &   1000 &         2 \\
	   2 &   2000 &         2 \\
	   3 &    450 &         2 \\
	   3 &    900 &         2 \\
	   4 &    250 &         2 \\
	   4 &    500 &         2 \\
	5.20 &    150 &         2 \\
	6.50 &    100 &         2 \\
	6.50 &    200 &         2 \\
	  10 &      5 &         2 \\
	  10 &     10 &         2 \\
	  10 &     20 &         2 \\
	  10 &     40 &         2 \\
	  10 &     80 &         2 \\
	  10 &    160 &         2 \\
	14.5 &     20 &         2 \\
	  20 &      1 &         4 \\
	  20 &      2 &         4 \\
	  20 &      5 &         4 \\
	  20 &     10 &         4 \\
	  20 &     20 &         4 \\
	  20 &     40 &         4 \\
	  20 &     80 &         4 \\
	  30 &     10 &         4 \\
	  30 &     20 &         4 \\
	  30 &     40 &         4 \\
	  40 &      1 &         4 \\
	  40 &      2 &         4 \\
	  40 &      5 &         4 \\
	  40 &     10 &         4 \\
	  40 &     20 &         4 \\
	  60 &     10 &         4 \\
	  80 &    0.5 &         8 \\
	  80 &      1 &         8 \\
	  80 &     10 &         8 \\
	\bottomrule
\end{tabular}
\begin{tabular}{|llll|}
	\toprule
	$n_h$ & $\kmax\eta$ &    $F_h$ & $\mathcal{{R}}$ \\
	\midrule
	  640 &        1.07 & 1.38e+01 &           63655 \\
	  640 &        1.09 & 5.37e+00 &            8858 \\
	 1280 &        1.29 & 5.10e+00 &           17761 \\
	  640 &        1.07 & 3.08e+00 &            3090 \\
	  640 &        1.09 & 1.66e+00 &             915 \\
	 1280 &        1.30 & 1.63e+00 &            1834 \\
	  640 &        1.10 & 1.08e+00 &             389 \\
	 1280 &        1.30 & 1.01e+00 &             772 \\
	  640 &        1.12 & 7.83e-01 &             204 \\
	 1280 &        1.33 & 7.64e-01 &             407 \\
	  640 &        1.13 & 5.86e-01 &             115 \\
	  640 &        1.11 & 4.58e-01 &            72.9 \\
	 1280 &        1.32 & 4.40e-01 &             144 \\
	  320 &        2.78 & 4.64e-01 &             3.5 \\
	  320 &        1.66 & 3.55e-01 &             6.8 \\
	  640 &        1.98 & 3.24e-01 &            13.4 \\
	  640 &        1.18 & 2.98e-01 &            26.6 \\
	 1280 &        1.40 & 2.87e-01 &            53.2 \\
	 1280 &        0.84 & 2.80e-01 &             106 \\
	  640 &        1.16 & 2.01e-01 &            12.3 \\
	  640 &        7.66 & 2.36e-01 &             0.4 \\
	  640 &        4.37 & 1.63e-01 &             0.9 \\
	  640 &        2.10 & 1.45e-01 &             2.7 \\
	 1280 &        2.46 & 1.39e-01 &             5.7 \\
	 1280 &        1.44 & 1.35e-01 &            11.9 \\
	 1280 &        0.85 & 1.38e-01 &            24.3 \\
	 1920 &        0.76 & 1.36e-01 &            49.6 \\
	 1280 &        1.31 & 6.12e-02 &             6.2 \\
	 1920 &        1.16 & 5.97e-02 &            12.7 \\
	 1280 &        0.46 & 6.24e-02 &            25.2 \\
	  640 &        2.48 & 3.01e-02 &             0.5 \\
	  640 &        1.42 & 2.58e-02 &             1.2 \\
	 1280 &        1.42 & 2.60e-02 &             3.2 \\
	 1280 &        0.84 & 2.38e-02 &             6.5 \\
	 2560 &        1.00 & 2.47e-02 &            13.0 \\
	 1280 &        0.45 & 6.54e-03 &             7.1 \\
	  640 &        1.37 & 4.61e-03 &             0.4 \\
	  640 &        0.80 & 2.94e-03 &             0.8 \\
	 2560 &        0.58 & 3.01e-03 &             7.1 \\
	\bottomrule
\end{tabular}
\begin{tabular}{|llll|}
	\toprule
	$n_h$ & $\kmax\eta$ &    $F_h$ & $\mathcal{{R}}$ \\
	\midrule
	  640 &        1.07 & 9.55e+00 &           62857 \\
	  640 &        1.10 & 3.40e+00 &            8472 \\
	 1280 &        1.31 & 3.19e+00 &           17019 \\
	  640 &        1.10 & 2.02e+00 &            2791 \\
	  640 &        1.16 & 1.02e+00 &             719 \\
	 1280 &        1.38 & 9.31e-01 &            1425 \\
	  640 &        1.18 & 7.07e-01 &             286 \\
	 1280 &        1.42 & 6.05e-01 &             544 \\
	  640 &        1.21 & 5.66e-01 &             151 \\
	 1280 &        1.47 & 4.43e-01 &             275 \\
	  640 &        1.21 & 4.34e-01 &            85.6 \\
	  640 &        1.18 & 3.76e-01 &            56.4 \\
	 1280 &        1.43 & 3.18e-01 &             104 \\
	  320 &        2.83 & 5.63e-01 &             3.2 \\
	  320 &        1.70 & 4.20e-01 &             6.2 \\
	  320 &        1.03 & 3.40e-01 &            11.7 \\
	  640 &        1.24 & 2.87e-01 &            22.2 \\
	  640 &        0.75 & 2.40e-01 &            41.0 \\
	 1280 &        0.91 & 2.04e-01 &            75.5 \\
	  640 &        1.22 & 2.01e-01 &            10.0 \\
	  640 &        8.00 & 2.95e-01 &             0.3 \\
	  640 &        4.73 & 2.06e-01 &             0.6 \\
	  640 &        2.25 & 1.73e-01 &             2.0 \\
	 1280 &        2.57 & 1.20e-01 &             4.7 \\
	 1280 &        1.55 & 8.75e-02 &             9.0 \\
	 1280 &        0.91 & 8.89e-02 &            18.7 \\
	 1280 &        0.55 & 7.53e-02 &            35.9 \\
	 1280 &        1.40 & 3.58e-02 &             4.7 \\
	 1280 &        0.83 & 3.14e-02 &             9.7 \\
	 1280 &        0.49 & 6.24e-02 &            19.7 \\
	  640 &        2.70 & 4.64e-02 &             0.4 \\
	  640 &        1.52 & 2.34e-02 &             0.9 \\
	 1280 &        1.50 & 3.83e-02 &             2.6 \\
	 1280 &        0.88 & 3.45e-02 &             5.3 \\
	 2560 &        1.06 & 3.52e-02 &            10.4 \\
	 1280 &        0.48 & 1.06e-02 &             5.2 \\
	  640 &        1.55 & 7.13e-03 &             0.2 \\
	  640 &        0.90 & 3.33e-03 &             0.5 \\
	 2560 &        0.63 & 8.09e-03 &             5.0 \\
	\bottomrule
	\end{tabular}
\label{table-better-simuls}
\end{table}

\newpage

\bibliography{main}

\begin{thebibliography}{74}%
\makeatletter
\providecommand \@ifxundefined [1]{%
 \@ifx{#1\undefined}
}%
\providecommand \@ifnum [1]{%
 \ifnum #1\expandafter \@firstoftwo
 \else \expandafter \@secondoftwo
 \fi
}%
\providecommand \@ifx [1]{%
 \ifx #1\expandafter \@firstoftwo
 \else \expandafter \@secondoftwo
 \fi
}%
\providecommand \natexlab [1]{#1}%
\providecommand \enquote  [1]{``#1''}%
\providecommand \bibnamefont  [1]{#1}%
\providecommand \bibfnamefont [1]{#1}%
\providecommand \citenamefont [1]{#1}%
\providecommand \href@noop [0]{\@secondoftwo}%
\providecommand \href [0]{\begingroup \@sanitize@url \@href}%
\providecommand \@href[1]{\@@startlink{#1}\@@href}%
\providecommand \@@href[1]{\endgroup#1\@@endlink}%
\providecommand \@sanitize@url [0]{\catcode `\\12\catcode `\$12\catcode
  `\&12\catcode `\#12\catcode `\^12\catcode `\_12\catcode `\%12\relax}%
\providecommand \@@startlink[1]{}%
\providecommand \@@endlink[0]{}%
\providecommand \url  [0]{\begingroup\@sanitize@url \@url }%
\providecommand \@url [1]{\endgroup\@href {#1}{\urlprefix }}%
\providecommand \urlprefix  [0]{URL }%
\providecommand \Eprint [0]{\href }%
\providecommand \doibase [0]{https://doi.org/}%
\providecommand \selectlanguage [0]{\@gobble}%
\providecommand \bibinfo  [0]{\@secondoftwo}%
\providecommand \bibfield  [0]{\@secondoftwo}%
\providecommand \translation [1]{[#1]}%
\providecommand \BibitemOpen [0]{}%
\providecommand \bibitemStop [0]{}%
\providecommand \bibitemNoStop [0]{.\EOS\space}%
\providecommand \EOS [0]{\spacefactor3000\relax}%
\providecommand \BibitemShut  [1]{\csname bibitem#1\endcsname}%
\let\auto@bib@innerbib\@empty
\bibitem [{\citenamefont {Zakharov}\ \emph {et~al.}(1992)\citenamefont
  {Zakharov}, \citenamefont {L’vov},\ and\ \citenamefont
  {Falkovich}}]{zakharov_kolmogorov_1992}%
  \BibitemOpen
  \bibfield  {author} {\bibinfo {author} {\bibfnamefont {V.~E.}\ \bibnamefont
  {Zakharov}}, \bibinfo {author} {\bibfnamefont {V.~S.}\ \bibnamefont
  {L’vov}},\ and\ \bibinfo {author} {\bibfnamefont {G.}~\bibnamefont
  {Falkovich}},\ }\href {https://doi.org/10.1007/978-3-642-50052-7} {\emph
  {\bibinfo {title} {Kolmogorov Spectra of Turbulence I}}},\ Springer Series in
  Nonlinear Dynamics\ (\bibinfo  {publisher} {Springer},\ \bibinfo {year}
  {1992})\BibitemShut {NoStop}%
\bibitem [{\citenamefont {Nazarenko}(2011)}]{nazarenko_wave_2011}%
  \BibitemOpen
  \bibfield  {author} {\bibinfo {author} {\bibfnamefont {S.}~\bibnamefont
  {Nazarenko}},\ }\href {https://doi.org/10.1007/978-3-642-15942-8} {\emph
  {\bibinfo {title} {Wave Turbulence}}},\ \bibinfo {series} {Lecture Notes in
  Physics}, Vol.\ \bibinfo {volume} {825}\ (\bibinfo  {publisher} {Springer},\
  \bibinfo {year} {2011})\BibitemShut {NoStop}%
\bibitem [{\citenamefont {Nazarenko}(2015)}]{nazarenko_wave_2015}%
  \BibitemOpen
  \bibfield  {author} {\bibinfo {author} {\bibfnamefont {S.}~\bibnamefont
  {Nazarenko}},\ }\bibfield  {title} {\bibinfo {title} {Wave turbulence},\
  }\href {https://doi.org/10.1080/00107514.2015.1015250} {\bibfield  {journal}
  {\bibinfo  {journal} {Contemporary Physics}\ }\textbf {\bibinfo {volume}
  {56}},\ \bibinfo {pages} {359} (\bibinfo {year} {2015})},\ \bibinfo {note}
  {publisher: Taylor \& Francis \_eprint:
  https://doi.org/10.1080/00107514.2015.1015250}\BibitemShut {NoStop}%
\bibitem [{\citenamefont {Falcon}\ and\ \citenamefont
  {Mordant}(2022)}]{falcon_experiments_2022}%
  \BibitemOpen
  \bibfield  {author} {\bibinfo {author} {\bibfnamefont {E.}~\bibnamefont
  {Falcon}}\ and\ \bibinfo {author} {\bibfnamefont {N.}~\bibnamefont
  {Mordant}},\ }\bibfield  {title} {\bibinfo {title} {Experiments in surface
  gravity–capillary wave turbulence},\ }\href
  {https://doi.org/10.1146/annurev-fluid-021021-102043} {\bibfield  {journal}
  {\bibinfo  {journal} {Annual Review of Fluid Mechanics}\ }\textbf {\bibinfo
  {volume} {54}},\ \bibinfo {pages} {1} (\bibinfo {year} {2022})},\ \bibinfo
  {note} {\_eprint:
  https://doi.org/10.1146/annurev-fluid-021021-102043}\BibitemShut {NoStop}%
\bibitem [{\citenamefont {Caillol}\ and\ \citenamefont
  {Zeitlin}(2000)}]{caillol_kinetic_2000}%
  \BibitemOpen
  \bibfield  {author} {\bibinfo {author} {\bibfnamefont {P.}~\bibnamefont
  {Caillol}}\ and\ \bibinfo {author} {\bibfnamefont {V.}~\bibnamefont
  {Zeitlin}},\ }\bibfield  {title} {\bibinfo {title} {Kinetic equations and
  stationary energy spectra of weakly nonlinear internal gravity waves},\
  }\href {https://doi.org/10.1016/S0377-0265(99)00043-3} {\bibfield  {journal}
  {\bibinfo  {journal} {Dynamics of Atmospheres and Oceans}\ }\textbf {\bibinfo
  {volume} {32}},\ \bibinfo {pages} {81} (\bibinfo {year} {2000})}\BibitemShut
  {NoStop}%
\bibitem [{\citenamefont {Galtier}(2003)}]{galtier_weak_2003}%
  \BibitemOpen
  \bibfield  {author} {\bibinfo {author} {\bibfnamefont {S.}~\bibnamefont
  {Galtier}},\ }\bibfield  {title} {\bibinfo {title} {Weak inertial-wave
  turbulence theory},\ }\href {https://doi.org/10.1103/PhysRevE.68.015301}
  {\bibfield  {journal} {\bibinfo  {journal} {Physical Review E}\ }\textbf
  {\bibinfo {volume} {68}},\ \bibinfo {pages} {015301} (\bibinfo {year}
  {2003})},\ \bibinfo {note} {publisher: American Physical Society}\BibitemShut
  {NoStop}%
\bibitem [{\citenamefont {Medvedev}\ and\ \citenamefont
  {Zeitlin}(2007)}]{medvedev_turbulence_2007}%
  \BibitemOpen
  \bibfield  {author} {\bibinfo {author} {\bibfnamefont {S.~B.}\ \bibnamefont
  {Medvedev}}\ and\ \bibinfo {author} {\bibfnamefont {V.}~\bibnamefont
  {Zeitlin}},\ }\bibfield  {title} {\bibinfo {title} {Turbulence of
  near-inertial waves in the continuously stratified fluid},\ }\href
  {https://doi.org/10.1016/j.physleta.2007.08.014} {\bibfield  {journal}
  {\bibinfo  {journal} {Physics Letters A}\ }\textbf {\bibinfo {volume}
  {371}},\ \bibinfo {pages} {221} (\bibinfo {year} {2007})}\BibitemShut
  {NoStop}%
\bibitem [{\citenamefont {Griffin}\ \emph {et~al.}(2022)\citenamefont
  {Griffin}, \citenamefont {Krstulovic}, \citenamefont {L’vov},\ and\
  \citenamefont {Nazarenko}}]{griffin_energy_2022}%
  \BibitemOpen
  \bibfield  {author} {\bibinfo {author} {\bibfnamefont {A.}~\bibnamefont
  {Griffin}}, \bibinfo {author} {\bibfnamefont {G.}~\bibnamefont {Krstulovic}},
  \bibinfo {author} {\bibfnamefont {V.~S.}\ \bibnamefont {L’vov}},\ and\
  \bibinfo {author} {\bibfnamefont {S.}~\bibnamefont {Nazarenko}},\ }\bibfield
  {title} {\bibinfo {title} {Energy spectrum of two-dimensional acoustic
  turbulence},\ }\href {https://doi.org/10.1103/PhysRevLett.128.224501}
  {\bibfield  {journal} {\bibinfo  {journal} {Physical Review Letters}\
  }\textbf {\bibinfo {volume} {128}},\ \bibinfo {pages} {224501} (\bibinfo
  {year} {2022})},\ \bibinfo {note} {publisher: American Physical
  Society}\BibitemShut {NoStop}%
\bibitem [{\citenamefont {Düring}\ \emph {et~al.}(2006)\citenamefont
  {Düring}, \citenamefont {Josserand},\ and\ \citenamefont
  {Rica}}]{during_weak_2006}%
  \BibitemOpen
  \bibfield  {author} {\bibinfo {author} {\bibfnamefont {G.}~\bibnamefont
  {Düring}}, \bibinfo {author} {\bibfnamefont {C.}~\bibnamefont {Josserand}},\
  and\ \bibinfo {author} {\bibfnamefont {S.}~\bibnamefont {Rica}},\ }\bibfield
  {title} {\bibinfo {title} {Weak turbulence for a vibrating plate: Can one
  hear a kolmogorov spectrum?},\ }\href
  {https://doi.org/10.1103/PhysRevLett.97.025503} {\bibfield  {journal}
  {\bibinfo  {journal} {Physical Review Letters}\ }\textbf {\bibinfo {volume}
  {97}},\ \bibinfo {pages} {025503} (\bibinfo {year} {2006})},\ \bibinfo {note}
  {publisher: American Physical Society}\BibitemShut {NoStop}%
\bibitem [{\citenamefont {Galtier}\ \emph {et~al.}(2000)\citenamefont
  {Galtier}, \citenamefont {Nazarenko}, \citenamefont {Newell},\ and\
  \citenamefont {Pouquet}}]{galtier_weak_2000}%
  \BibitemOpen
  \bibfield  {author} {\bibinfo {author} {\bibfnamefont {S.}~\bibnamefont
  {Galtier}}, \bibinfo {author} {\bibfnamefont {S.~V.}\ \bibnamefont
  {Nazarenko}}, \bibinfo {author} {\bibfnamefont {A.~C.}\ \bibnamefont
  {Newell}},\ and\ \bibinfo {author} {\bibfnamefont {A.}~\bibnamefont
  {Pouquet}},\ }\bibfield  {title} {\bibinfo {title} {A weak turbulence theory
  for incompressible magnetohydrodynamics},\ }\href
  {https://doi.org/10.1017/S0022377899008284} {\bibfield  {journal} {\bibinfo
  {journal} {Journal of Plasma Physics}\ }\textbf {\bibinfo {volume} {63}},\
  \bibinfo {pages} {447} (\bibinfo {year} {2000})},\ \bibinfo {note}
  {publisher: Cambridge University Press}\BibitemShut {NoStop}%
\bibitem [{\citenamefont {L’vov}\ and\ \citenamefont
  {Nazarenko}(2010)}]{lvov_weak_2010}%
  \BibitemOpen
  \bibfield  {author} {\bibinfo {author} {\bibfnamefont {V.~S.}\ \bibnamefont
  {L’vov}}\ and\ \bibinfo {author} {\bibfnamefont {S.}~\bibnamefont
  {Nazarenko}},\ }\bibfield  {title} {\bibinfo {title} {Weak turbulence of
  kelvin waves in superfluid he},\ }\href {https://doi.org/10.1063/1.3499242}
  {\bibfield  {journal} {\bibinfo  {journal} {Low Temperature Physics}\
  }\textbf {\bibinfo {volume} {36}},\ \bibinfo {pages} {785} (\bibinfo {year}
  {2010})},\ \bibinfo {note} {publisher: American Institute of
  Physics}\BibitemShut {NoStop}%
\bibitem [{\citenamefont {Dyachenko}\ \emph {et~al.}(1992)\citenamefont
  {Dyachenko}, \citenamefont {Newell}, \citenamefont {Pushkarev},\ and\
  \citenamefont {Zakharov}}]{dyachenko_optical_1992}%
  \BibitemOpen
  \bibfield  {author} {\bibinfo {author} {\bibfnamefont {S.}~\bibnamefont
  {Dyachenko}}, \bibinfo {author} {\bibfnamefont {A.~C.}\ \bibnamefont
  {Newell}}, \bibinfo {author} {\bibfnamefont {A.}~\bibnamefont {Pushkarev}},\
  and\ \bibinfo {author} {\bibfnamefont {V.~E.}\ \bibnamefont {Zakharov}},\
  }\bibfield  {title} {\bibinfo {title} {Optical turbulence: weak turbulence,
  condensates and collapsing filaments in the nonlinear schrödinger
  equation},\ }\href {https://doi.org/10.1016/0167-2789(92)90090-A} {\bibfield
  {journal} {\bibinfo  {journal} {Physica D: Nonlinear Phenomena}\ }\textbf
  {\bibinfo {volume} {57}},\ \bibinfo {pages} {96} (\bibinfo {year}
  {1992})}\BibitemShut {NoStop}%
\bibitem [{\citenamefont {Galtier}\ and\ \citenamefont
  {Nazarenko}(2017)}]{galtier_turbulence_2017}%
  \BibitemOpen
  \bibfield  {author} {\bibinfo {author} {\bibfnamefont {S.}~\bibnamefont
  {Galtier}}\ and\ \bibinfo {author} {\bibfnamefont {S.~V.}\ \bibnamefont
  {Nazarenko}},\ }\bibfield  {title} {\bibinfo {title} {Turbulence of weak
  gravitational waves in the early universe},\ }\href
  {https://doi.org/10.1103/PhysRevLett.119.221101} {\bibfield  {journal}
  {\bibinfo  {journal} {Physical Review Letters}\ }\textbf {\bibinfo {volume}
  {119}},\ \bibinfo {pages} {221101} (\bibinfo {year} {2017})},\ \bibinfo
  {note} {publisher: American Physical Society}\BibitemShut {NoStop}%
\bibitem [{\citenamefont {L'vov}\ and\ \citenamefont
  {Nazarenko}(2010)}]{lvov_discrete_2010}%
  \BibitemOpen
  \bibfield  {author} {\bibinfo {author} {\bibfnamefont {V.~S.}\ \bibnamefont
  {L'vov}}\ and\ \bibinfo {author} {\bibfnamefont {S.}~\bibnamefont
  {Nazarenko}},\ }\bibfield  {title} {\bibinfo {title} {Discrete and mesoscopic
  regimes of finite-size wave turbulence},\ }\href
  {https://doi.org/10.1103/PhysRevE.82.056322} {\bibfield  {journal} {\bibinfo
  {journal} {Phys. Rev. E}\ }\textbf {\bibinfo {volume} {82}},\ \bibinfo
  {pages} {056322} (\bibinfo {year} {2010})}\BibitemShut {NoStop}%
\bibitem [{\citenamefont {Biven}\ \emph {et~al.}(2001)\citenamefont {Biven},
  \citenamefont {Nazarenko},\ and\ \citenamefont
  {Newell}}]{biven_breakdown_2001}%
  \BibitemOpen
  \bibfield  {author} {\bibinfo {author} {\bibfnamefont {L.}~\bibnamefont
  {Biven}}, \bibinfo {author} {\bibfnamefont {S.~V.}\ \bibnamefont
  {Nazarenko}},\ and\ \bibinfo {author} {\bibfnamefont {A.~C.}\ \bibnamefont
  {Newell}},\ }\bibfield  {title} {\bibinfo {title} {Breakdown of wave
  turbulence and the onset of intermittency},\ }\href
  {https://doi.org/10.1016/S0375-9601(01)00016-0} {\bibfield  {journal}
  {\bibinfo  {journal} {Physics Letters A}\ }\textbf {\bibinfo {volume}
  {280}},\ \bibinfo {pages} {28} (\bibinfo {year} {2001})}\BibitemShut
  {NoStop}%
\bibitem [{\citenamefont {Miquel}\ and\ \citenamefont
  {Mordant}(2011)}]{miquel_nonstationary_2011}%
  \BibitemOpen
  \bibfield  {author} {\bibinfo {author} {\bibfnamefont {B.}~\bibnamefont
  {Miquel}}\ and\ \bibinfo {author} {\bibfnamefont {N.}~\bibnamefont
  {Mordant}},\ }\bibfield  {title} {\bibinfo {title} {Nonstationary wave
  turbulence in an elastic plate},\ }\href
  {https://doi.org/10.1103/PhysRevLett.107.034501} {\bibfield  {journal}
  {\bibinfo  {journal} {Physical Review Letters}\ }\textbf {\bibinfo {volume}
  {107}},\ \bibinfo {pages} {034501} (\bibinfo {year} {2011})},\ \bibinfo
  {note} {publisher: American Physical Society}\BibitemShut {NoStop}%
\bibitem [{\citenamefont {Yokoyama}\ and\ \citenamefont
  {Takaoka}(2014)}]{yokoyama_identification_2014}%
  \BibitemOpen
  \bibfield  {author} {\bibinfo {author} {\bibfnamefont {N.}~\bibnamefont
  {Yokoyama}}\ and\ \bibinfo {author} {\bibfnamefont {M.}~\bibnamefont
  {Takaoka}},\ }\bibfield  {title} {\bibinfo {title} {Identification of a
  separation wave number between weak and strong turbulence spectra for a
  vibrating plate},\ }\href {https://doi.org/10.1103/PhysRevE.89.012909}
  {\bibfield  {journal} {\bibinfo  {journal} {Physical Review E}\ }\textbf
  {\bibinfo {volume} {89}},\ \bibinfo {pages} {012909} (\bibinfo {year}
  {2014})},\ \bibinfo {note} {publisher: American Physical Society}\BibitemShut
  {NoStop}%
\bibitem [{\citenamefont {Pan}\ and\ \citenamefont
  {Yue}(2014)}]{pan_direct_2014}%
  \BibitemOpen
  \bibfield  {author} {\bibinfo {author} {\bibfnamefont {Y.}~\bibnamefont
  {Pan}}\ and\ \bibinfo {author} {\bibfnamefont {D.~K.}\ \bibnamefont {Yue}},\
  }\bibfield  {title} {\bibinfo {title} {Direct numerical investigation of
  turbulence of capillary waves},\ }\href
  {https://doi.org/10.1103/PhysRevLett.113.094501} {\bibfield  {journal}
  {\bibinfo  {journal} {Physical Review Letters}\ }\textbf {\bibinfo {volume}
  {113}},\ \bibinfo {pages} {094501} (\bibinfo {year} {2014})},\ \bibinfo
  {note} {publisher: American Physical Society}\BibitemShut {NoStop}%
\bibitem [{\citenamefont {Zhu}\ \emph {et~al.}(2022)\citenamefont {Zhu},
  \citenamefont {Semisalov}, \citenamefont {Krstulovic},\ and\ \citenamefont
  {Nazarenko}}]{zhu_testing_2022}%
  \BibitemOpen
  \bibfield  {author} {\bibinfo {author} {\bibfnamefont {Y.}~\bibnamefont
  {Zhu}}, \bibinfo {author} {\bibfnamefont {B.}~\bibnamefont {Semisalov}},
  \bibinfo {author} {\bibfnamefont {G.}~\bibnamefont {Krstulovic}},\ and\
  \bibinfo {author} {\bibfnamefont {S.}~\bibnamefont {Nazarenko}},\ }\bibfield
  {title} {\bibinfo {title} {Testing wave turbulence theory for the
  gross-pitaevskii system},\ }\href
  {https://doi.org/10.1103/PhysRevE.106.014205} {\bibfield  {journal} {\bibinfo
   {journal} {Physical Review E}\ }\textbf {\bibinfo {volume} {106}},\ \bibinfo
  {pages} {014205} (\bibinfo {year} {2022})},\ \bibinfo {note} {publisher:
  American Physical Society}\BibitemShut {NoStop}%
\bibitem [{\citenamefont {Yokoyama}\ and\ \citenamefont
  {Takaoka}(2019)}]{yokoyama_energy-based_2019}%
  \BibitemOpen
  \bibfield  {author} {\bibinfo {author} {\bibfnamefont {N.}~\bibnamefont
  {Yokoyama}}\ and\ \bibinfo {author} {\bibfnamefont {M.}~\bibnamefont
  {Takaoka}},\ }\bibfield  {title} {\bibinfo {title} {Energy-based analysis and
  anisotropic spectral distribution of internal gravity waves in strongly
  stratified turbulence},\ }\href
  {https://doi.org/10.1103/PhysRevFluids.4.104602} {\bibfield  {journal}
  {\bibinfo  {journal} {Physical Review Fluids}\ }\textbf {\bibinfo {volume}
  {4}},\ \bibinfo {pages} {104602} (\bibinfo {year} {2019})},\ \bibinfo {note}
  {publisher: American Physical Society}\BibitemShut {NoStop}%
\bibitem [{\citenamefont {Yokoyama}\ and\ \citenamefont
  {Takaoka}(2021)}]{yokoyama_energy-flux_2021}%
  \BibitemOpen
  \bibfield  {author} {\bibinfo {author} {\bibfnamefont {N.}~\bibnamefont
  {Yokoyama}}\ and\ \bibinfo {author} {\bibfnamefont {M.}~\bibnamefont
  {Takaoka}},\ }\bibfield  {title} {\bibinfo {title} {Energy-flux vector in
  anisotropic turbulence: application to rotating turbulence},\ }\href
  {https://doi.org/10.1017/jfm.2020.860} {\bibfield  {journal} {\bibinfo
  {journal} {Journal of Fluid Mechanics}\ }\textbf {\bibinfo {volume} {908}},\
  \bibinfo {pages} {A17} (\bibinfo {year} {2021})},\ \bibinfo {note}
  {publisher: Cambridge University Press}\BibitemShut {NoStop}%
\bibitem [{\citenamefont {Waite}(2011)}]{waite_stratified_2011}%
  \BibitemOpen
  \bibfield  {author} {\bibinfo {author} {\bibfnamefont {M.~L.}\ \bibnamefont
  {Waite}},\ }\bibfield  {title} {\bibinfo {title} {Stratified turbulence at
  the buoyancy scale},\ }\href {https://doi.org/10.1063/1.3599699} {\bibfield
  {journal} {\bibinfo  {journal} {Physics of Fluids}\ }\textbf {\bibinfo
  {volume} {23}},\ \bibinfo {pages} {066602} (\bibinfo {year} {2011})},\
  \bibinfo {note} {publisher: American Institute of Physics}\BibitemShut
  {NoStop}%
\bibitem [{\citenamefont {Staquet}\ and\ \citenamefont
  {Sommeria}(2002)}]{staquet_internal_2002}%
  \BibitemOpen
  \bibfield  {author} {\bibinfo {author} {\bibfnamefont {C.}~\bibnamefont
  {Staquet}}\ and\ \bibinfo {author} {\bibfnamefont {J.}~\bibnamefont
  {Sommeria}},\ }\bibfield  {title} {\bibinfo {title} {{INTERNAL} {GRAVITY}
  {WAVES}: From instabilities to turbulence},\ }\href
  {https://doi.org/10.1146/annurev.fluid.34.090601.130953} {\bibfield
  {journal} {\bibinfo  {journal} {Annual Review of Fluid Mechanics}\ }\textbf
  {\bibinfo {volume} {34}},\ \bibinfo {pages} {559} (\bibinfo {year} {2002})},\
  \bibinfo {note} {\_eprint:
  https://doi.org/10.1146/annurev.fluid.34.090601.130953}\BibitemShut {NoStop}%
\bibitem [{\citenamefont {Vallis}(2017)}]{vallis_atmospheric_2017}%
  \BibitemOpen
  \bibfield  {author} {\bibinfo {author} {\bibfnamefont {G.~K.}\ \bibnamefont
  {Vallis}},\ }\href {https://doi.org/10.1017/9781107588417} {\emph {\bibinfo
  {title} {Atmospheric and Oceanic Fluid Dynamics: Fundamentals and Large-Scale
  Circulation}}},\ \bibinfo {edition} {2nd}\ ed.\ (\bibinfo  {publisher}
  {Cambridge University Press},\ \bibinfo {year} {2017})\BibitemShut {NoStop}%
\bibitem [{\citenamefont {{MacKinnon}}\ \emph {et~al.}(2017)\citenamefont
  {{MacKinnon}}, \citenamefont {Zhao}, \citenamefont {Whalen}, \citenamefont
  {Waterhouse}, \citenamefont {Trossman}, \citenamefont {Sun}, \citenamefont
  {Laurent}, \citenamefont {Simmons}, \citenamefont {Polzin}, \citenamefont
  {Pinkel}, \citenamefont {Pickering}, \citenamefont {Norton}, \citenamefont
  {Nash}, \citenamefont {Musgrave}, \citenamefont {Merchant}, \citenamefont
  {Melet}, \citenamefont {Mater}, \citenamefont {Legg}, \citenamefont {Large},
  \citenamefont {Kunze}, \citenamefont {Klymak}, \citenamefont {Jochum},
  \citenamefont {Jayne}, \citenamefont {Hallberg}, \citenamefont {Griffies},
  \citenamefont {Diggs}, \citenamefont {Danabasoglu}, \citenamefont
  {Chassignet}, \citenamefont {Buijsman}, \citenamefont {Bryan}, \citenamefont
  {Briegleb}, \citenamefont {Barna}, \citenamefont {Arbic}, \citenamefont
  {Ansong},\ and\ \citenamefont {Alford}}]{mackinnon_climate_2017}%
  \BibitemOpen
  \bibfield  {author} {\bibinfo {author} {\bibfnamefont {J.~A.}\ \bibnamefont
  {{MacKinnon}}}, \bibinfo {author} {\bibfnamefont {Z.}~\bibnamefont {Zhao}},
  \bibinfo {author} {\bibfnamefont {C.~B.}\ \bibnamefont {Whalen}}, \bibinfo
  {author} {\bibfnamefont {A.~F.}\ \bibnamefont {Waterhouse}}, \bibinfo
  {author} {\bibfnamefont {D.~S.}\ \bibnamefont {Trossman}}, \bibinfo {author}
  {\bibfnamefont {O.~M.}\ \bibnamefont {Sun}}, \bibinfo {author} {\bibfnamefont
  {L.~C.~S.}\ \bibnamefont {Laurent}}, \bibinfo {author} {\bibfnamefont
  {H.~L.}\ \bibnamefont {Simmons}}, \bibinfo {author} {\bibfnamefont
  {K.}~\bibnamefont {Polzin}}, \bibinfo {author} {\bibfnamefont
  {R.}~\bibnamefont {Pinkel}}, \bibinfo {author} {\bibfnamefont
  {A.}~\bibnamefont {Pickering}}, \bibinfo {author} {\bibfnamefont {N.~J.}\
  \bibnamefont {Norton}}, \bibinfo {author} {\bibfnamefont {J.~D.}\
  \bibnamefont {Nash}}, \bibinfo {author} {\bibfnamefont {R.}~\bibnamefont
  {Musgrave}}, \bibinfo {author} {\bibfnamefont {L.~M.}\ \bibnamefont
  {Merchant}}, \bibinfo {author} {\bibfnamefont {A.~V.}\ \bibnamefont {Melet}},
  \bibinfo {author} {\bibfnamefont {B.}~\bibnamefont {Mater}}, \bibinfo
  {author} {\bibfnamefont {S.}~\bibnamefont {Legg}}, \bibinfo {author}
  {\bibfnamefont {W.~G.}\ \bibnamefont {Large}}, \bibinfo {author}
  {\bibfnamefont {E.}~\bibnamefont {Kunze}}, \bibinfo {author} {\bibfnamefont
  {J.~M.}\ \bibnamefont {Klymak}}, \bibinfo {author} {\bibfnamefont
  {M.}~\bibnamefont {Jochum}}, \bibinfo {author} {\bibfnamefont {S.~R.}\
  \bibnamefont {Jayne}}, \bibinfo {author} {\bibfnamefont {R.~W.}\ \bibnamefont
  {Hallberg}}, \bibinfo {author} {\bibfnamefont {S.~M.}\ \bibnamefont
  {Griffies}}, \bibinfo {author} {\bibfnamefont {S.}~\bibnamefont {Diggs}},
  \bibinfo {author} {\bibfnamefont {G.}~\bibnamefont {Danabasoglu}}, \bibinfo
  {author} {\bibfnamefont {E.~P.}\ \bibnamefont {Chassignet}}, \bibinfo
  {author} {\bibfnamefont {M.~C.}\ \bibnamefont {Buijsman}}, \bibinfo {author}
  {\bibfnamefont {F.~O.}\ \bibnamefont {Bryan}}, \bibinfo {author}
  {\bibfnamefont {B.~P.}\ \bibnamefont {Briegleb}}, \bibinfo {author}
  {\bibfnamefont {A.}~\bibnamefont {Barna}}, \bibinfo {author} {\bibfnamefont
  {B.~K.}\ \bibnamefont {Arbic}}, \bibinfo {author} {\bibfnamefont {J.~K.}\
  \bibnamefont {Ansong}},\ and\ \bibinfo {author} {\bibfnamefont {M.~H.}\
  \bibnamefont {Alford}},\ }\bibfield  {title} {\bibinfo {title} {Climate
  process team on internal wave–driven ocean mixing},\ }\href
  {https://doi.org/10.1175/BAMS-D-16-0030.1} {\bibfield  {journal} {\bibinfo
  {journal} {Bulletin of the American Meteorological Society}\ }\textbf
  {\bibinfo {volume} {98}},\ \bibinfo {pages} {2429} (\bibinfo {year}
  {2017})},\ \bibinfo {note} {publisher: American Meteorological Society
  Section: Bulletin of the American Meteorological Society}\BibitemShut
  {NoStop}%
\bibitem [{\citenamefont {Gregg}\ \emph {et~al.}(2018)\citenamefont {Gregg},
  \citenamefont {D'Asaro}, \citenamefont {Riley},\ and\ \citenamefont
  {Kunze}}]{gregg_mixing_2018}%
  \BibitemOpen
  \bibfield  {author} {\bibinfo {author} {\bibfnamefont {M.}~\bibnamefont
  {Gregg}}, \bibinfo {author} {\bibfnamefont {E.}~\bibnamefont {D'Asaro}},
  \bibinfo {author} {\bibfnamefont {J.}~\bibnamefont {Riley}},\ and\ \bibinfo
  {author} {\bibfnamefont {E.}~\bibnamefont {Kunze}},\ }\bibfield  {title}
  {\bibinfo {title} {Mixing efficiency in the ocean},\ }\href
  {https://doi.org/10.1146/annurev-marine-121916-063643} {\bibfield  {journal}
  {\bibinfo  {journal} {Annual Review of Marine Science}\ }\textbf {\bibinfo
  {volume} {10}},\ \bibinfo {pages} {443} (\bibinfo {year} {2018})},\ \bibinfo
  {note} {\_eprint:
  https://doi.org/10.1146/annurev-marine-121916-063643}\BibitemShut {NoStop}%
\bibitem [{\citenamefont {Billant}\ and\ \citenamefont
  {Chomaz}(2001)}]{billant_self-similarity_2001}%
  \BibitemOpen
  \bibfield  {author} {\bibinfo {author} {\bibfnamefont {P.}~\bibnamefont
  {Billant}}\ and\ \bibinfo {author} {\bibfnamefont {J.-M.}\ \bibnamefont
  {Chomaz}},\ }\bibfield  {title} {\bibinfo {title} {Self-similarity of
  strongly stratified inviscid flows},\ }\href
  {https://doi.org/10.1063/1.1369125} {\bibfield  {journal} {\bibinfo
  {journal} {Physics of Fluids}\ }\textbf {\bibinfo {volume} {13}},\ \bibinfo
  {pages} {1645} (\bibinfo {year} {2001})},\ \bibinfo {note} {publisher:
  American Institute of Physics}\BibitemShut {NoStop}%
\bibitem [{\citenamefont {Waite}\ and\ \citenamefont
  {Bartello}(2004)}]{waite_stratified_2004}%
  \BibitemOpen
  \bibfield  {author} {\bibinfo {author} {\bibfnamefont {M.~L.}\ \bibnamefont
  {Waite}}\ and\ \bibinfo {author} {\bibfnamefont {P.}~\bibnamefont
  {Bartello}},\ }\bibfield  {title} {\bibinfo {title} {Stratified turbulence
  dominated by vortical motion},\ }\href
  {https://doi.org/10.1017/S0022112004000977} {\bibfield  {journal} {\bibinfo
  {journal} {Journal of Fluid Mechanics}\ }\textbf {\bibinfo {volume} {517}},\
  \bibinfo {pages} {281} (\bibinfo {year} {2004})},\ \bibinfo {note}
  {publisher: Cambridge University Press}\BibitemShut {NoStop}%
\bibitem [{\citenamefont {Waite}\ and\ \citenamefont
  {Bartello}(2006)}]{waite_stratified_2006}%
  \BibitemOpen
  \bibfield  {author} {\bibinfo {author} {\bibfnamefont {M.~L.}\ \bibnamefont
  {Waite}}\ and\ \bibinfo {author} {\bibfnamefont {P.}~\bibnamefont
  {Bartello}},\ }\bibfield  {title} {\bibinfo {title} {Stratified turbulence
  generated by internal gravity waves},\ }\href
  {https://doi.org/10.1017/S0022112005007111} {\bibfield  {journal} {\bibinfo
  {journal} {Journal of Fluid Mechanics}\ }\textbf {\bibinfo {volume} {546}},\
  \bibinfo {pages} {313} (\bibinfo {year} {2006})},\ \bibinfo {note}
  {publisher: Cambridge University Press}\BibitemShut {NoStop}%
\bibitem [{\citenamefont {Lindborg}(2006)}]{lindborg_energy_2006}%
  \BibitemOpen
  \bibfield  {author} {\bibinfo {author} {\bibfnamefont {E.}~\bibnamefont
  {Lindborg}},\ }\bibfield  {title} {\bibinfo {title} {The energy cascade in a
  strongly stratified fluid},\ }\href
  {https://doi.org/10.1017/S0022112005008128} {\bibfield  {journal} {\bibinfo
  {journal} {Journal of Fluid Mechanics}\ }\textbf {\bibinfo {volume} {550}},\
  \bibinfo {pages} {207} (\bibinfo {year} {2006})},\ \bibinfo {note}
  {publisher: Cambridge University Press}\BibitemShut {NoStop}%
\bibitem [{\citenamefont {Brethouwer}\ \emph {et~al.}(2007)\citenamefont
  {Brethouwer}, \citenamefont {Billant}, \citenamefont {Lindborg},\ and\
  \citenamefont {Chomaz}}]{brethouwer_scaling_2007}%
  \BibitemOpen
  \bibfield  {author} {\bibinfo {author} {\bibfnamefont {G.}~\bibnamefont
  {Brethouwer}}, \bibinfo {author} {\bibfnamefont {P.}~\bibnamefont {Billant}},
  \bibinfo {author} {\bibfnamefont {E.}~\bibnamefont {Lindborg}},\ and\
  \bibinfo {author} {\bibfnamefont {J.-M.}\ \bibnamefont {Chomaz}},\ }\bibfield
   {title} {\bibinfo {title} {Scaling analysis and simulation of strongly
  stratified turbulent flows},\ }\href
  {https://doi.org/10.1017/S0022112007006854} {\bibfield  {journal} {\bibinfo
  {journal} {Journal of Fluid Mechanics}\ }\textbf {\bibinfo {volume} {585}},\
  \bibinfo {pages} {343} (\bibinfo {year} {2007})},\ \bibinfo {note}
  {publisher: Cambridge University Press}\BibitemShut {NoStop}%
\bibitem [{\citenamefont {Kimura}\ and\ \citenamefont
  {Herring}(2012)}]{kimura_energy_2012}%
  \BibitemOpen
  \bibfield  {author} {\bibinfo {author} {\bibfnamefont {Y.}~\bibnamefont
  {Kimura}}\ and\ \bibinfo {author} {\bibfnamefont {J.~R.}\ \bibnamefont
  {Herring}},\ }\bibfield  {title} {\bibinfo {title} {Energy spectra of stably
  stratified turbulence},\ }\href {https://doi.org/10.1017/jfm.2011.546}
  {\bibfield  {journal} {\bibinfo  {journal} {Journal of Fluid Mechanics}\
  }\textbf {\bibinfo {volume} {698}},\ \bibinfo {pages} {19} (\bibinfo {year}
  {2012})},\ \bibinfo {note} {publisher: Cambridge University
  Press}\BibitemShut {NoStop}%
\bibitem [{\citenamefont {Bartello}\ and\ \citenamefont
  {Tobias}(2013)}]{bartello_sensitivity_2013}%
  \BibitemOpen
  \bibfield  {author} {\bibinfo {author} {\bibfnamefont {P.}~\bibnamefont
  {Bartello}}\ and\ \bibinfo {author} {\bibfnamefont {S.~M.}\ \bibnamefont
  {Tobias}},\ }\bibfield  {title} {\bibinfo {title} {Sensitivity of stratified
  turbulence to the buoyancy reynolds number},\ }\href
  {https://doi.org/10.1017/jfm.2013.170} {\bibfield  {journal} {\bibinfo
  {journal} {Journal of Fluid Mechanics}\ }\textbf {\bibinfo {volume} {725}},\
  \bibinfo {pages} {1} (\bibinfo {year} {2013})},\ \bibinfo {note} {publisher:
  Cambridge University Press}\BibitemShut {NoStop}%
\bibitem [{\citenamefont {Brunner-Suzuki}\ \emph {et~al.}(2014)\citenamefont
  {Brunner-Suzuki}, \citenamefont {Sundermeyer},\ and\ \citenamefont
  {Lelong}}]{brunner-suzuki_upscale_2014}%
  \BibitemOpen
  \bibfield  {author} {\bibinfo {author} {\bibfnamefont {A.-M. E.~G.}\
  \bibnamefont {Brunner-Suzuki}}, \bibinfo {author} {\bibfnamefont {M.~A.}\
  \bibnamefont {Sundermeyer}},\ and\ \bibinfo {author} {\bibfnamefont {M.-P.}\
  \bibnamefont {Lelong}},\ }\bibfield  {title} {\bibinfo {title} {Upscale
  energy transfer by the vortical mode and internal waves},\ }\href
  {https://doi.org/10.1175/JPO-D-12-0149.1} {\bibfield  {journal} {\bibinfo
  {journal} {Journal of Physical Oceanography}\ }\textbf {\bibinfo {volume}
  {44}},\ \bibinfo {pages} {2446} (\bibinfo {year} {2014})},\ \bibinfo {note}
  {publisher: American Meteorological Society Section: Journal of Physical
  Oceanography}\BibitemShut {NoStop}%
\bibitem [{\citenamefont {Augier}\ \emph {et~al.}(2015)\citenamefont {Augier},
  \citenamefont {Billant},\ and\ \citenamefont
  {Chomaz}}]{augier_stratified_2015}%
  \BibitemOpen
  \bibfield  {author} {\bibinfo {author} {\bibfnamefont {P.}~\bibnamefont
  {Augier}}, \bibinfo {author} {\bibfnamefont {P.}~\bibnamefont {Billant}},\
  and\ \bibinfo {author} {\bibfnamefont {J.-M.}\ \bibnamefont {Chomaz}},\
  }\bibfield  {title} {\bibinfo {title} {Stratified turbulence forced with
  columnar dipoles: numerical study},\ }\href
  {https://doi.org/10.1017/jfm.2015.76} {\bibfield  {journal} {\bibinfo
  {journal} {Journal of Fluid Mechanics}\ }\textbf {\bibinfo {volume} {769}},\
  \bibinfo {pages} {403} (\bibinfo {year} {2015})},\ \bibinfo {note}
  {publisher: Cambridge University Press}\BibitemShut {NoStop}%
\bibitem [{\citenamefont {Maffioli}(2017)}]{maffioli_vertical_2017}%
  \BibitemOpen
  \bibfield  {author} {\bibinfo {author} {\bibfnamefont {A.}~\bibnamefont
  {Maffioli}},\ }\bibfield  {title} {\bibinfo {title} {Vertical spectra of
  stratified turbulence at large horizontal scales},\ }\href
  {https://doi.org/10.1103/PhysRevFluids.2.104802} {\bibfield  {journal}
  {\bibinfo  {journal} {Physical Review Fluids}\ }\textbf {\bibinfo {volume}
  {2}},\ \bibinfo {pages} {104802} (\bibinfo {year} {2017})},\ \bibinfo {note}
  {publisher: American Physical Society}\BibitemShut {NoStop}%
\bibitem [{\citenamefont {Lvov}\ and\ \citenamefont
  {Tabak}(2001)}]{lvov_hamiltonian_2001}%
  \BibitemOpen
  \bibfield  {author} {\bibinfo {author} {\bibfnamefont {Y.~V.}\ \bibnamefont
  {Lvov}}\ and\ \bibinfo {author} {\bibfnamefont {E.~G.}\ \bibnamefont
  {Tabak}},\ }\bibfield  {title} {\bibinfo {title} {Hamiltonian formalism and
  the garrett-munk spectrum of internal waves in the ocean},\ }\href
  {https://doi.org/10.1103/PhysRevLett.87.168501} {\bibfield  {journal}
  {\bibinfo  {journal} {Physical Review Letters}\ }\textbf {\bibinfo {volume}
  {87}},\ \bibinfo {pages} {168501} (\bibinfo {year} {2001})},\ \bibinfo {note}
  {publisher: American Physical Society}\BibitemShut {NoStop}%
\bibitem [{\citenamefont {Dematteis}\ and\ \citenamefont
  {Lvov}(2021)}]{dematteis_downscale_2021}%
  \BibitemOpen
  \bibfield  {author} {\bibinfo {author} {\bibfnamefont {G.}~\bibnamefont
  {Dematteis}}\ and\ \bibinfo {author} {\bibfnamefont {Y.~V.}\ \bibnamefont
  {Lvov}},\ }\bibfield  {title} {\bibinfo {title} {Downscale energy fluxes in
  scale-invariant oceanic internal wave turbulence},\ }\href
  {https://doi.org/10.1017/jfm.2021.99} {\bibfield  {journal} {\bibinfo
  {journal} {Journal of Fluid Mechanics}\ }\textbf {\bibinfo {volume} {915}},\
  \bibinfo {pages} {A129} (\bibinfo {year} {2021})},\ \bibinfo {note}
  {publisher: Cambridge University Press}\BibitemShut {NoStop}%
\bibitem [{\citenamefont {Dematteis}\ \emph {et~al.}(2022)\citenamefont
  {Dematteis}, \citenamefont {Polzin},\ and\ \citenamefont
  {Lvov}}]{dematteis_origins_2022}%
  \BibitemOpen
  \bibfield  {author} {\bibinfo {author} {\bibfnamefont {G.}~\bibnamefont
  {Dematteis}}, \bibinfo {author} {\bibfnamefont {K.}~\bibnamefont {Polzin}},\
  and\ \bibinfo {author} {\bibfnamefont {Y.~V.}\ \bibnamefont {Lvov}},\
  }\bibfield  {title} {\bibinfo {title} {On the origins of the oceanic
  ultraviolet catastrophe},\ }\href {https://doi.org/10.1175/JPO-D-21-0121.1}
  {\bibfield  {journal} {\bibinfo  {journal} {Journal of Physical
  Oceanography}\ }\textbf {\bibinfo {volume} {52}},\ \bibinfo {pages} {597}
  (\bibinfo {year} {2022})},\ \bibinfo {note} {publisher: American
  Meteorological Society Section: Journal of Physical Oceanography}\BibitemShut
  {NoStop}%
\bibitem [{\citenamefont {Maffioli}\ \emph {et~al.}(2020)\citenamefont
  {Maffioli}, \citenamefont {Delache},\ and\ \citenamefont
  {Godeferd}}]{maffioli_signature_2020}%
  \BibitemOpen
  \bibfield  {author} {\bibinfo {author} {\bibfnamefont {A.}~\bibnamefont
  {Maffioli}}, \bibinfo {author} {\bibfnamefont {A.}~\bibnamefont {Delache}},\
  and\ \bibinfo {author} {\bibfnamefont {F.~S.}\ \bibnamefont {Godeferd}},\
  }\bibfield  {title} {\bibinfo {title} {Signature and energetics of internal
  gravity waves in stratified turbulence},\ }\href
  {https://doi.org/10.1103/PhysRevFluids.5.114802} {\bibfield  {journal}
  {\bibinfo  {journal} {Physical Review Fluids}\ }\textbf {\bibinfo {volume}
  {5}},\ \bibinfo {pages} {114802} (\bibinfo {year} {2020})},\ \bibinfo {note}
  {publisher: American Physical Society}\BibitemShut {NoStop}%
\bibitem [{\citenamefont {Lam}\ \emph {et~al.}(2020)\citenamefont {Lam},
  \citenamefont {Delache},\ and\ \citenamefont
  {Godeferd}}]{lam_partitioning_2020}%
  \BibitemOpen
  \bibfield  {author} {\bibinfo {author} {\bibfnamefont {H.}~\bibnamefont
  {Lam}}, \bibinfo {author} {\bibfnamefont {A.}~\bibnamefont {Delache}},\ and\
  \bibinfo {author} {\bibfnamefont {F.~S.}\ \bibnamefont {Godeferd}},\
  }\bibfield  {title} {\bibinfo {title} {Partitioning waves and eddies in
  stably stratified turbulence},\ }\href
  {https://doi.org/10.3390/atmos11040420} {\bibfield  {journal} {\bibinfo
  {journal} {Atmosphere}\ }\textbf {\bibinfo {volume} {11}},\ \bibinfo {pages}
  {420} (\bibinfo {year} {2020})},\ \bibinfo {note} {number: 4 Publisher:
  Multidisciplinary Digital Publishing Institute}\BibitemShut {NoStop}%
\bibitem [{\citenamefont {Lam}\ \emph {et~al.}(2021)\citenamefont {Lam},
  \citenamefont {Delache},\ and\ \citenamefont {Godeferd}}]{lam_energy_2021}%
  \BibitemOpen
  \bibfield  {author} {\bibinfo {author} {\bibfnamefont {H.}~\bibnamefont
  {Lam}}, \bibinfo {author} {\bibfnamefont {A.}~\bibnamefont {Delache}},\ and\
  \bibinfo {author} {\bibfnamefont {F.~S.}\ \bibnamefont {Godeferd}},\
  }\bibfield  {title} {\bibinfo {title} {Energy balance and mixing between
  waves and eddies in stably stratified turbulence},\ }\href
  {https://doi.org/10.1017/jfm.2021.589} {\bibfield  {journal} {\bibinfo
  {journal} {Journal of Fluid Mechanics}\ }\textbf {\bibinfo {volume} {923}},\
  \bibinfo {pages} {A31} (\bibinfo {year} {2021})},\ \bibinfo {note}
  {publisher: Cambridge University Press}\BibitemShut {NoStop}%
\bibitem [{\citenamefont {Brouzet}(2016)}]{brouzet_internal_2016}%
  \BibitemOpen
  \bibfield  {author} {\bibinfo {author} {\bibfnamefont {C.}~\bibnamefont
  {Brouzet}},\ }\href {https://theses.hal.science/tel-01361201} {\bibinfo
  {title} {Internal wave attractors : from geometrical focusing to non-linear
  energy cascade and mixing}} (\bibinfo {year} {2016})\BibitemShut {NoStop}%
\bibitem [{\citenamefont {Davis}\ \emph {et~al.}(2020)\citenamefont {Davis},
  \citenamefont {Jamin}, \citenamefont {Deleuze}, \citenamefont {Joubaud},\
  and\ \citenamefont {Dauxois}}]{davis_succession_2020}%
  \BibitemOpen
  \bibfield  {author} {\bibinfo {author} {\bibfnamefont {G.}~\bibnamefont
  {Davis}}, \bibinfo {author} {\bibfnamefont {T.}~\bibnamefont {Jamin}},
  \bibinfo {author} {\bibfnamefont {J.}~\bibnamefont {Deleuze}}, \bibinfo
  {author} {\bibfnamefont {S.}~\bibnamefont {Joubaud}},\ and\ \bibinfo {author}
  {\bibfnamefont {T.}~\bibnamefont {Dauxois}},\ }\bibfield  {title} {\bibinfo
  {title} {Succession of resonances to achieve internal wave turbulence},\
  }\href {https://doi.org/10.1103/PhysRevLett.124.204502} {\bibfield  {journal}
  {\bibinfo  {journal} {Physical Review Letters}\ }\textbf {\bibinfo {volume}
  {124}},\ \bibinfo {pages} {204502} (\bibinfo {year} {2020})},\ \bibinfo
  {note} {publisher: American Physical Society}\BibitemShut {NoStop}%
\bibitem [{\citenamefont {Rodda}\ \emph {et~al.}(2022)\citenamefont {Rodda},
  \citenamefont {Savaro}, \citenamefont {Davis}, \citenamefont {Reneuve},
  \citenamefont {Augier}, \citenamefont {Sommeria}, \citenamefont {Valran},
  \citenamefont {Viboud},\ and\ \citenamefont
  {Mordant}}]{rodda_experimental_2022}%
  \BibitemOpen
  \bibfield  {author} {\bibinfo {author} {\bibfnamefont {C.}~\bibnamefont
  {Rodda}}, \bibinfo {author} {\bibfnamefont {C.}~\bibnamefont {Savaro}},
  \bibinfo {author} {\bibfnamefont {G.}~\bibnamefont {Davis}}, \bibinfo
  {author} {\bibfnamefont {J.}~\bibnamefont {Reneuve}}, \bibinfo {author}
  {\bibfnamefont {P.}~\bibnamefont {Augier}}, \bibinfo {author} {\bibfnamefont
  {J.}~\bibnamefont {Sommeria}}, \bibinfo {author} {\bibfnamefont
  {T.}~\bibnamefont {Valran}}, \bibinfo {author} {\bibfnamefont
  {S.}~\bibnamefont {Viboud}},\ and\ \bibinfo {author} {\bibfnamefont
  {N.}~\bibnamefont {Mordant}},\ }\bibfield  {title} {\bibinfo {title}
  {Experimental observations of internal wave turbulence transition in a
  stratified fluid},\ }\href {https://doi.org/10.1103/PhysRevFluids.7.094802}
  {\bibfield  {journal} {\bibinfo  {journal} {Physical Review Fluids}\ }\textbf
  {\bibinfo {volume} {7}},\ \bibinfo {pages} {094802} (\bibinfo {year}
  {2022})},\ \bibinfo {note} {publisher: American Physical Society}\BibitemShut
  {NoStop}%
\bibitem [{\citenamefont {Reun}\ \emph {et~al.}(2018)\citenamefont {Reun},
  \citenamefont {Favier},\ and\ \citenamefont {Bars}}]{reun_parametric_2018}%
  \BibitemOpen
  \bibfield  {author} {\bibinfo {author} {\bibfnamefont {T.~L.}\ \bibnamefont
  {Reun}}, \bibinfo {author} {\bibfnamefont {B.}~\bibnamefont {Favier}},\ and\
  \bibinfo {author} {\bibfnamefont {M.~L.}\ \bibnamefont {Bars}},\ }\bibfield
  {title} {\bibinfo {title} {Parametric instability and wave turbulence driven
  by tidal excitation of internal waves},\ }\href
  {https://doi.org/10.1017/jfm.2018.18} {\bibfield  {journal} {\bibinfo
  {journal} {Journal of Fluid Mechanics}\ }\textbf {\bibinfo {volume} {840}},\
  \bibinfo {pages} {498} (\bibinfo {year} {2018})},\ \bibinfo {note}
  {publisher: Cambridge University Press}\BibitemShut {NoStop}%
\bibitem [{\citenamefont {Meyrand}\ \emph {et~al.}(2016)\citenamefont
  {Meyrand}, \citenamefont {Galtier},\ and\ \citenamefont
  {Kiyani}}]{meyrand_direct_2016}%
  \BibitemOpen
  \bibfield  {author} {\bibinfo {author} {\bibfnamefont {R.}~\bibnamefont
  {Meyrand}}, \bibinfo {author} {\bibfnamefont {S.}~\bibnamefont {Galtier}},\
  and\ \bibinfo {author} {\bibfnamefont {K.~H.}\ \bibnamefont {Kiyani}},\
  }\bibfield  {title} {\bibinfo {title} {Direct evidence of the transition from
  weak to strong magnetohydrodynamic turbulence},\ }\href
  {https://doi.org/10.1103/PhysRevLett.116.105002} {\bibfield  {journal}
  {\bibinfo  {journal} {Physical Review Letters}\ }\textbf {\bibinfo {volume}
  {116}},\ \bibinfo {pages} {105002} (\bibinfo {year} {2016})},\ \bibinfo
  {note} {publisher: American Physical Society}\BibitemShut {NoStop}%
\bibitem [{\citenamefont {Le~Reun}\ \emph {et~al.}(2017)\citenamefont
  {Le~Reun}, \citenamefont {Favier}, \citenamefont {Barker},\ and\
  \citenamefont {Le~Bars}}]{le_reun_inertial_2017}%
  \BibitemOpen
  \bibfield  {author} {\bibinfo {author} {\bibfnamefont {T.}~\bibnamefont
  {Le~Reun}}, \bibinfo {author} {\bibfnamefont {B.}~\bibnamefont {Favier}},
  \bibinfo {author} {\bibfnamefont {A.~J.}\ \bibnamefont {Barker}},\ and\
  \bibinfo {author} {\bibfnamefont {M.}~\bibnamefont {Le~Bars}},\ }\bibfield
  {title} {\bibinfo {title} {Inertial wave turbulence driven by elliptical
  instability},\ }\href {https://doi.org/10.1103/PhysRevLett.119.034502}
  {\bibfield  {journal} {\bibinfo  {journal} {Physical Review Letters}\
  }\textbf {\bibinfo {volume} {119}},\ \bibinfo {pages} {034502} (\bibinfo
  {year} {2017})},\ \bibinfo {note} {publisher: American Physical
  Society}\BibitemShut {NoStop}%
\bibitem [{\citenamefont {Reun}\ \emph {et~al.}(2019)\citenamefont {Reun},
  \citenamefont {Favier},\ and\ \citenamefont {Bars}}]{reun_experimental_2019}%
  \BibitemOpen
  \bibfield  {author} {\bibinfo {author} {\bibfnamefont {T.~L.}\ \bibnamefont
  {Reun}}, \bibinfo {author} {\bibfnamefont {B.}~\bibnamefont {Favier}},\ and\
  \bibinfo {author} {\bibfnamefont {M.~L.}\ \bibnamefont {Bars}},\ }\bibfield
  {title} {\bibinfo {title} {Experimental study of the nonlinear saturation of
  the elliptical instability: inertial wave turbulence versus geostrophic
  turbulence},\ }\href {https://doi.org/10.1017/jfm.2019.646} {\bibfield
  {journal} {\bibinfo  {journal} {Journal of Fluid Mechanics}\ }\textbf
  {\bibinfo {volume} {879}},\ \bibinfo {pages} {296} (\bibinfo {year}
  {2019})},\ \bibinfo {note} {publisher: Cambridge University
  Press}\BibitemShut {NoStop}%
\bibitem [{\citenamefont {Brunet}\ \emph {et~al.}(2020)\citenamefont {Brunet},
  \citenamefont {Gallet},\ and\ \citenamefont {Cortet}}]{brunet_shortcut_2020}%
  \BibitemOpen
  \bibfield  {author} {\bibinfo {author} {\bibfnamefont {M.}~\bibnamefont
  {Brunet}}, \bibinfo {author} {\bibfnamefont {B.}~\bibnamefont {Gallet}},\
  and\ \bibinfo {author} {\bibfnamefont {P.-P.}\ \bibnamefont {Cortet}},\
  }\bibfield  {title} {\bibinfo {title} {Shortcut to geostrophy in wave-driven
  rotating turbulence: The quartetic instability},\ }\href
  {https://doi.org/10.1103/PhysRevLett.124.124501} {\bibfield  {journal}
  {\bibinfo  {journal} {Physical Review Letters}\ }\textbf {\bibinfo {volume}
  {124}},\ \bibinfo {pages} {124501} (\bibinfo {year} {2020})},\ \bibinfo
  {note} {publisher: American Physical Society}\BibitemShut {NoStop}%
\bibitem [{\citenamefont {Monsalve}\ \emph {et~al.}(2020)\citenamefont
  {Monsalve}, \citenamefont {Brunet}, \citenamefont {Gallet},\ and\
  \citenamefont {Cortet}}]{monsalve_quantitative_2020}%
  \BibitemOpen
  \bibfield  {author} {\bibinfo {author} {\bibfnamefont {E.}~\bibnamefont
  {Monsalve}}, \bibinfo {author} {\bibfnamefont {M.}~\bibnamefont {Brunet}},
  \bibinfo {author} {\bibfnamefont {B.}~\bibnamefont {Gallet}},\ and\ \bibinfo
  {author} {\bibfnamefont {P.-P.}\ \bibnamefont {Cortet}},\ }\bibfield  {title}
  {\bibinfo {title} {Quantitative experimental observation of weak
  inertial-wave turbulence},\ }\href
  {https://doi.org/10.1103/PhysRevLett.125.254502} {\bibfield  {journal}
  {\bibinfo  {journal} {Physical Review Letters}\ }\textbf {\bibinfo {volume}
  {125}},\ \bibinfo {pages} {254502} (\bibinfo {year} {2020})},\ \bibinfo
  {note} {publisher: American Physical Society}\BibitemShut {NoStop}%
\bibitem [{\citenamefont {Lanchon}\ \emph {et~al.}(2023)\citenamefont
  {Lanchon}, \citenamefont {Mora}, \citenamefont {Monsalve},\ and\
  \citenamefont {Cortet}}]{lanchon_internal_2023}%
  \BibitemOpen
  \bibfield  {author} {\bibinfo {author} {\bibfnamefont {N.}~\bibnamefont
  {Lanchon}}, \bibinfo {author} {\bibfnamefont {D.~O.}\ \bibnamefont {Mora}},
  \bibinfo {author} {\bibfnamefont {E.}~\bibnamefont {Monsalve}},\ and\
  \bibinfo {author} {\bibfnamefont {P.-P.}\ \bibnamefont {Cortet}},\ }\bibfield
   {title} {\bibinfo {title} {Internal wave turbulence in a stratified fluid
  with and without eigenmodes of the experimental domain},\ }\href
  {https://doi.org/10.1103/PhysRevFluids.8.054802} {\bibfield  {journal}
  {\bibinfo  {journal} {Phys. Rev. Fluids}\ }\textbf {\bibinfo {volume} {8}},\
  \bibinfo {pages} {054802} (\bibinfo {year} {2023})}\BibitemShut {NoStop}%
\bibitem [{\citenamefont {Müller}\ \emph {et~al.}(1986)\citenamefont
  {Müller}, \citenamefont {Holloway}, \citenamefont {Henyey},\ and\
  \citenamefont {Pomphrey}}]{muller_nonlinear_1986}%
  \BibitemOpen
  \bibfield  {author} {\bibinfo {author} {\bibfnamefont {P.}~\bibnamefont
  {Müller}}, \bibinfo {author} {\bibfnamefont {G.}~\bibnamefont {Holloway}},
  \bibinfo {author} {\bibfnamefont {F.}~\bibnamefont {Henyey}},\ and\ \bibinfo
  {author} {\bibfnamefont {N.}~\bibnamefont {Pomphrey}},\ }\bibfield  {title}
  {\bibinfo {title} {Nonlinear interactions among internal gravity waves},\
  }\href {https://doi.org/10.1029/RG024i003p00493} {\bibfield  {journal}
  {\bibinfo  {journal} {Reviews of Geophysics}\ }\textbf {\bibinfo {volume}
  {24}},\ \bibinfo {pages} {493} (\bibinfo {year} {1986})},\ \bibinfo {note}
  {\_eprint:
  https://onlinelibrary.wiley.com/doi/pdf/10.1029/{RG}024i003p00493}\BibitemShut
  {NoStop}%
\bibitem [{\citenamefont {Lvov}\ \emph {et~al.}(2012)\citenamefont {Lvov},
  \citenamefont {Polzin},\ and\ \citenamefont {Yokoyama}}]{lvov_resonant_2012}%
  \BibitemOpen
  \bibfield  {author} {\bibinfo {author} {\bibfnamefont {Y.~V.}\ \bibnamefont
  {Lvov}}, \bibinfo {author} {\bibfnamefont {K.~L.}\ \bibnamefont {Polzin}},\
  and\ \bibinfo {author} {\bibfnamefont {N.}~\bibnamefont {Yokoyama}},\
  }\bibfield  {title} {\bibinfo {title} {Resonant and near-resonant internal
  wave interactions},\ }\href {https://doi.org/10.1175/2011JPO4129.1}
  {\bibfield  {journal} {\bibinfo  {journal} {Journal of Physical
  Oceanography}\ }\textbf {\bibinfo {volume} {42}},\ \bibinfo {pages} {669}
  (\bibinfo {year} {2012})},\ \bibinfo {note} {publisher: American
  Meteorological Society Section: Journal of Physical Oceanography}\BibitemShut
  {NoStop}%
\bibitem [{\citenamefont {Lvov}\ \emph {et~al.}(2010)\citenamefont {Lvov},
  \citenamefont {Polzin}, \citenamefont {Tabak},\ and\ \citenamefont
  {Yokoyama}}]{lvov_oceanic_2010}%
  \BibitemOpen
  \bibfield  {author} {\bibinfo {author} {\bibfnamefont {Y.~V.}\ \bibnamefont
  {Lvov}}, \bibinfo {author} {\bibfnamefont {K.~L.}\ \bibnamefont {Polzin}},
  \bibinfo {author} {\bibfnamefont {E.~G.}\ \bibnamefont {Tabak}},\ and\
  \bibinfo {author} {\bibfnamefont {N.}~\bibnamefont {Yokoyama}},\ }\bibfield
  {title} {\bibinfo {title} {Oceanic internal-wave field: Theory of
  scale-invariant spectra},\ }\href {https://doi.org/10.1175/2010JPO4132.1}
  {\bibfield  {journal} {\bibinfo  {journal} {Journal of Physical
  Oceanography}\ }\textbf {\bibinfo {volume} {40}},\ \bibinfo {pages} {2605}
  (\bibinfo {year} {2010})},\ \bibinfo {note} {publisher: American
  Meteorological Society Section: Journal of Physical Oceanography}\BibitemShut
  {NoStop}%
\bibitem [{\citenamefont {{McComas}}\ and\ \citenamefont
  {Bretherton}(1977)}]{mccomas_resonant_1977}%
  \BibitemOpen
  \bibfield  {author} {\bibinfo {author} {\bibfnamefont {C.~H.}\ \bibnamefont
  {{McComas}}}\ and\ \bibinfo {author} {\bibfnamefont {F.~P.}\ \bibnamefont
  {Bretherton}},\ }\bibfield  {title} {\bibinfo {title} {Resonant interaction
  of oceanic internal waves},\ }\href {https://doi.org/10.1029/JC082i009p01397}
  {\bibfield  {journal} {\bibinfo  {journal} {Journal of Geophysical Research
  (1896-1977)}\ }\textbf {\bibinfo {volume} {82}},\ \bibinfo {pages} {1397}
  (\bibinfo {year} {1977})},\ \bibinfo {note} {\_eprint:
  https://onlinelibrary.wiley.com/doi/pdf/10.1029/{JC}082i009p01397}\BibitemShut
  {NoStop}%
\bibitem [{\citenamefont {{MacKinnon}}\ \emph {et~al.}(2013)\citenamefont
  {{MacKinnon}}, \citenamefont {Alford}, \citenamefont {Sun}, \citenamefont
  {Pinkel}, \citenamefont {Zhao},\ and\ \citenamefont
  {Klymak}}]{mackinnon_parametric_2013}%
  \BibitemOpen
  \bibfield  {author} {\bibinfo {author} {\bibfnamefont {J.~A.}\ \bibnamefont
  {{MacKinnon}}}, \bibinfo {author} {\bibfnamefont {M.~H.}\ \bibnamefont
  {Alford}}, \bibinfo {author} {\bibfnamefont {O.}~\bibnamefont {Sun}},
  \bibinfo {author} {\bibfnamefont {R.}~\bibnamefont {Pinkel}}, \bibinfo
  {author} {\bibfnamefont {Z.}~\bibnamefont {Zhao}},\ and\ \bibinfo {author}
  {\bibfnamefont {J.}~\bibnamefont {Klymak}},\ }\bibfield  {title} {\bibinfo
  {title} {Parametric subharmonic instability of the internal tide at
  29${}^\circ$n},\ }\href {https://doi.org/10.1175/JPO-D-11-0108.1} {\bibfield
  {journal} {\bibinfo  {journal} {Journal of Physical Oceanography}\ }\textbf
  {\bibinfo {volume} {43}},\ \bibinfo {pages} {17} (\bibinfo {year} {2013})},\
  \bibinfo {note} {publisher: American Meteorological Society Section: Journal
  of Physical Oceanography}\BibitemShut {NoStop}%
\bibitem [{\citenamefont {Nazarenko}\ and\ \citenamefont
  {Schekochihin}(2011)}]{nazarenko_critical_2011}%
  \BibitemOpen
  \bibfield  {author} {\bibinfo {author} {\bibfnamefont {S.~V.}\ \bibnamefont
  {Nazarenko}}\ and\ \bibinfo {author} {\bibfnamefont {A.~A.}\ \bibnamefont
  {Schekochihin}},\ }\bibfield  {title} {\bibinfo {title} {Critical balance in
  magnetohydrodynamic, rotating and stratified turbulence: towards a universal
  scaling conjecture},\ }\href {https://doi.org/10.1017/S002211201100067X}
  {\bibfield  {journal} {\bibinfo  {journal} {Journal of Fluid Mechanics}\
  }\textbf {\bibinfo {volume} {677}},\ \bibinfo {pages} {134} (\bibinfo {year}
  {2011})},\ \bibinfo {note} {publisher: Cambridge University
  Press}\BibitemShut {NoStop}%
\bibitem [{\citenamefont {Craya}(1957)}]{craya_contribution_1957}%
  \BibitemOpen
  \bibfield  {author} {\bibinfo {author} {\bibfnamefont {A.}~\bibnamefont
  {Craya}},\ }\href {https://tel.archives-ouvertes.fr/tel-00684659} {\bibinfo
  {title} {Contribution à l'analyse de la turbulence associée à des vitesses
  moyennes}} (\bibinfo {year} {1957})\BibitemShut {NoStop}%
\bibitem [{\citenamefont {Lvov}\ and\ \citenamefont
  {Yokoyama}(2009)}]{lvov_nonlinear_2009}%
  \BibitemOpen
  \bibfield  {author} {\bibinfo {author} {\bibfnamefont {Y.~V.}\ \bibnamefont
  {Lvov}}\ and\ \bibinfo {author} {\bibfnamefont {N.}~\bibnamefont
  {Yokoyama}},\ }\bibfield  {title} {\bibinfo {title} {Nonlinear wave–wave
  interactions in stratified flows: Direct numerical simulations},\ }\href
  {https://doi.org/10.1016/j.physd.2009.01.016} {\bibfield  {journal} {\bibinfo
   {journal} {Physica D: Nonlinear Phenomena}\ }\textbf {\bibinfo {volume}
  {238}},\ \bibinfo {pages} {803} (\bibinfo {year} {2009})}\BibitemShut
  {NoStop}%
\bibitem [{\citenamefont {Mohanan}\ \emph
  {et~al.}(2019{\natexlab{a}})\citenamefont {Mohanan}, \citenamefont {Bonamy},
  \citenamefont {Linares},\ and\ \citenamefont
  {Augier}}]{mohanan_fluidsim_2019}%
  \BibitemOpen
  \bibfield  {author} {\bibinfo {author} {\bibfnamefont {A.~V.}\ \bibnamefont
  {Mohanan}}, \bibinfo {author} {\bibfnamefont {C.}~\bibnamefont {Bonamy}},
  \bibinfo {author} {\bibfnamefont {M.~C.}\ \bibnamefont {Linares}},\ and\
  \bibinfo {author} {\bibfnamefont {P.}~\bibnamefont {Augier}},\ }\bibfield
  {title} {\bibinfo {title} {{FluidSim}: Modular, object-oriented python
  package for high-performance {CFD} simulations},\ }\href
  {https://doi.org/10.5334/jors.239} {\bibfield  {journal} {\bibinfo  {journal}
  {Journal of Open Research Software}\ }\textbf {\bibinfo {volume} {7}},\
  \bibinfo {pages} {14} (\bibinfo {year} {2019}{\natexlab{a}})},\ \bibinfo
  {note} {number: 1 Publisher: Ubiquity Press}\BibitemShut {NoStop}%
\bibitem [{\citenamefont {Augier}\ \emph {et~al.}(2019)\citenamefont {Augier},
  \citenamefont {Mohanan},\ and\ \citenamefont {Bonamy}}]{fluiddyn}%
  \BibitemOpen
  \bibfield  {author} {\bibinfo {author} {\bibfnamefont {P.}~\bibnamefont
  {Augier}}, \bibinfo {author} {\bibfnamefont {A.~V.}\ \bibnamefont
  {Mohanan}},\ and\ \bibinfo {author} {\bibfnamefont {C.}~\bibnamefont
  {Bonamy}},\ }\bibfield  {title} {\bibinfo {title} {{FluidDyn}: A python
  open-source framework for research and teaching in fluid dynamics by
  simulations, experiments and data processing},\ }\bibfield  {journal}
  {\bibinfo  {journal} {Journal of Open Research Software}\ }\textbf {\bibinfo
  {volume} {7}},\ \href {https://doi.org/10.5334/jors.237} {10.5334/jors.237}
  (\bibinfo {year} {2019})\BibitemShut {NoStop}%
\bibitem [{\citenamefont {Mohanan}\ \emph
  {et~al.}(2019{\natexlab{b}})\citenamefont {Mohanan}, \citenamefont {Bonamy},\
  and\ \citenamefont {Augier}}]{fluidfft}%
  \BibitemOpen
  \bibfield  {author} {\bibinfo {author} {\bibfnamefont {A.~V.}\ \bibnamefont
  {Mohanan}}, \bibinfo {author} {\bibfnamefont {C.}~\bibnamefont {Bonamy}},\
  and\ \bibinfo {author} {\bibfnamefont {P.}~\bibnamefont {Augier}},\
  }\bibfield  {title} {\bibinfo {title} {{FluidFFT}: Common {API} ({C}$++$ and
  {P}ython) for fast fourier transform {HPC} libraries},\ }\bibfield  {journal}
  {\bibinfo  {journal} {Journal of Open Research Software}\ }\textbf {\bibinfo
  {volume} {7}},\ \href {https://doi.org/10.5334/jors.238} {10.5334/jors.238}
  (\bibinfo {year} {2019}{\natexlab{b}})\BibitemShut {NoStop}%
\bibitem [{\citenamefont {de~Bruyn~Kops}\ and\ \citenamefont
  {Riley}(1998)}]{deBruynKops1998}%
  \BibitemOpen
  \bibfield  {author} {\bibinfo {author} {\bibfnamefont {S.}~\bibnamefont
  {de~Bruyn~Kops}}\ and\ \bibinfo {author} {\bibfnamefont {J.}~\bibnamefont
  {Riley}},\ }\bibfield  {title} {\bibinfo {title} {Direct numerical simulation
  of laboratory experiments in isotropic turbulence},\ }\href@noop {}
  {\bibfield  {journal} {\bibinfo  {journal} {Physics of Fluids}\ }\textbf
  {\bibinfo {volume} {10}},\ \bibinfo {pages} {2125} (\bibinfo {year}
  {1998})}\BibitemShut {NoStop}%
\bibitem [{\citenamefont {Smith}\ and\ \citenamefont
  {Waleffe}(2002)}]{smith_generation_2002}%
  \BibitemOpen
  \bibfield  {author} {\bibinfo {author} {\bibfnamefont {L.~M.}\ \bibnamefont
  {Smith}}\ and\ \bibinfo {author} {\bibfnamefont {F.}~\bibnamefont
  {Waleffe}},\ }\bibfield  {title} {\bibinfo {title} {Generation of slow large
  scales in forced rotating stratified turbulence},\ }\href
  {https://doi.org/10.1017/S0022112001006309} {\bibfield  {journal} {\bibinfo
  {journal} {Journal of Fluid Mechanics}\ }\textbf {\bibinfo {volume} {451}},\
  \bibinfo {pages} {145} (\bibinfo {year} {2002})},\ \bibinfo {note}
  {publisher: Cambridge University Press}\BibitemShut {NoStop}%
\bibitem [{\citenamefont {Laval}\ \emph {et~al.}(2003)\citenamefont {Laval},
  \citenamefont {{McWilliams}},\ and\ \citenamefont
  {Dubrulle}}]{laval_forced_2003}%
  \BibitemOpen
  \bibfield  {author} {\bibinfo {author} {\bibfnamefont {J.-P.}\ \bibnamefont
  {Laval}}, \bibinfo {author} {\bibfnamefont {J.~C.}\ \bibnamefont
  {{McWilliams}}},\ and\ \bibinfo {author} {\bibfnamefont {B.}~\bibnamefont
  {Dubrulle}},\ }\bibfield  {title} {\bibinfo {title} {Forced stratified
  turbulence: Successive transitions with reynolds number},\ }\href
  {https://doi.org/10.1103/PhysRevE.68.036308} {\bibfield  {journal} {\bibinfo
  {journal} {Physical Review E}\ }\textbf {\bibinfo {volume} {68}},\ \bibinfo
  {pages} {036308} (\bibinfo {year} {2003})},\ \bibinfo {note} {publisher:
  American Physical Society}\BibitemShut {NoStop}%
\bibitem [{\citenamefont {Godeferd}\ \emph {et~al.}(2010)\citenamefont
  {Godeferd}, \citenamefont {Delache},\ and\ \citenamefont
  {Cambon}}]{godeferd_toroidalpoloidal_2010}%
  \BibitemOpen
  \bibfield  {author} {\bibinfo {author} {\bibfnamefont {F.~S.}\ \bibnamefont
  {Godeferd}}, \bibinfo {author} {\bibfnamefont {A.}~\bibnamefont {Delache}},\
  and\ \bibinfo {author} {\bibfnamefont {C.}~\bibnamefont {Cambon}},\
  }\bibfield  {title} {\bibinfo {title} {Toroidal/poloidal modes dynamics in
  anisotropic turbulence},\ }in\ \href
  {https://doi.org/10.1007/978-3-642-14139-3_18} {\emph {\bibinfo {booktitle}
  {Turbulence and Interactions}}},\ \bibinfo {series and number} {Notes on
  Numerical Fluid Mechanics and Multidisciplinary Design},\ \bibinfo {editor}
  {edited by\ \bibinfo {editor} {\bibfnamefont {M.}~\bibnamefont {Deville}},
  \bibinfo {editor} {\bibfnamefont {T.-H.}\ \bibnamefont {Lê}},\ and\ \bibinfo
  {editor} {\bibfnamefont {P.}~\bibnamefont {Sagaut}}}\ (\bibinfo  {publisher}
  {Springer},\ \bibinfo {year} {2010})\ pp.\ \bibinfo {pages}
  {151--158}\BibitemShut {NoStop}%
\bibitem [{\citenamefont {Nikurashin}\ and\ \citenamefont
  {Legg}(2011)}]{nikurashin_legg_mechanism_2011}%
  \BibitemOpen
  \bibfield  {author} {\bibinfo {author} {\bibfnamefont {M.}~\bibnamefont
  {Nikurashin}}\ and\ \bibinfo {author} {\bibfnamefont {S.}~\bibnamefont
  {Legg}},\ }\bibfield  {title} {\bibinfo {title} {A mechanism for local
  dissipation of internal tides generated at rough topography},\ }\href
  {https://doi.org/https://doi.org/10.1175/2010JPO4522.1} {\bibfield  {journal}
  {\bibinfo  {journal} {Journal of Physical Oceanography}\ }\textbf {\bibinfo
  {volume} {41}},\ \bibinfo {pages} {378 } (\bibinfo {year}
  {2011})}\BibitemShut {NoStop}%
\bibitem [{\citenamefont {Bartello}(1995)}]{bartello_geostrophic_1995}%
  \BibitemOpen
  \bibfield  {author} {\bibinfo {author} {\bibfnamefont {P.}~\bibnamefont
  {Bartello}},\ }\bibfield  {title} {\bibinfo {title} {Geostrophic adjustment
  and inverse cascades in rotating stratified turbulence},\ }\href
  {https://doi.org/https://doi.org/10.1175/1520-0469(1995)052<4410:GAAICI>2.0.CO;2}
  {\bibfield  {journal} {\bibinfo  {journal} {Journal of Atmospheric Sciences}\
  }\textbf {\bibinfo {volume} {52}},\ \bibinfo {pages} {4410 } (\bibinfo {year}
  {1995})}\BibitemShut {NoStop}%
\bibitem [{\citenamefont {Linares}(2020)}]{linares_numerical_2020}%
  \BibitemOpen
  \bibfield  {author} {\bibinfo {author} {\bibfnamefont {M.~C.}\ \bibnamefont
  {Linares}},\ }\href {https://tel.archives-ouvertes.fr/tel-02612797} {\bibinfo
  {title} {Numerical study of 2d stratified turbulence forced by internal
  gravity waves}} (\bibinfo {year} {2020})\BibitemShut {NoStop}%
\bibitem [{\citenamefont {Cerri}\ \emph {et~al.}(2022)\citenamefont {Cerri},
  \citenamefont {Passot}, \citenamefont {Laveder}, \citenamefont {Sulem},\ and\
  \citenamefont {Kunz}}]{cerri_turbulent_2022}%
  \BibitemOpen
  \bibfield  {author} {\bibinfo {author} {\bibfnamefont {S.~S.}\ \bibnamefont
  {Cerri}}, \bibinfo {author} {\bibfnamefont {T.}~\bibnamefont {Passot}},
  \bibinfo {author} {\bibfnamefont {D.}~\bibnamefont {Laveder}}, \bibinfo
  {author} {\bibfnamefont {P.-L.}\ \bibnamefont {Sulem}},\ and\ \bibinfo
  {author} {\bibfnamefont {M.~W.}\ \bibnamefont {Kunz}},\ }\href
  {https://doi.org/10.48550/arXiv.2207.04301} {\bibinfo {title} {Turbulent
  regimes in collisions of 3d alfv{\textbackslash}'en-wave packets}} (\bibinfo
  {year} {2022}),\ \Eprint {https://arxiv.org/abs/2207.04301 [astro-ph,
  physics:physics]} {2207.04301 [astro-ph, physics:physics]} \BibitemShut
  {NoStop}%
\bibitem [{\citenamefont {Garrett}\ and\ \citenamefont
  {Munk}(1979)}]{garrett_internal_1979}%
  \BibitemOpen
  \bibfield  {author} {\bibinfo {author} {\bibfnamefont {C.}~\bibnamefont
  {Garrett}}\ and\ \bibinfo {author} {\bibfnamefont {W.}~\bibnamefont {Munk}},\
  }\bibfield  {title} {\bibinfo {title} {Internal waves in the ocean},\ }\href
  {https://doi.org/10.1146/annurev.fl.11.010179.002011} {\bibfield  {journal}
  {\bibinfo  {journal} {Annual Review of Fluid Mechanics}\ }\textbf {\bibinfo
  {volume} {11}},\ \bibinfo {pages} {339} (\bibinfo {year} {1979})},\ \bibinfo
  {note} {\_eprint:
  https://doi.org/10.1146/annurev.fl.11.010179.002011}\BibitemShut {NoStop}%
\bibitem [{\citenamefont {Herbert}\ \emph {et~al.}(2016)\citenamefont
  {Herbert}, \citenamefont {Marino}, \citenamefont {Rosenberg},\ and\
  \citenamefont {Pouquet}}]{herbert_waves_2016}%
  \BibitemOpen
  \bibfield  {author} {\bibinfo {author} {\bibfnamefont {C.}~\bibnamefont
  {Herbert}}, \bibinfo {author} {\bibfnamefont {R.}~\bibnamefont {Marino}},
  \bibinfo {author} {\bibfnamefont {D.}~\bibnamefont {Rosenberg}},\ and\
  \bibinfo {author} {\bibfnamefont {A.}~\bibnamefont {Pouquet}},\ }\bibfield
  {title} {\bibinfo {title} {Waves and vortices in the inverse cascade regime
  of stratified turbulence with or without rotation},\ }\href
  {https://doi.org/10.1017/jfm.2016.581} {\bibfield  {journal} {\bibinfo
  {journal} {Journal of Fluid Mechanics}\ }\textbf {\bibinfo {volume} {806}},\
  \bibinfo {pages} {165} (\bibinfo {year} {2016})},\ \bibinfo {note}
  {publisher: Cambridge University Press}\BibitemShut {NoStop}%
\bibitem [{\citenamefont {Waite}(2013)}]{waite_potential_2013}%
  \BibitemOpen
  \bibfield  {author} {\bibinfo {author} {\bibfnamefont {M.~L.}\ \bibnamefont
  {Waite}},\ }\bibfield  {title} {\bibinfo {title} {Potential enstrophy in
  stratified turbulence},\ }\href {https://doi.org/10.1017/jfm.2013.150}
  {\bibfield  {journal} {\bibinfo  {journal} {Journal of Fluid Mechanics}\
  }\textbf {\bibinfo {volume} {722}},\ \bibinfo {pages} {R4} (\bibinfo {year}
  {2013})}\BibitemShut {NoStop}%
\end{thebibliography}%
\end{document}